\documentclass[a4paper,11pt]{article}
\usepackage[a4paper,left=0.99in, right=0.99in,top=1.2in, bottom=1.2in]{geometry}

\usepackage{amsmath}
\usepackage{amssymb}
\usepackage{hyperref}
\usepackage{graphicx}
\usepackage{color}
\usepackage{cite}
\usepackage[OT4]{fontenc}
\usepackage[utf8x]{inputenc}
\usepackage{xspace}   
\usepackage{fixmath}
\usepackage{bm,braket}
\usepackage{slashed}

\definecolor{red}{rgb}{1,0,0}

\newcommand{\unitop}{{\boldmath{1\!\!1}}}

\newcommand{\shat}{{\hat s}}
\newcommand{\ttbar}{\ensuremath{t \bar t}\xspace}
\newcommand{\qcd}{\text{QCD}}

\newcommand{\nbar}{\ensuremath{\bar{n}}}
\newcommand{\qqbar}{{\ensuremath{q \bar q}}\xspace}
\newcommand{\calP}{{\cal P}}
\newcommand{\calM}{{\cal M}}
\newcommand{\calI}{{\cal I}}
\newcommand{\calW}{{\cal W}}
\newcommand{\calR}{{\cal R}}

\newcommand{\as}{{\alpha_s}}
\newcommand{\asp}{\frac{\as}{4\pi}}
\newcommand{\qbar}{{\bar q}}
\newcommand{\bfS}{\bm{S}}

\newcommand{\iibar}{{i \bar i}}
\newcommand{\ie}{{\it i.e. }}
\newcommand{\cf}{{\it c.f. }}
\newcommand{\eg}{{\it e.g. }}
\newcommand{\bfZ}{\bm{Z}}
\newcommand{\bare}{\text{bare}}
\newcommand{\msbar}{\ensuremath{\overline{\text{MS}}}\xspace}
\newcommand{\LT}{L_\perp}
\newcommand{\xT}{x_\perp}
\newcommand{\bfgamma}{\bm{\gamma}}
\newcommand{\bfI}{\bm{1}}
\newcommand{\bfGamma}{\bm{\Gamma}}
\newcommand{\order}[1]{{{\cal O}\left(#1\right)}}
\newcommand{\idop}{{1\hspace{-4pt} 1}}
\newcommand{\bfw}{\bm{w}}

\newcommand{\kp}{k_+}

\newcommand{\lp}{l_+}
\newcommand{\Li}{{\text{Li}}}
\newcommand{\Litwo}{{\text{Li}_2}}
\newcommand{\Lithree}{{\text{Li}_3}}
\newcommand{\gdp}{\text{Graph}(n_i, k, l)}

\def\cT{\bm{\mathrm{T}}}
\def\cm{\mathcal{M}}
\def\la{\langle}
\def\ra{\rangle}
\newcommand{\im}{\ensuremath{\text{Im}}\xspace}
\def \pipj{(p_i\cdot p_j)}
\def \pip{(p_i\cdot p)}
\def \pjp{(p_j\cdot p)}
\def \ai{\alpha_i}
\def \aj{\alpha_j}
\newcommand{\iep}{\ensuremath{i \epsilon}}
\newcommand{\betaIJ}{\beta_{IJ}}
\newcommand{\vIJ}{v_{IJ}}
\newcommand{\sIJ}{s_{IJ}}

\numberwithin{equation}{section}

\title{
\vskip 40pt
NNLO soft function for top quark pair production \\
at small transverse momentum
}
\author{
  Ren\'e \'Angeles-Mart\'inez$^{1}$,  Michał Czakon$^{2}$ and
  Sebastian Sapeta$^{1}$ \\\\
  $^1$ {\it Institute of Nuclear Physics, Polish Academy of Sciences} \\
       {\it ul.\ Radzikowskiego 152, 31-342 Krak\'ow, Poland}\\\\
  $^2$ {\it Institute for Theoretical Particle Physics and Cosmology}\\
       {\it RWTH Aachen University, Aachen, Germany}
}

\date{}

\begin{document}
\maketitle

\thispagestyle{empty}

\vspace{-28em}
\begin{flushright}
  IFJPAN-IV-2018-15\\
  TTK-18-35
\end{flushright}

\vspace{25em}

\begin{abstract}
The cross section for top quark pair production factorizes at small transverse
momentum of the heavy quark pair, $q_T$. One of the key ingredients that appears
in the factorization formula is the soft function, which mediates soft gluon
exchanges between particles and gives rise to colour correlations.
We present the complete result for the small-$q_T$ soft
function at the next-to-next-to-leading order. This is the last missing
element needed to calculate the NNLO cross section for top quark pair
production by means of the $q_T$ slicing method.
In order to evaluate divergent integrals appearing in the calculation, we
develop methods based on sector decomposition and differential equations. 
We present an extensive validation of our framework. In particular, we recover
results predicted by the renormalization group, which constitutes a~direct
demonstration of validity of the small-$q_T$ factorization at NNLO. 
We provide complete results for the real and imaginary part of the soft
function, which are ready for application in the calculation of the $t\bar t$
cross section at NNLO.
\end{abstract}

\newpage
\setcounter{tocdepth}{2}
\tableofcontents

\section{Introduction}

One of the most important classes of measurements studied at the Large Hadron
Collider~(LHC) concern processes which involve the production of the top quark. Of
particular interest is top-anti-top production, which is relevant both in studies of
properties of the Standard Model, as well as in searches for new physics, as it
forms significant backgrounds to many signatures~\cite{Husemann:2017eka}.

In the Standard Model, the top quark is the heaviest particle and it does not
form bound states, but decays immediately to the $W$ boson and the bottom quark. 
As the heaviest quark, it also couples most strongly to the Higgs boson, and
therefore plays an important role in the mechanism of electroweak symmetry
breaking.
Top quark pair production enters into pole mass extraction
as well as determination of gluon PDFs. Precise predictions for top production
allow one also to study rare decays, like those happening through
flavour-changing neutral currents, which are predicted to be very small in the
Standard Model. The value of the top mass also has an impact on 
the question of stability of the vacuum~\cite{Buttazzo:2013uya}.

The experiments at the LHC have observed millions of top
quarks and more will be detected in the upcoming runs, including the one
with high luminosity.
The increasing accuracy of experimental data from the LHC is in many
cases superior to that of theoretical predictions and this presents a challenge
for theory.

The cross section for top pair production is currently known up to
next-to-next-to-leading order~(NNLO) in Quantum Chromodynamics~(QCD), both total
and differential~\cite{Czakon:2012pz, Czakon:2012zr,
Czakon:2013goa, Czakon:2015owf, Czakon:2016ckf}. 
This, single, complete, NNLO result has been obtained with the
approach based on STRIPPER~\cite{Czakon:2010td, Czakon:2011ve, Czakon:2014oma}.
There exist also several partial results including NNLO corrections to the
off-diagonal channels obtained with the $q_T$ slicing
method~\cite{Bonciani:2015sha}, leading-colour NNLO correction to the \qqbar
channel calculated with the antenna subtraction~\cite{Abelof:2015lna}, as well as
approximate NNLO correction including semi-leptonic decays in the narrow-width
approximation~\cite{Broggio:2014yca}.
Besides, the electroweak corrections for top pair production are known up to
NLO~\cite{Beenakker:1993yr,  Bernreuther:2006vg}, and they were also combined
with the NNLO QCD corrections~\cite{Czakon:2017wor}.

In addition to the fixed-order results, a number of resummed cross sections for
our process of interest have been calculated. In particular, soft
gluons have been resummed at threshold up to next-to-next-to-leading
logarithmic~(NNLL) accuracy~\cite{Czakon:2009zw, Beneke:2009rj, Ahrens:2010zv,
Kidonakis:2010dk, Cacciari:2011hy}. Also, combined resummation of soft and
small-mass logarithms at NNLL has been
performed~\cite{Pecjak:2016nee, Czakon:2018nun}.  Soft and Coulomb gluons have
also been resummed simultaneously~\cite{Piclum:2018ndt}.
Finally, the top quark production cross section has been resummed in the
small-$q_T$ limit up to NNLL~\cite{Li:2013mia, Catani:2014qha, Catani:2018mei}.

Given the complexity of the NNLO calculation for top pair production, a second,
independent result is highly desirable. One of the most promising methods
that could be used to obtain it is the $q_T$ slicing
approach~\cite{Catani:2007vq}. 

Consider the process $  h_1 + h_2 \to F\, {\textstyle (q_T)} + X$,
where two hadrons, $h_1$ and $h_2$, collide and produce an object $F$, which is
registered in a detector, together with an undetected QCD radiation $X$. Then,
the cross section at order N$^m$LO can be written as
a sum of two components
\begin{equation}
  \frac{\sigma_{{\rm N}^m{\rm LO}}^{F}}{d\Phi} = 
  \int^{q_{T{\rm cut}}}_0 
  \!\!\!\!\! d q_T\,
  \frac{d\sigma^{F}_{{\rm N}^m{\rm LO}}}{d\Phi dq_T} +
  \int_{q_{T{\rm cut}}}^\infty 
  \!\!\!\!\! d q_T\,
  \frac{d\sigma^{F + {\rm jet}}_{{\rm N}^{m-1}{\rm LO}}}{d\Phi dq_T}\,,
  \label{eq:qT-slicing}
\end{equation}
each of which is separately finite.
The advantage of this approach is that the second term in the above equation,
which represents resolved emissions, is required only at the N$^{m-1}$LO
accuracy, and it is already known in most relevant cases.  
On the contrary, the first term in Eq.~(\ref{eq:qT-slicing}), which combines
virtual and unresolved real corrections, is usually unknown. However, it is
needed only in the small-$q_T$ approximation.

In order to calculate the latter, we can use the Soft Collinear Effective
Theory~(SCET)~\cite{Becher:2014oda}, in which the cross section factorizes at
small $q_T$. 
SCET is an effective theory derived from QCD by expanding diagrams around
low-energy scales related to soft and collinear particle emissions. It is based
on the strategy of regions~\cite{Jantzen:2011nz} and it leads to representing a
single QCD field by a set of fields corresponding to collinear, anti-collinear
and soft radiation. The hard degrees of freedom are integrated out into Wilson
coefficients, which are then used to adjust couplings of the effective
theory. The new fields decouple in the Lagrangian and this separation largely
facilitates proofs of factorization theorems. 

One of such factorizations~\cite{Li:2013mia} lies at the basis of the formalism
used in our calculation, and the only missing piece needed to use it to evaluate
the cross section for top pair production at the next-to-next-to-leading order
is the NNLO, small-$q_T$ soft function. 
The result for the latter is presented in this work.

Our result, together with the framework and tools developed to obtain it, form
key  elements of an alternative calculation of the complete NNLO cross section
for the top pair production. They also make up an essential step towards
extending the $q_T$ slicing  method to N$^3$LO.

We note that our calculation
shares many features with that of the NNLO soft function for top pair production
in the threshold limit~\cite{Ferroglia:2012uy, Wang:2018vgu}. However, the
result for the latter is not of direct use in the context of $q_T$ slicing.

The paper is organized as follows. In Section~\ref{sec:formalism}, we introduce,
in detail, all concepts relevant for our calculation. In particular, we define
the variables, discuss the factorization and define the small-$q_T$ soft
function. We elaborate on the colour algebra and introduce the idea of
multiplicative renormalization of the soft function. In Section~\ref{sec:NLOsf},
we present the LO and NLO soft function, the latter up to the order $\epsilon$,
which is necessary for renormalization of the NNLO soft function.
Section~\ref{sec:methods} is devoted to a detailed description of the
calculation of the bare NNLO soft function. We discuss diagrams which
have to be included and we explain the methods which we developed to evaluate
all the divergent integrals. Finally, in Section~\ref{sec:results}, we present
the results for the complete, small-$q_T$ NNLO soft function up to the order
$\epsilon^0$. 
There, we also validate our framework by comparing the results from direct
calculation to predictions from the renormalization group, and by comparing
the results for a sub-class of NNLO graphs obtained with two two different
methods.
In that section, we also show the results for the NNLO soft function after
renormalization.
Our findings are summarized in Section~\ref{sec:summary}.

\section{The formalism}
\label{sec:formalism}

We consider the hadronic process
\begin{equation}
  h_1(P_1) + h_2(P_2) \to t(p_3) + \bar t(p_4) + X(p_X)\,,
  \label{eq:procdef}
\end{equation}
with the leading-order, partonic subprocesses
\begin{eqnarray}
  q(p_1) + \bar q(p_2) & \to & t(p_3) + \bar t(p_4)\,, \\
  g(p_1) + g(p_2)      & \to & t(p_3) + \bar t(p_4)\,,
\end{eqnarray}
where $p_1 = \xi_1 P_1$ and $p_2 = \xi_2 P_2$.
In Eq.~(\ref{eq:procdef}), $X$ represents undetected QCD radiation.

\subsection{Kinematics and notation}
\label{sec:kinematics}
We define the following variables
\begin{equation}
  {\renewcommand{\arraystretch}{2.0}%
  \begin{array}{ccc}
  s = (P_1 + P_2)^2\,,   &\hspace{30pt} & \hat s = (p_1 + p_2)^2\,, \\
  u_1 = (p_1-p_4) - m_t^2\,,&\hspace{30pt} & t_1 = (p_1-p_3)^2 - m_t^2\,,\\
  M^2 = (p_3 + p_4)^2\,, &\hspace{30pt} &  
  y = \displaystyle \frac12 \ln\frac{\xi_1}{\xi_2}\,,
  \label{eq:kin-var}
  \end{array}
  }
\end{equation}
where $q_T$ is the transverse momentum of the \ttbar pair, $y$ is its rapidity,
and $m_t$ is the top quark mass.
The small-transverse momentum limit is defined as
\begin{equation}
  \shat, M^2, |t_1|, |u_1|, m_t^2 \gg q_T^2 \gg \Lambda^2_\qcd\,.
\end{equation}

We carry out our calculations consistently in $d = 4 -2 \epsilon$ dimensions.
It is convenient to introduce the following vectors
\begin{equation}
  n = (1,0_\perp^{(d-2)},1)\,,\quad 
  \nbar = (1,0_\perp^{(d-2)},-1)\,, \qquad n\cdot \nbar = 2\,,
  \qquad n^2 = \nbar^2 = 0\,,
  \label{eq:nnbar-defs}
\end{equation}
which point towards directions of the colliding partons, and
\begin{eqnarray}
  k^\mu 
  & =  &
  n\cdot k \frac{\nbar^\mu}{2} + \nbar \cdot k \frac{n^\mu}{2} + 
  k_\perp^\mu \\
  & \equiv & 
  k^- \frac{\nbar^\mu}{2} +  k^+ \frac{n^\mu}{2} + k_\perp^\mu\,.
\end{eqnarray}
Notice that $k_\perp$ is a $d$-vector, although with pure $d-2$-dimensional
transverse part.

The momenta of the incoming, $p_1$ and $p_2$, and the outgoing, $p_3$ and $p_4$,
partons, can be written as
%
%
\begin{subequations}
  \label{eq:external-momenta}
  \begin{align}
    \label{eq:light-momenta}
    p_{1}^\mu &= \frac{\sqrt{\hat s}}{2}\, n\,,     \quad \qquad \qquad
    p_{2}^\mu  = \frac{\sqrt{\hat s}}{2}\, \nbar\,,
    \\[0.3em]
    p_i^\mu & = m_t\, v_i^\mu + k_i^\mu\,,\qquad  \
    v_i^2 = 1\,,\qquad i=3,4\,,
    \label{eq:hq-momenta}
  \end{align}
\end{subequations}
where $k_i^\mu$ is a residual momentum which scales like the soft mode 
$k_i^\mu \sim \lambda = q_T/M \ll 1$.
The total $d$-momentum of the \ttbar pair reads
\begin{equation}
  q = p_3 + p_4\,.
\end{equation}
We also introduce 
\begin{equation}
  \beta_t  =  \sqrt{1-\frac{4 m_t^2}{M^2}}\,,
  \qquad
  \text{and}
  \qquad
  x_s  = \frac{1-\beta_t}{1+\beta_t}\,.
  \label{eq:beta-def}
\end{equation}
The variable $\beta_t$ is related to the relative velocity of the
top and anti-top (or, equivalently, velocity of the top quark in the \ttbar rest
frame)~\cite{Ferroglia:2009ii} 
%
%
\begin{equation}
  |\vec v_t - \vec v_{\bar t}| = 2 \beta_t\,.
\end{equation}

We will also work in coordinate space, where the position of the \ttbar pair is
given by 
\begin{equation}
  x^\mu = 
  x^- \frac{\nbar^\mu}{2} +  x^+ \frac{n^\mu}{2} + x_\perp^\mu\,,
\end{equation}
where $x_\perp$ is a $d$-dimensional vector with purely transverse part and the
length $x_T^2 = -x_\perp^2$.
It is useful to express its norm through the following logarithm
\begin{equation}
  L_\perp = \ln\frac{x_T^2\mu^2}{4\, e^{-2\gamma_E}}\,.
  \label{eq:LTdef}
\end{equation}

The soft function is invariant with respect to rescalings $v_i \to
\kappa_i\, v_i$, where $\kappa_i$s are arbitrary constants. It turns out to be
convenient to use this property and redefine the vectors $v_3$ and $v_4$ with
slightly different normalization
\begin{equation}
  \tilde v_3 = \sqrt{1-\beta_t^2}\; v_3\,,
  \qquad \qquad
  \tilde v_4 = \sqrt{1-\beta_t^2}\; v_4\,.
  \label{eq:v3v4-tilde-def}
\end{equation}

Unless stated otherwise, we shall use the following parametrizations for the
$d$-vectors
\begin{subequations}
  \label{eq:kparam}
  \begin{align}
  \tilde v_3 & =  (1,0_\perp^{(d-3)},\beta_t\sin\theta,\beta_t\cos\theta)\,, \\
  \tilde v_4 & =  (1,0_\perp^{(d-3)},-\beta_t\sin\theta,-\beta_t\cos\theta)\,,\\
  k & =  k_0\, (1,0_\perp^{(d-4)},\sin\theta_1\sin\theta_2,
               \sin\theta_1\cos\theta_2,\cos\theta_1)\,, \\
  l & =  l_0\, (1,0_\perp^{(d-4)},\sin\chi_1\sin\chi_2,
               \sin\chi_1\cos\chi_2,\cos\chi_1)\,, 
  \end{align}
\end{subequations}
where $k$ and $l$ are the momenta of the two soft gluons radiated at NNLO.
The scalar products involving the above vectors read
\allowdisplaybreaks
\begin{eqnarray}
  \tilde v_3^2  =  \tilde v_4^2 & = & 1-\beta_t^2\,,
  \\
  \tilde v_3 \cdot \tilde v_4  & = & 1+\beta_t^2\,, \\
  n\cdot \tilde v_3 = \nbar \cdot \tilde v_4 & = & 1-\beta_t \cos\theta\,, \\
  \nbar\cdot \tilde v_3 = n \cdot \tilde v_4 & = & 1+\beta_t \cos\theta\,, \\
  n \cdot k & = & k_0\, (1-\cos\theta_1)\,, \\
  \nbar \cdot k & = & k_0\, (1+\cos\theta_1)\,, \\
  n \cdot l & = & l_0\, (1-\cos\chi_1)\,, \\
  \nbar \cdot l & = & l_0\, (1+\cos\chi_1)\,, \\
  \tilde v_3 \cdot k & = & k_0\, (1-\beta_t\sin\theta_1\cos\theta_2\sin\theta -
  \beta_t\cos\theta_1\cos\theta)\,, 
  \label{eq:tv3k} \\
  \tilde v_4 \cdot k & = & k_0\, (1+\beta_t\sin\theta_1\cos\theta_2\sin\theta +
  \beta_t\cos\theta_1\cos\theta)\,,
  \label{eq:tv4k} \\
  \tilde v_3 \cdot l & = & l_0\, (1-\beta_t\sin\chi_1\cos\chi_2\sin\theta-
  \beta_t\cos\chi_1\cos\theta)\,, \\
  \tilde v_4 \cdot l & = & l_0\, (1+\beta_t\sin\chi_1\cos\chi_2\sin\theta+
  \beta_t\cos\chi_1\cos\theta)\,.
\end{eqnarray}
We see, in particular, that
\begin{subequations}
  \begin{align}
   k_0 & = \frac12\left(n\cdot k + \nbar \cdot k\right) 
         = \frac12\left(k_+ + k_-\right)
         = \frac12\left(\tilde v_3\cdot k+\tilde v_4 \cdot k\right)\,,
   \\[0.5em]
   l_0 & = \frac12\left(n\cdot l + \nbar \cdot l\right) 
         = \frac12\left(l_+ + l_-\right)
         = \frac12\left(\tilde v_3\cdot l+\tilde v_4 \cdot l\right)\,.
  \end{align}
\end{subequations}

\subsection{Small-$q_T$ factorization}

At small transverse momenta of the top quark pair, $q_T$, the cross section
factorizes according to the formula~\cite{Li:2013mia}
\begin{align}
  &\frac{d^4\sigma}{dq_T^2 \, dy \, dM \, d\cos\theta} = \frac{8\pi\beta_t}{3s
M} \frac{1}{2} \int x_Tdx_T \, \frac{d\phi}{2\pi} \, J_0(x_Tq_T) \, 
  \nonumber
  \\
  \times
  \bigg\{ 
  &\left( \frac{x_T^2M^2}{4e^{-2\gamma_E}} \right)^{-F_{gg}(x_T^2,\mu)} 4
B^{\mu\rho}_{g/h_1}(\xi_1,x_\perp,\mu) \,
B^{\nu\sigma}_{g/h_2}(\xi_2,x_\perp,\mu) \, \mathrm{Tr} \big[
\bm{H}^{\mu\nu\rho\sigma}_{gg}(M,m_t,v_3,\mu) \, \bm{W}_{gg}(x_\perp,\mu) \big]
  \nonumber
  \\
  &+ \left( \frac{x_T^2M^2}{4e^{-2\gamma_E}} \right)^{-F_{q\bar{q}}(x_T^2,\mu)}
B_{q/h_1}(\xi_1,x_T^2,\mu) \, B_{\bar{q}/h_2}(\xi_2,x_T^2,\mu) \, \mathrm{Tr}
\big[ \bm{H}_{q\bar{q}}(M,m_t,\cos\theta,\mu) \, \bm{W}_{q\bar{q}}(x_\perp,\mu)
\big] \nonumber
  \\
  &+ (q \leftrightarrow \bar{q}) \bigg\} \,,
  \label{eq:factorization-scet}
\end{align}
which is a convolution of the beam functions, $B_{q/h_i}$,
$B^{\mu\rho}_{g/h_i}$, the hard functions, 
$\bm{H}_{q\bar{q}}$, 
$\bm{H}^{\mu\nu\rho\sigma}_{gg}$,
the soft functions, 
$\bm{W}_{q\bar{q}}$, $\bm{W}_{gg}$, and the anomaly exponents 
$F_{q\bar{q}}(x_T^2,\mu)$, $F_{gg}(x_T^2,\mu)$.
Above, $y$ corresponds to the rapidity of the top-anti-top quark system, defined
in Eq.~(\ref{eq:kin-var}), $\theta$ to the scattering angle of the top quark in
the \ttbar rest frame, and $\phi$ is the relative azimuthal angle between
$x_\perp$ an $v_3$.
 
We note that the gluon beam functions depend, in general, on the vector
$x_\perp$, which means that they are sensitive both to its length and to its
azimuthal position in the transverse plane. On the contrary, the quark beam
functions depend only on the magnitude of the transverse position vector.

The functions appearing in Eq.~(\ref{eq:factorization-scet}) capture
contributions of gluon emissions from different regions of phase space. In the
light-cone parametrization, $k^\mu = (k_+, k_-, k_\perp)$, the momenta
corresponding to each function scale as
\begin{equation}
  {\renewcommand{\arraystretch}{1.3}%
  \begin{array}{l@{\qquad \qquad}l@{\qquad \qquad}l}
   \text{collinear}  
       & k_i^\mu \sim (1,\lambda^2,\lambda)\, M^2   & B_{i/h_1},  \\
   \text{anti-collinear} 
       & k_i^\mu \sim (\lambda^2, 1, \lambda)\, M^2 & B_{i/h_2},  \\
   \text{hard}           
       & k_i^\mu \sim (1,1,1)\, M^2                 & \bm{H}_\iibar,    \\
   \text{soft}           
       & k_i^\mu \sim (\lambda,\lambda, \lambda)\, M^2 & \bm{S}_\iibar \,.   
  \end{array}
  }
\end{equation}

The beam functions are process-independent and they are currently known  up to
NNLO~\cite{Gehrmann:2012ze,Gehrmann:2014yya}.
The hard and the soft functions are not universal, hence, they have to be
calculated on a process-by-process basis.  The hard function can be extracted
from Refs.~\cite{Czakon:2008zk, Baernreuther:2013caa}.  The small-$q_T$ soft
function has been calculated up to NLO in Refs.~\cite{Li:2013mia,
Catani:2014qha}.

Each function defined on the right hand side of
Eq.~(\ref{eq:factorization-scet}) is separately divergent when calculated
directly from diagrammatic definitions, like the ones that shall be discussed in
Section~\ref{sec:methods}.  These divergencies correspond to the soft and
collinear limits and they must cancel between the hard, soft and beam functions,
as the entire cross section has to be finite. 

It turns out to be useful to remove divergences also at the level of the
functions entering the factorization formula~(\ref{eq:factorization-scet}). This
can be achieved by the procedure of multiplicative renormalization, and it will
be discussed in detail, for the case of the soft function, in
Section~\ref{sec:renormalization}.
As usual in the procedure of renormalization, the renormalized object acquires
dependence on an arbitrary parameter $\mu$, which has to vanish at the level of
the cross section. This implies that the renormalized versions of the functions
entering the factorization formula (\ref{eq:factorization-scet}) must satisfy
certain evolution equations that govern their $\mu$ dependence.

\subsection{The soft function}

\begin{figure}[t]
  \begin{center}
    \includegraphics[width=0.30\textwidth]{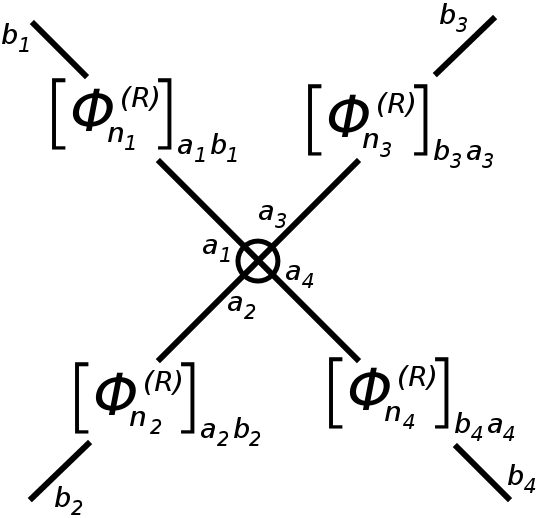}
  \end{center}
  \caption{
  Representation of the amplitude involving four Wilson lines.
  }
  \label{fig:SFamp}
\end{figure}

The general definition for the position-space, small-$q_T$ soft function
entering the factorization formula~(\ref{eq:factorization-scet})
reads~\cite{Li:2013mia}
\begin{equation}
  \mathbold{W} (x_\perp,\mu) = \frac{1}{d_R} 
  \langle 0| 
  \mathbold{\bar T} [\mathbold{O}_s^\dagger (x_\perp)] 
  \mathbold{T} [\mathbold{O}_s(0)] 
  |0 \rangle\,,
  \label{eq:Wdef}
\end{equation}
where $\mathbold{\bar T}$ and $\mathbold{T}$ represent time and anti-time
ordering~\cite{Becher:2007ty}. The normalization factors are 
\begin{equation}
  \begin{array}{ll}
  d_R  = N\,,     & \quad \quad \text{in the $\qqbar$ channel,} 
  \\[0.5em]
  d_R  = N^2-1\,, & \quad \quad \text{in the $gg$ channel\,,}
  \end{array}
\end{equation}
where $N$ is the number of colours, hence, in QCD, $N=3$.
It turns out to be convenient to insert the sum over all final states (both
discrete and continuous quantum numbers) into the definition (\ref{eq:Wdef}) and
obtain
\begin{align}
  \mathbold{W} (x_\perp,\mu) 
  &= 
  \frac{1}{d_R} \ \underset{\!\!X}{\int\kern-1.5em\sum} \,
  \langle 0| \mathbold{\bar T} [\mathbold{O}_s^\dagger (x_\perp)] | X \rangle
  \langle X | \mathbold{T} [\mathbold{O}_s(0)] |0 \rangle \\
  & =
  \frac{1}{d_R} \ \underset{\!\!X}{\int\kern-1.5em\sum} e^{i P_\perp \cdot
  x_\perp}\,
  \langle 0| \mathbold{\bar T} [\mathbold{O}_s^\dagger (0)] | X \rangle
  \langle X | \mathbold{T} [\mathbold{O}_s(0)] |0 \rangle \,,
  \label{eq:Wdef2}
\end{align}
where $P_\perp$ is the total transverse momentum carried by gluons or massless
quarks (which recoil against the \ttbar system) in the state~$|X\rangle$.
The amplitude  $\langle X | \mathbold{\bar T} [\mathbold{O}_s (0)]
| 0 \rangle$ is shown in Fig.~\ref{fig:SFamp}.  
%
%

The operator $\mathbold{O}_s$ is defined as a product of Wilson lines, which
mediate soft gluon exchanges between particles and give rise to colour
correlations.
The Wilson line for a particle in representation $R$, moving along the straight
line with $d$-momentum $n_i$ between the points $x+a n_i$ and $x+b n_i$, is
defined as~\cite{Czakon:2013hxa}
\begin{equation}
  \left[\mathbold{\Phi}_{n_i}^{(R)} (x; b, a)\right]_{a_i b_i} =
  \left\{
  \calP 
  \exp 
  \left( i g_s^0 \int_a^b\!\! dt\, n_i \cdot 
  A^c(x+t\, n_i)\, \cT^{(R)\, c}
  \right)
  \right\}_{a_i b_i}
  \,,
\end{equation}
where $\calP$ denotes path ordering. The indices $a_i$ and $b_i$ arise since the
Wilson line is an operator in colour space and it accounts for colour evolution
of a particle due to gluon emissions.
The operator $\mathbold{O}_s$ for top quark pair production in the
\qqbar channel is given by
\begin{equation}
  \mathbold{O}_s(x) = 
  \left[\mathbold{\Phi}_{v_3}^{(3)} (x; \infty, 0)\right]_{b_3 a_3} 
  \left[\mathbold{\Phi}_{v_4}^{(\bar 3)} (x; \infty, 0)\right]_{b_4 a_4} 
  \left[\mathbold{\Phi}_{n}^{(3)} (x; 0, -\infty)\right]_{a_1 b_1} 
  \left[\mathbold{\Phi}_{\bar n}^{(\bar 3)} (x; 0, -\infty)\right]_{a_2 b_2} \,,
  \label{eq:Osqqdef}
\end{equation}
and in the $gg$ channel
\begin{equation}
  \mathbold{O}_s(x) = 
  \left[\mathbold{\Phi}_{v_3}^{(3)} (x; \infty, 0)\right]_{b_3 a_3} 
  \left[\mathbold{\Phi}_{v_4}^{(\bar 3)} (x; \infty, 0)\right]_{b_4 a_4} 
  \left[\mathbold{\Phi}_{n}^{(8)} (x; 0, -\infty)\right]_{a_1 b_1} 
  \left[\mathbold{\Phi}_{\bar n}^{(8)} (x; 0, -\infty)\right]_{a_2 b_2} \,.
  \label{eq:Osggdef}
\end{equation}

\subsection{Azimuthal averaging}

As we see in Eq.~(\ref{eq:factorization-scet}), the factorization formula
involves integration over $\phi$, the azimuthal angle between $x_\perp$ an
$v_3$. If we restrict ourselves to the NNLO cross section for top quark pair
production, the NNLO soft function will enter the factorization
formula~(\ref{eq:factorization-scet}) multiplied by the leading order hard
function and the leading order beam functions, both of which are Lorentz
scalars.  Hence, the integration over the azimuthal angle can be pulled directly
to the soft function. This motivates us to define the averaged soft function
\begin{equation}
  \bfS_{i\bar i}(x_\perp) = 
  \int \frac{d\Omega_{d-3}}{S_{d-3}} \bm{W}_{i\bar i}(x_\perp)\,,
  \label{eq:sf-avd}
\end{equation}
with $d\Omega_{d-3}$ being a rotationally-invariant measure on the unit
$(d-3)$-sphere, defined recursively as
\begin{equation}
  \int_{S_1^{d-3}}\!\!\! d\Omega_{d-3} =
  \int_{S_1^{d-3}}\!\!\! d\Omega(\theta_1,\theta_2,\ldots,\theta_{d-3}) =
  \int_0^\pi\!\! d\theta_1 \sin^{d-4}\theta_1
  \int_{S_1^{d-4}}\!\!\! d\Omega(\theta_2,\ldots,\theta_{d-3})\,,
  \label{eq:rot-inv-measure}
\end{equation}
with
\begin{equation}
  \theta_1,\ldots,\theta_{d-4} \in [0,\pi]\,,\qquad \qquad
  \theta_{d-3} \in [0,2\pi]\,,
\end{equation}
while $S_{d-3}$ is this sphere's surface, which can be calculated from the
general formula
\begin{equation}
  S_{d}= \frac{2\pi^{(d+1)/2}}{\Gamma[(d+1)/2]}\,.
  \label{eq:Sd}
\end{equation}
We note that the averaging of Eq.~(\ref{eq:sf-avd}) is defined in $d-3$
dimensions as the entire calculation of the NNLO soft function is performed
consistently in $d$ dimensions.  That implies that all external and internal
momenta, as well as polarizations, are $d$-dimensional.

The averaged soft function can be expanded as perturbation series 
\begin{equation}
  \bfS_{i\bar i}(x_\perp) = 
  \sum_{n=0}^\infty \bfS_{i\bar i}^{(n)}(x_\perp) \left(\asp\right)^n\,.
  \label{eq:sf-expansion}
\end{equation}

Finally, we mention that in order to compute the complete NNLO cross section for
\ttbar production in the $gg$ channel, one also needs a version of the NLO soft
function which is averaged azimuthally together with the $x_\perp^\mu
x_\perp^\nu$ Lorentz structure which comes from the NLO gluon beam
function~\cite{Gehrmann:2014yya}.
This component can be obtained using the unaveraged NLO soft function of
Ref.~\cite{Catani:2014qha}.

\subsection{Colour space}

The soft function discussed so far is an abstract operator in colour space. It
turns out to be useful to represent it as a matrix, with the elements 
\begin{equation}
 S_{IJ} = \langle c_I | \bfS | c_J \rangle\,.
\end{equation}
We choose the basis vectors, $|c_I\rangle = \big(c_I^{i\bar i}\big)_{\{a\}}$,
following Ref.~\cite{Ferroglia:2012uy}.
In the $q\qbar$ channel, they read
%
%
%
\begin{equation}
  \left(c_1^{q\bar{q}}\right)_{\{a\}} = \delta_{a_1a_2} \,\delta_{a_3a_4}\,,
  \qquad
  \left(c_2^{q\bar{q}}\right)_{\{a\}} = t_{a_2a_1}^c \, t_{a_3a_4}^c\,,
  \label{eq:colour-basis-qq}
\end{equation}
and in the $gg$ channel
\begin{equation}
    \left(c_1^{gg}\right)_{\{a\}} = \delta^{a_1a_2} \, \delta_{a_3a_4} \, , 
    \qquad
    \left(c_2^{gg}\right)_{\{a\}} = i f^{a_1a_2c} \, t^{c}_{a_3a_4} \, , 
    \qquad
    \left(c_3^{gg}\right)_{\{a\}} = d^{a_1 a_2 c} \, t^c_{a_3a_4} \, .
  \label{eq:colour-basis-gg}
\end{equation}
The inner product is defined as a sum over all colour indices
\begin{equation}
  \langle c_I | c_J \rangle = 
  \sum_{\{a\}} 
  \left( c_I \right)^*_{a_1 a_2 a_3 a_4}
  \left( c_J \right)_{a_1 a_2 a_3 a_4}\,.
\end{equation}
The above basis vectors are orthogonal but not
orthonormal~\cite{Ferroglia:2012uy}.

The NLO and NNLO contributions to the soft function can be represented as
\begin{eqnarray}
  \bfS^{(1)}_\text{bare}(x_\perp) & = &
  \sum_{i,j} \mathbold{w}_{ij}^{(1)}\, I_{ij}(\xT)\,, 
  \label{S1bareNLO}
  \\
  \bfS^{(2)}_\text{bare}(x_\perp) & = &
  \sum_{i,j} \mathbold{w}^{(1)}_{ij}\, I_{ij}^{(1)}(\xT) +
  \sum_{i,j,k,l} \mathbold{w}^{(2S)}_{ijkl}\, I_{ijkl}^{(2)}(\xT) +
  \sum_{i,j,k} \mathbold{w}^{(2A)}_{ijk}\, I_{ijk}^{(2)}(\xT)\,,
  \label{S1bareNNLO}
\end{eqnarray}
where $\mathbold{w}$ are the colour matrices defined as
%
%
\begin{subequations}
  \begin{align}
   \bm{w}_{ij}^{(1)} & =
   \frac{1}{d_R} \langle c_I | \cT_i\cdot\cT_j | c_J \rangle\,,
   \label{eq:w1def}
   \\[0.5em]
   \bm{w}_{ijkl}^{(2S)} & =
   \frac{1}{d_R} \langle c_I | \left\{\cT_i\cdot\cT_j, \cT_k\cdot\cT_l\right\} 
   | c_J \rangle\,,
   \label{eq:w2def}
   \\[0.5em]
   \bm{w}_{ijk}^{(2A)} & =
   \frac{1}{d_R} \langle c_I | \left[\cT_i\cdot\cT_k, \cT_j\cdot\cT_k\right] 
   | c_J \rangle\,.
   \label{eq:w3def}
  \end{align}
  \label{eq:w1w2def}
\end{subequations}
The Hermitian, colour operators $\cT_i$ satisfy the following
relations~\cite{Catani:1996vz}
\begin{equation}
\cT_i \cdot \cT_j=\cT_j\cdot\cT_i = T^c_iT^c_j, \quad
\cT_i \cdot \cT_i\,  = \cT_i^2=C_i\, \unitop =C_{a_i}\, \unitop, \quad
\sum_{i}\cT_i|\cm_{n}\ra=0 \,,
\end{equation}
where $\unitop$ is the identity operator, $a_i \in \{q, \qbar, g\}$, and
$C_g = C_A,\, C_q = C_{\bar{q}} = C_F$.
The operators $\cT_i$ act on vectors in the colour space as follows
\begin{eqnarray}
\la c_1,\dots,c_i,\dots,c_n,c|\cT_i|b_1,\dots,b_i,\dots,b_n\ra &=&
\la c_1,\dots,c_i,\dots,c_n|T^c_i|b_1,\dots,b_i,\dots,b_n\ra
\nonumber\\ &=& \delta_{c_1b_1} \dots T^c_{c_i b_i}\dots\delta_{c_nb_n} \; .
\end{eqnarray}
The matrix elements of the $i^\text{th}$ parton operator, $\cT_i$, are given by
$\left(\cT^c_i\right)_{c_1c_2} = if^{c_1cc_2}$, for an initial-state and
final-state gluon, $\left(\cT^c_i\right)_{c_1 c_2}= t^c_{c_1 c_2} (= -t^c_{c_2
c_1})$, for a final-state quark (anti-quark), and $\left(\cT^c_i\right)_{c_1 c_2} =
-t^c_{c_2 c_1} (= t^c_{c_1 c_2})$, for an initial-state quark (anti-quark).

\subsection{Fourier transform}

In practice, it is easier to calculate the soft function in momentum space. The
relevant scalar integrals, which appear in Eqs.~(\ref{S1bareNLO}) and
(\ref{S1bareNNLO}), can be transferred to momentum space by means of the Fourier
transform 
\begin{subequations}
   \begin{eqnarray}
     \label{eq:fourier-transformd}
     \tilde I_{\{i\}} (q_\perp)& = &
     \frac{1}{\left(2\pi\right)^{\frac{d-2}{2}}}
     \int d^{d-2} x_\perp I_{\{i\}} (x_\perp) \, e^{i x_\perp \cdot q_\perp}\,,
     \\
     \label{eq:fourier-transform-inv}
     I_{\{i\}} (x_\perp)& = &
     \frac{1}{\left(2\pi\right)^{\frac{d-2}{2}}}
     \int d^{d-2} q_\perp \tilde I_{\{i\}} (q_\perp) \, e^{-i x_\perp \cdot
     q_\perp}\,,
   \end{eqnarray}
\end{subequations}
where $\{i\}$ represents the two, three or four particle indices.
The structure of the $I_{\{i\}}(x_\perp)$ integrals, for all types of diagrams
encountered at NNLO, is
\begin{equation}
  I_{\{i\}}(x_\perp) = 
\int d^d p
  \, e^{-i x_\perp \cdot p_\perp}\,
  \times (...)\,,
  \label{eq:Igenform}
\end{equation}
where the ellipsis depend on details of the graph, \ie the direction of the
Wilson lines and the soft parton emissions, and $p$ denotes the total momentum
of the latter.

By plugging Eq.~(\ref{eq:Igenform}) to Eq.(\ref{eq:fourier-transformd}), we
obtain
\begin{align}
  \tilde I_{\{i\}}(q_\perp)  & = 
 (2\pi)^{\frac d2-1}
\int d^d p
 \,  
  \delta^{(d-2)}(p_\perp-q_\perp)
  \times (\dots) \,. \label{eq:defmspace}
\end{align}
And by applying the azimuthal averaging of Eq.~(\ref{eq:sf-avd}) we get
\begin{align}
\int \frac{d\Omega_{d-3}}{ S_{d-3}}\,  \tilde I_{\{i\}}(q_\perp) 
  & =
  \frac{(2\pi)^{\frac d2-1}}{ S_{d-3}}
  \int d^{d}p\,  d\Omega_{d-3}\,
  \delta^{(d-2)}(p_\perp-q_\perp)
  \times (\dots) 
  \label{eq:avmagic1}
  \\
  & = 
  \frac{(2\pi)^{\frac d2-1}}{ S_{d-3}}
  \int \frac{d^dp}{q_T^{d-3}}\, \delta(p_T-q_T)
  \times (\dots) \,.
  \label{eq:avmagic2}
\end{align}
In the above, we used rotational invariance of the measures and the fact that
the order of the integrations can be changed. This leads to a reinterpretation of the
angles in $d\Omega_{d-3}$ as the azimuthal angles of $q_\perp$ and that allows
one to go from (\ref{eq:avmagic1})  to (\ref{eq:avmagic2}) with the help of the
identity
\begin{equation}
  \int d\Omega_{d-3}\, \delta^{(d-2)} (p_\perp-q_\perp) =
  \frac{1}{q_T^{d-3}}\, \delta(p_T-q_T)\,.
\end{equation}
Now, we rescale the momenta associated to the emissions by $q_T$, in particular $\tilde{p}^\mu=q_T p^\mu$ , and get
\begin{align}
\int \frac{d\Omega_{d-3}}{ S_{d-3}}\,  \tilde I_{\{i\}}(q_T) 
  & = 
  \frac{1}{q_T^r}
  \frac{\left(2\pi\right)^{\frac{d}{2}-1}}{ S_{d-3}}
  \int d^d\tilde p\, \delta(\tilde p_T-1)
  \times (\dots) \,.
\end{align}
Above, the overall power of $q_T$ has been denoted as $r$. It is a sum of the
contribution from the $d-2$-dimensional delta function and a genuine
contribution from the graph part.
The rescaling factorizes all the $q_T$ dependence. As a consequence, all
integrals in momentum space are proportional to the factor $1/q_T^r$, with the
power $r$, which depends on the order of perturbative expansion. Hence, the
transformation from momentum to position space, by means of the Fourier
Transform~(FT), will amount to multiplication by a factor given by the compact
expression 
%
%
\begin{align}
  \text{FT}\left[\frac{1}{q_T^r}\right]^{d-2} 
   & =
   2^{-2 + 3\epsilon + r}
   \sqrt{\pi}\,
   \left(\frac{e^{-\gamma_E+\frac12\LT}}{\mu}\right)^{-2+2\epsilon+r}
   \!\!\!\!\!\!
   \frac{\Gamma(2-2\epsilon-r)}{
   \Gamma\left(\frac{3}{2}- \epsilon -\frac{r}{2}\right)
   \Gamma\left(\frac{r}{2}\right)}.
   \label{eq:FTfinal}
\end{align}
At NLO, $r=2+\alpha$, and at NNLO, $r=2+2\alpha+2\epsilon$ for double-cut
diagrams and $r=2+\alpha+2\epsilon$, for single-cut diagrams, where $\alpha$ is
the analytic regulator discussed in the next section. An important feature of
(\ref{eq:FTfinal}), both at NLO and NNLO, is that its expansion begins at
order $1/\epsilon$ 
\begin{equation}
  \text{FT}\left[\frac{1}{q_T^{2+\alpha}}\right]^{d-2}  = 
  -\frac{1}{2\epsilon} + \order{\alpha/\epsilon^2}\,,
  \qquad
  \text{FT}\left[\frac{1}{q_T^{2+2\alpha+2\epsilon}}\right]^{d-2}   =
  -\frac{1}{4\epsilon} + \order{\alpha/\epsilon^2}\,.
  \label{eq:FTexpand}
\end{equation}
Hence, to obtain a result in position space at the order
$\epsilon^0$, one needs to calculate the momentum-space soft function up to
$\epsilon^1$.

\subsection{Analytic regulator}

The phase space integrals $\tilde I_{\{i\}}$ turn out to be divergent not only
when the gluons become soft, but also in the limit where the light-cone
components of gluons' momenta tend to zero or infinity. Through the relation
\begin{equation}
  y_g = \frac12 \ln \frac{k_+}{k_-}\,,
\end{equation}
this limit occurs when the gluon rapidity $y_g \to \pm \infty$. That is why,
these are called ``rapidity divergencies''. They arise because, in SCET, we
approximate the full QCD Feynman integrals following the expansion by regions.
Yet, we integrate each expression over the full phase space of gluons' momenta.
We note that this does not give rise to double counting~\cite{Becher:2014oda}.
Nevertheless, it forces us to introduce another regulator to handle the
integrals.

The reason why contributions from different regions do not overlap even when
integrated from $-\infty$ to $+\infty$ is that integrals in each region depend
on a single scale and expanding them further (\eg soft integrals in the
collinear region) leads to scaleless integrals, which vanish.

In our calculation, we chose to adopt the analytic regulator prescription of
Ref.~\cite{Becher:2011dz}, which amounts to the following replacement of the
integration measure
\begin{equation}
 \int d^d k\, \delta^{+}(k^2)  \to
 \int d^d k \left(\frac{\nu}{n\cdot k}\right)^\alpha \delta^{+}(k^2)\,,
\end{equation}
where $\nu$ is a free parameter introduced for dimensional reasons (an analogue
of $\mu$ in dimensional regularization) and we denote $\delta^+(k^2) =
\delta(k^2) \theta(k_0)$.
The regulator $\alpha$ becomes necessary at intermediate stages of the
calculation. Since rapidity divergencies do not appear in full QCD, the result
for the complete cross section is finite in the limit $\alpha \to 0$.
In the case of the analytic regulator, the poles in $\alpha$ cancel not only at
the level of the cross section but even at the level of the soft function. This
comes from the fact that the soft function for the Drell-Yan process is equal to
one, which implies that the product of the beam functions is
$\alpha$-independent (but not the beam functions themselves, see
Ref.~\cite{Gehrmann:2014yya}). As beam functions are universal, the same,
$\alpha$-independent product of the beam functions is used in the process of
\ttbar production. This means that the only dependence on $\alpha$ can occur in
the soft function. Therefore, all $\alpha$ poles have to vanish within the 
latter. We shall use this feature as one of validations of our calculation.

\subsection{Renormalization}
\label{sec:renormalization}

The renormalized soft function of the small-$q_T$ factorization satisfies the
following renormalization group evolution~(RGE) equation~\cite{Zhu:2012ts}
%
%
\begin{equation}
  \frac{d}{d\ln \mu} \bfS_\iibar (\mu) =
  - \bfgamma^{s \dagger}_\iibar \,\bfS_\iibar (\mu)  
  - \bfS_\iibar (\mu)\, \bfgamma^{s}_\iibar \,,
  \label{eq:SF-RGE-main}
\end{equation}
where
\begin{equation}
  \bfgamma^{s}_\iibar = \bfgamma^{h}_\iibar - 2 \gamma^{i} \bfI\,,
\end{equation}
and $\bfgamma^{h}_\iibar$ is defined as a non-$\Gamma_\text{cusp}$ part of the
full anomalous dimension matrix $\bfGamma$~\cite{Ahrens:2010zv}, while
$\gamma^{i}$ is the massless-particle anomalous dimension (and enters RGE
equations for beam functions in Drell-Yan and Higgs
production~\cite{Becher:2010tm, Becher:2012yn}). To make the notation lighter
we shall omit the index $\iibar$, keeping in mind that the soft
function and the anomalous dimension are different in the \qqbar and $gg$
channels.

The soft anomalous dimension matrix $\bfgamma^{s}$ is related to the soft
renormalization factor (also a matrix in colour space), $\bfZ_s$, as follows
\begin{equation}
  \bfgamma^{s} = - \bfZ_s^{-1}\frac{d\bfZ_s}{d\ln \mu}\,,
  \label{eq:softAD}
\end{equation}
and $\bfZ_s$ absorbs all UV divergences such that
\begin{equation}
  \bfS(\mu) = \bfZ^\dagger_s(\mu,\epsilon) \bfS_\bare(\epsilon)
  \bfZ_s(\mu,\epsilon)\,.
\end{equation}
Each quantity in the above equation has a perturbative expansion, either in the
renormalized coupling, 
$a_s = \alpha_s/(4\pi)$, 
or in the bare coupling,
$a_s^0 = \alpha_s^0/(4\pi)$, 
and the relation between the two is
\begin{equation}
  a_s^0(\epsilon) = 
  \left(\frac{\mu^2 e^{\gamma_E}}{4\pi}\right)^\epsilon Z_\alpha a_s
  \equiv \xi_\as(\epsilon,\mu)\, Z_\alpha\, a_s (\mu)\,,
  \label{eq:alpha}
\end{equation}
where the \msbar renormalization constant reads
\begin{equation}
  Z_\alpha = 
  1- \frac{\beta_0 \as}{4\pi \epsilon} + \ldots =
  1- \frac{\beta_0}{\epsilon} a_s(\mu) + \ldots\,,
  \label{eq:Zalpha}
\end{equation}
and $\beta_0$ is the one-loop coefficient of the QCD $\beta$ function
given in Eq.~(\ref{eq:beta0def}).
Hence, substitution of Eq.~(\ref{eq:sf-expansion}) for the bare and renormalized
soft function, as well as Eqs.~(\ref{eq:alpha}) and~(\ref{eq:Zalpha}) to
Eq.~(\ref{eq:SF-RGE-main}) leads to the following, order-by-order relations
\begin{eqnarray}
  \bfS^{(0)} & = &  \bfS^{(0)}_\bare \,,
  \\[0.5em]
  \bfS^{(1)} & = &
  \bfZ^{\dagger (1)}_s \bfS^{(0)}_\bare  + \bfS^{(0)}_\bare \bfZ^{(1)}_s  +
  \xi_\as \bfS^{(1)}_\bare \,,
  \\[0.5em]
  \bfS^{(2)} & = &
  \bfZ^{\dagger (2)}_s \bfS^{(0)}_\bare  + \bfS^{(0)}_\bare \bfZ^{(2)}_s  +
  \bfZ^{\dagger (1)}_s \bfS^{(0)}_\bare \bfZ^{(1)}_s 
  \nonumber \\[0.5em]
  & &
  + \bfZ^{\dagger (1)}_s \xi_\as \bfS^{(1)}_\bare  + 
  \xi_\as \bfS^{(1)}_\bare \bfZ^{(1)}_s  +
  \xi_\as^2 \bfS^{(2)}_\bare 
  -\frac{\beta_0}{\epsilon}\, \xi_\as \bfS^{(1)}_\bare
  \,.
  \label{eq:S2ren}
\end{eqnarray}
The quantities on the l.h.s. are finite in the limit $\epsilon \to 0$. The
$\bfZ^{(i)}_s$ factors have only singular terms with poles in $\epsilon$, while
the bare functions, $\bfS^{(1)}_\bare$ and $\bfS^{(2)}_\bare$, have both
singular and finite parts.

For notational simplicity, in what follows we will absorb the $\xi_\as$
prefactors into the definitions of the bare soft functions and change the
notation according to
\begin{equation}
  \xi_\as^n \bfS^{(n)}_\bare \to
  \bfS^{(n)}_\bare\,.
\end{equation}

At the order $a_s^2$, from Eq.~(\ref{eq:S2ren}), we get
\begin{eqnarray}
  \underbrace{\bfS^{(2)}}_{\text{finite part only}}   
  &  =  &
  \overbrace{\bfZ^{\dagger (2)}_s \bfS^{(0)}_\bare  + 
   \bfS^{(0)}_\bare \bfZ^{(2)}_s  +
   \bfZ^{\dagger (1)}_s \bfS^{(0)}_\bare \bfZ^{(1)}_s}^{\text{(I) pole part only}} 
  \nonumber \\
  & &
  \quad +  \quad 
  \underbrace{\bfZ^{\dagger (1)}_s \bfS^{(1)}_\bare  + 
  \bfS^{(1)}_\bare \bfZ^{(1)}_s  +
  \bfS^{(2)}_\bare
 -\frac{\beta_0}{\epsilon}\, \bfS^{(1)}_\bare
  }_{\text{(II) finite + pole part}} \,.
  \label{eq:renS2}
\end{eqnarray}
Part (I) on the r.h.s. has only terms singular in $\epsilon$, which come
from the fact that the $\bfZ_s$ factors are defined in the \msbar scheme. 

These pole terms have to cancel against the singular terms of part (II),
which can be used in the following way:
knowing $\bfZ_s^{(1)}$ and $\bfZ_s^{(2)}$, as well as $\bfS^{(0)}_\bare$ and
$\bfS^{(1)}_\bare$, allows one to cross-check all the singular terms of
the $\bfS^{(2)}_\bare$ function obtained from direct calculation.
The explicit form of the $\bfZ_s$ factor, together with the relevant anomalous
dimensions are given in Appendix~\ref{app:AD}.

\subsection[Determination of $L_\perp$-dependent terms of the soft
function from RGE]{Determination of $\mathbold{L_\perp}$-dependent terms of the soft function from RGE}

\label{sec:RGevolution}

The RGE equation (\ref{eq:SF-RGE-main}) can be written as
\begin{equation}
  \frac{d}{d L_\perp} \bfS(\mu) =
  - \frac{1}{2} \left[
  \bfgamma^{s \dagger}\,\bfS(\mu)  
  + \bfS(\mu)\, \bfgamma^{s}
  \right]\,,
  \label{eq:SF-RGE-Lp}
\end{equation}
where $\LT$ was defined in Eq.~(\ref{eq:LTdef}).
Both the soft function and the anomalous dimension have perturbative expansions
\begin{eqnarray}
  \bfS  & = &
  \bfS^{(0)} + a_s \bfS^{(1)} + a_s^2 \bfS^{(2)} + \ldots\,,
  \label{eq:Sren-exp}
  \\
  \bfgamma_{s} & = &
  a_s \left(\bfgamma_{s,0} + a_s \bfgamma_{s,1}+ \ldots\right)\,.
  \label{eq:gammas-exp}
\end{eqnarray}
All quantities in the above equations are renormalized and they are defined in
$d$ dimensions.
 
Because the anomalous dimension matrix starts at the order $a_s$,
Eq.~(\ref{eq:SF-RGE-Lp}) can be solved iteratively. We just need to plug
Eqs.~(\ref{eq:Sren-exp}) and (\ref{eq:gammas-exp}) into 
Eq.~(\ref{eq:SF-RGE-Lp}) and  remember that the renormalized, 4-dimensional
coupling $a_s$ also depends on $\ln\mu$, which implies
\begin{equation}
  \frac{d a_s}{dL_\perp} = \frac12 \frac{d a_s}{d\ln\mu} = 
  \frac12 \frac{\partial a_s}{\partial \ln\mu} = 
  \frac12 \frac{\beta(a_s)}{4\pi} = 
  - \beta_0\, a_s^2 + \order{a_s^3}\,,
\end{equation}
where we used the expansion of the QCD $\beta$ function given in
Eq.~(\ref{eq:betaexp}).

After collecting terms at each order, we
arrive at the following differential equations
\begin{eqnarray}
  \frac{d}{dL_\perp} \bfS^{(0)} & = &  0\,,
  \\
  \frac{d}{dL_\perp} \bfS^{(1)} & = & 
  -\frac12\left[
  \bfS^{(0)}\, \bfgamma_{s,0} + \bfgamma_{s,0}^\dagger\, \bfS^{(0)} 
  \right]\,,
  \\
  \frac{d}{dL_\perp} \bfS^{(2)} & = & 
  -\frac12\left[
  \bfS^{(1)}\, \bfgamma_{s,0} + \bfgamma_{s,0}^\dagger\, \bfS^{(1)}  
  - 2 \beta_0 \bfS^{(1)}  +
  \bfS^{(0)}\, \bfgamma_{s,1} + \bfgamma_{s,1}^\dagger\, \bfS^{(0)} 
  \right]\,.
\end{eqnarray}
 
Hence, knowing the soft anomalous dimension to order $a_s^2$, and the soft
function to order~$a_s^1$ allows one to determine all pieces of the soft
function at order $a_s^2$ except the constant, \ie $\LT$-independent term.
Specifically, at order $a_s^2$, we get
%
%
\begin{eqnarray}
  \bfS^{(2)} 
  & = & 
  -\frac12\Bigg\{
  \frac12\left[
  \bfS^{(1)}_{L_\perp}\, \bfgamma_{s,0} + 
  \bfgamma_{s,0}^\dagger\, \bfS^{(1)}_{L_\perp}  
  - 2 \beta_0 \bfS^{(1)}_{L_\perp} 
  \right] L_\perp^2
   \\
  & &
  \hspace{25pt} + \left[
  \bfS^{(1)}_{\slashed{L}_\perp}\, \bfgamma_{s,0} + 
  \bfgamma_{s,0}^\dagger\, \bfS^{(1)}_{\slashed{L}_\perp}  
  - 2 \beta_0 \bfS^{(1)}_{\slashed{L}_\perp} +
  \bfS^{(0)}\, \bfgamma_{s,1} + \bfgamma_{s,1}^\dagger\, \bfS^{(0)} 
  \right] L_\perp
  \Bigg\} 
  + \text{const}\,,
  \nonumber
\end{eqnarray}
where $\bfS^{(1)}_{L_\perp}$ and $\bfS^{(1)}_{\slashed{L}_\perp}$ denote,
respectively, the $L_\perp$-dependent and $L_\perp$-independent pieces of the
$\bfS^{(1)}$ soft function.

\section{NLO soft function}
\label{sec:NLOsf}

The leading order soft function corresponds to the case without radiation. It is
given by the following, constant matrices~\cite{Li:2013mia, Ferroglia:2012uy}
\begin{equation}
 \bfS^{(0)}_{q\qbar}  =  
 \left( \begin{array}{cc}
   N & 0 \\
   0 & \frac{C_F}{2} 
 \end{array} \right)\,,
 \qquad
 \bfS^{(0)}_{gg}  =  
 \left( \begin{array}{ccc}
   N & 0           & 0 \\
   0 & \frac{N}{2} & 0 \\
   0 & 0           & \frac{N^2-4}{2N}
 \end{array} \right)\,,
\end{equation}
respectively for the $\qqbar$ and $gg$ channels.
Because the LO soft function is not divergent, the above result corresponds both to the bare and the renormalized case. Hence, we see that  $\bfZ_{i\bar
i}^{(0)} =  \idop$.

The next-to-leading order, bare soft function, is given by
\begin{equation}
  \bfS_{i\bar i}^{(1), \bare} = 
  \sum_{i,j = 1,\ldots,4} \!\!\! \bm{w}_{ij}^{(1)} I_{ij}\,,
  \label{eq:S-W-Iij}
\end{equation}
where the colour structure is encoded in the $\bm{w}_{ij}^{(1)}$
matrices defined as
\begin{equation}
  \Big(\bm{w}_{ij}^{(1)}\Big)_{IJ} =
  \frac{1}{d_R} \langle c_I |\bm{T}_i \cdot \bm{T}_j | c_J \rangle\,,
  \label{eq:w1matrixdef}
\end{equation}
with the basis vectors $| c_J \rangle$ introduced in
Eqs.~(\ref{eq:colour-basis-qq}) and~(\ref{eq:colour-basis-gg}).
For the sake of notational simplicity, in what follows, we suppress the index
$\iibar$ in the colour matrices. It can always be inferred either from the
context, or based on the size  of a matrix.
The explicit expressions for colour matrices (\ref{eq:w1matrixdef}) are given in
Appendix~\ref{app:colmat}.

\begin{figure}[t]
  \begin{center}
    \hspace{40pt}
    \includegraphics[width=0.30\textwidth]{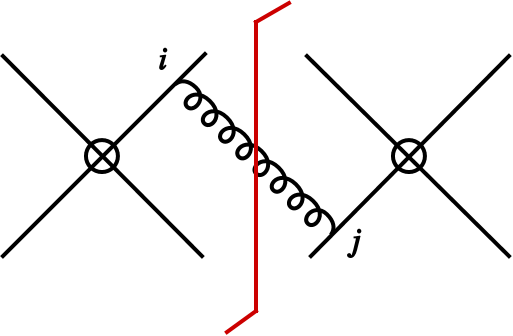}
  \end{center}
  \caption{Feynman diagram for the NLO soft function.  The solid lines pointing
  towards the cut represent the Wilson lines of the top quarks. The lines
  pointing away from the cut correspond to the Wilson lines of the incoming,
  massless partons, quarks or gluons. The diagram represents a class of graphs
  with all possible gluon connections between the Wilson lines.
  }
  \label{fig:sf-nlo}
\end{figure}

$I_{ij}$s in Eq.~(\ref{eq:S-W-Iij}) correspond to phase-space integrals,
represented diagrammatically in Fig.~\ref{fig:sf-nlo}.
The solid lines pointing towards the cut represent the 
top quarks. The lines pointing away from the cut correspond to the
incoming, massless partons.
The momentum-space versions of the integrals represented in
Fig.~\ref{fig:sf-nlo}, obtained with the Fourier
transform~(\ref{eq:fourier-transformd}), read
\begin{equation}
  \tilde I_{ij} =  
  (4\pi)^2  \left(\frac{\mu^2 e^{\gamma_E}}{4\pi}\right)^\epsilon 
    \frac{(2\pi)^{\frac{d}{2}-1}}{ S_{d-3}\, q_T^{d-3}} 
  \int \frac{d^d k}{(2\pi)^{d-1}}
  \left(\frac{\nu}{n\cdot k}\right)^\alpha
  \delta(k^2)\, \theta(k^0)\,
  \delta(k_T - q_T)
  \frac{(-v_i \cdot v_j)}{v_i \cdot k\; v_j \cdot k}\,,
  \label{eq:soft-func-nlo-mom-space-ddim}
\end{equation}
where $\left(\frac{\mu^2 e^{\gamma_E}}{4\pi}\right)^\epsilon$ comes from
renormalization of the strong coupling, see Eq.~(\ref{eq:alpha}), and the remaining
prefactors arise from the Fourier transform and azimuthal averaging.
Only the real-type diagrams contribute to the NLO soft function as the virtual
graphs are scaleless and vanish.

\begin{figure}[t]
  \begin{center}
    \includegraphics[width=0.99\textwidth]{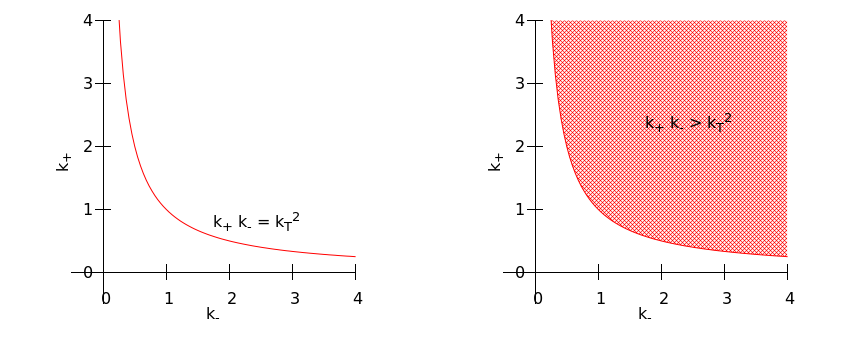}
  \end{center}
  \caption{
  Integration regions over the light-cone components of the on-shell (left) and
  off-shell gluon. The former corresponds to the integrals of the NLO soft
  function defined in Section~\ref{sec:NLOsf}, the latter to bubble integrals of
  the NNLO soft function discussed in Section~\ref{sec:debubble}.
  }
  \label{fig:light-cone-phase-space}
\end{figure}

The integration measure can be written as $d^d k = d k_+ d k_- d^{d-2} k_\perp$
and the $k_+$ and $k_-$ components are integrated from minus to plus infinity.
However, the phase space of integration of these light-cone momenta is
restricted by $\delta(k^2)\, \theta(k^0)$ appearing in
Eq.~(\ref{eq:soft-func-nlo-mom-space-ddim}).  This can be easily seen by
rewriting the above condition as $\delta(k_+ k_- - k_T^2)\, \theta(k_+ + k_-)$.
The delta function fixes $k_+ k_- = k_T^2 > 0$, hence the light-cone components
must be both positive or negative.  And the theta function, chooses them to be
both positive. All in all, the integration over $d k_+ d k_-$ is restricted to
the line depicted in Fig.~\ref{fig:light-cone-phase-space} (left).

The NLO soft function has been calculated up to order $\epsilon^0$ in
Refs.~\cite{Li:2013mia, Catani:2014qha}. However, as can be seen from
Eq.~(\ref{eq:renS2}), to renormalize the NNLO soft function, we need to
know the NLO soft function up to the order $\epsilon^1$. This order has been
calculated in Ref.~\cite{AntoniaMTh}, yet, using a slightly different definition
of azimuthal averaging. We have adjusted this result to our definition, as well
as fully cross checked it using a sector decomposition-based approach (described
in detail in Section~\ref{sec:secdec}), finding a perfect agreement.

Hence, the final result for the position-space NLO soft function up to order
$\epsilon^1$ reads
%
%
\allowdisplaybreaks
\begin{align}
\bfS_{i\bar i}^{(1)}  & =  
-4\, \bfw_{13}^{(1)}
\left[\text{Li}_2\left(1-\frac{t_1 u_1}{m_t^2 M^2}\right)
-2 L_\perp \ln \frac{-t_1}{m_t M}\right]
-4\, \bfw_{23}^{(1)}
\left[\text{Li}_2\left(1-\frac{t_1 u_1}{m_t^2 M^2}\right)
-2 L_\perp \ln \frac{-u_1}{m_t M}\right]
\nonumber \\
&
+4\, \bfw_{33}^{(1)}
\left[L_\perp+\ln \left(\frac{t_1 u_1}{m_t^2 M^2} \right)\right]
-2\, \bfw_{34}^{(1)}
\frac{1+\beta_t ^2}{\beta_t} \Big[L_\perp \ln x_s + f_{34} \Big]
\nonumber \\
&
+
4\, \epsilon\, \bfw_{13} 
\left[L_\perp^2 \ln \frac{-t_1}{m_t M}
-L_\perp \text{Li}_2\left(1-\frac{t_1 u_1}{m_t^2 M^2 }\right)+
\frac{\pi ^2}{6} \ln \frac{-t_1}{m_t M}
-\text{Li}_3\left(1-\frac{t_1 u_1}{m_t^2 M^2}\right)\right]
\nonumber \\
&
+
4\, \epsilon\, \bfw_{23} 
\left[L_\perp^2 \ln \frac{-u_1}{m_t M}
- L_\perp \text{Li}_2\left(1-\frac{t_1 u_1}{m_t^2 M^2 }\right)+
\frac{\pi^2}{6} \ln \frac{-u_1}{m_t M}
-\text{Li}_3\left(1-\frac{t_1 u_1}{m_t^2 M^2}\right)\right]
\nonumber \\
&
+
\epsilon\, \bfw_{33} 
\left[2 L_\perp^2 + 4  L_\perp \ln \frac{t_1 u_1}{m_t^2 M^2}
-4\, \text{Li}_2\left(1-\frac{t_1 u_1}{m_t^2 M^2 }\right)
+\frac{\pi^2}{3} \right]
\nonumber \\
&
-
\epsilon\, \bfw_{34} \frac{1+\beta_t ^2}{\beta_t} 
\Big[L_\perp^2 \ln x_s+ 2\, f_{34} L_\perp + 2\, J_{34} 
+\frac{\pi^2}{6} \ln x_s\Big]\,,
\label{eq:NLOSFres}
\end{align}
where
\begin{align}
  f_{34} & = 
  -\Litwo\left(-x_s\tan^2\frac{\theta}{2}\right)
  +\Litwo\left(-\frac{1}{x_s}\tan^2\frac{\theta}{2}\right)
  +4\ln x_s \ln\cos\frac{\theta}{2}\,,
  \\
  J_{34} & = -2 \int_0^{\beta_t} \frac{db}{b^2-1}
  \text{Li}_2\left(\frac{b^2 \sin ^2 \theta}{b^2-1}\right)\,,
\end{align}
and the result for $J_{34}$ can be expressed in terms of ordinary
polylogarithms.

\section{NNLO soft function: methods of calculation}
\label{sec:methods}

To calculate the next-to-next-to leading order contribution to the bare soft
function, one needs to sum several groups of diagrams, each multiplied by a
proper colour factor. The relevant master formula can be derived directly using
definitions given in Eqs.~(\ref{eq:Wdef})-(\ref{eq:Osggdef}), expanding the Wilson
lines and truncating the series at $\order{\as^2}$.
The diagrams can be grouped according to the number of distinct Wilson lines
connected by the gluons, which can be two, three or four. They can also be
classified based on how many lines are cut. And this can be two, one or none.
The bare soft function at NNLO reads
\begin{equation}
  \label{eq:bareSFNNLOdef}
  \bfS^{(2)}_\text{bare} = \bfS_{\text{2-cut}, gg} +
            \bfS_{\text{2-cut}, \qqbar} + 
            \bfS_{\text{1-cut}} +
            \bfS_{\text{0-cut}}\,.
\end{equation}
We shall now discuss groups of diagrams contributing to each of the terms in
the above equation. Then, we will describe the methods used to calculate them.

\subsubsection*{Double-cut diagrams}

\begin{figure}[t]
  \begin{center}
    \includegraphics[width=0.30\textwidth]{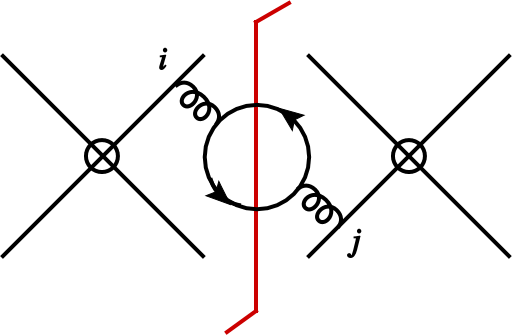}
    \hfill
    \includegraphics[width=0.30\textwidth]{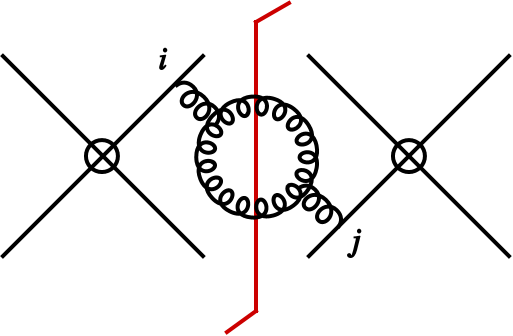}
    \hfill
    \includegraphics[width=0.30\textwidth]{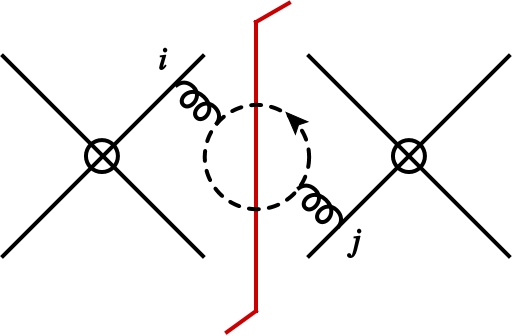}
  \end{center}
  \caption{Bubble diagrams contributing the NNLO soft function.
  }
  \label{fig:bubble-graphs}
\end{figure}

The gluons in double-cut diagrams can connect two, three or four Wilson lines.
Amongst the two-Wilson-line graphs, one can distinguish a special class of
\emph{bubble diagrams}, depicted in Fig.~\ref{fig:bubble-graphs}. Besides gluons,
they may also involve quarks and, as we work in the Feynman gauge, also the ghost bubble. 
Even though they belong to the class of two-cut diagrams, they are easier to
calculate than most of the graphs in this group. The
integrals corresponding to the bubble diagrams, and methods applied to evaluate
them, shall be discussed in detail in Section~\ref{sec:debubble}. The bubble graphs come with colour matrices which
are identical to those appearing in the NLO soft function, \ie $\bfw^{(1)}$, given
in Appendix~\ref{app:colmat}.

\begin{figure}[t]
  \begin{center}
    \includegraphics[width=0.30\textwidth]{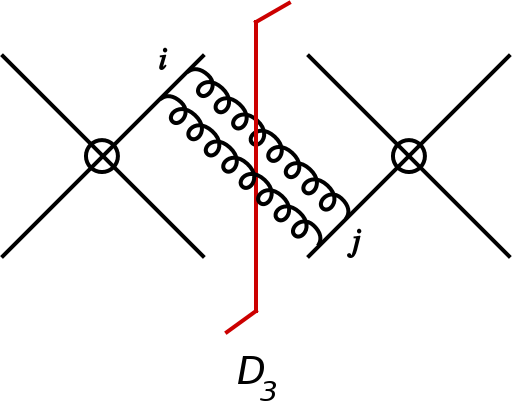}
    \hfill
    \includegraphics[width=0.30\textwidth]{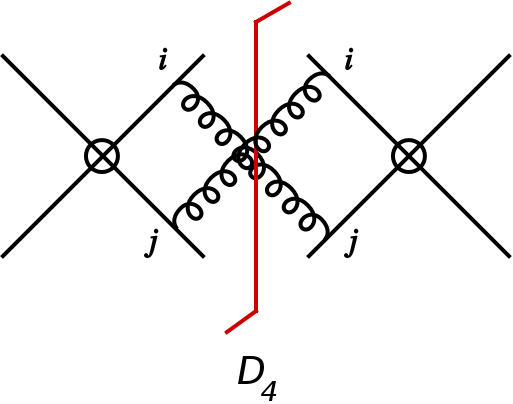}
    \hfill
    \includegraphics[width=0.30\textwidth]{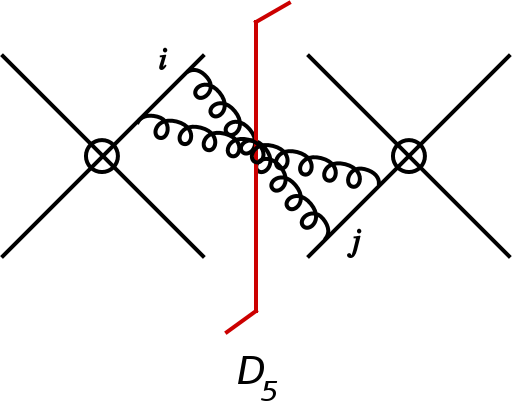}
    \hfill
    \includegraphics[width=0.30\textwidth]{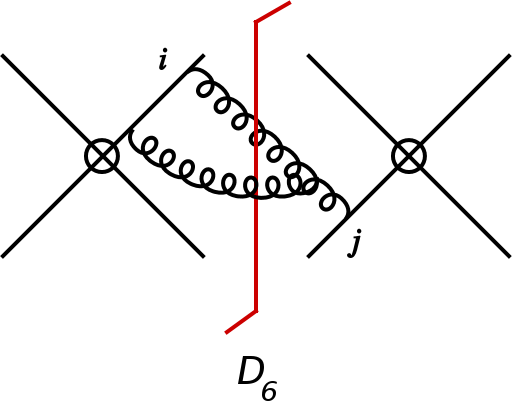}
    \hspace{30pt}
    \includegraphics[width=0.30\textwidth]{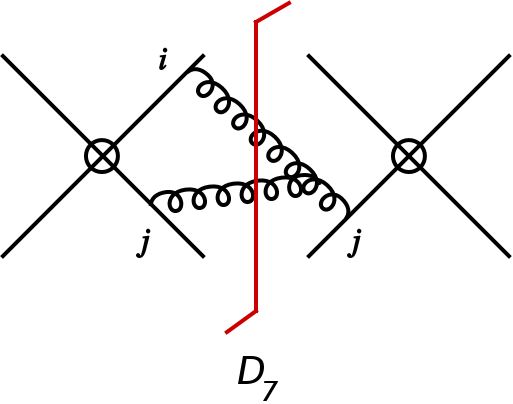}
  \end{center}
  \caption{Two-Wilson-line, double-cut diagrams required for the calculation of
  the NNLO soft function.
  }
  \label{fig:2WL2cut}
\end{figure}

The second group consists of non-bubble graphs in which the gluons attach to
only two distinct Wilson lines: $i$ an $j$.
These graphs are shown in Fig.~\ref{fig:2WL2cut}.  We see that they
include both abelian and non-abelian structures.

\begin{figure}[t]
  \begin{center}
    \includegraphics[width=0.30\textwidth]{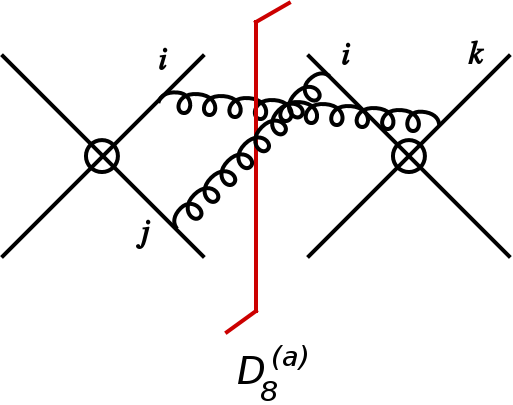}
    \hfill
    \includegraphics[width=0.30\textwidth]{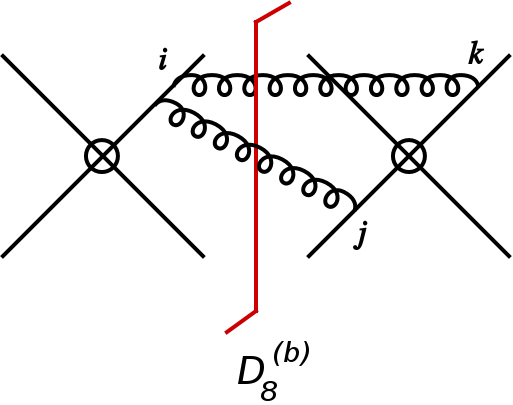}
    \hfill
    \includegraphics[width=0.30\textwidth]{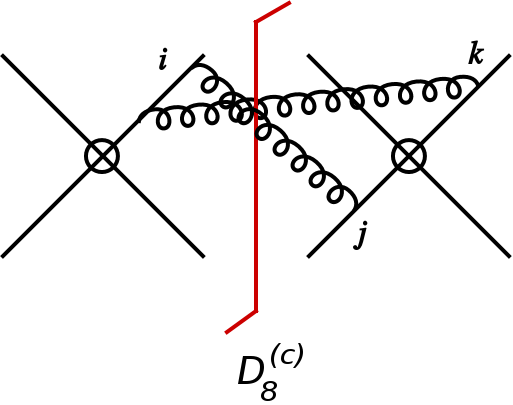}
  \end{center}
  \caption{Three-Wilson-line, double-cut diagrams required for the calculation
  of the NNLO soft function.
  }
  \label{fig:3WL2cut}
\end{figure}

The third group is formed by the three-Wilson-line, double-cut  diagrams shown
in Fig.~\ref{fig:3WL2cut}. 
The sum of the integral corresponding to $D^{(a)}_8$ and
its complex conjugate produce an integral which is a convolution of two NLO
expressions. Exactly the same expression is obtained when $D^{(b)}_8$ and
$D^{(c)}_8$ are added to their complex conjugates. And all the sums 
$D^{(i)}_8 + D^{(i)*}_8 $, for $i = a,b$ or $c$, are multiplied by the same colour factor
 $\{\cT_i^a, \cT_i^b\} \cT_j^a \cT_k^b$. 
The above observations were also made in the related calculation of the NNLO
soft function for top pair production in the threshold
limit~\cite{Ferroglia:2012uy, Wang:2018vgu}.

\begin{figure}[t]
  \begin{minipage}[t]{0.48\textwidth}
  \begin{center}
    \includegraphics[width=0.625\textwidth]{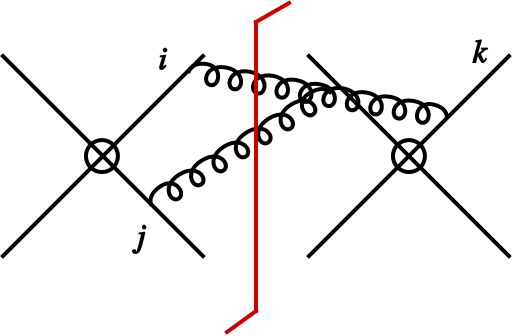}
  \end{center}
  \caption{Three-Wilson-line, double cut diagram appearing in the calculation of
  the NNLO soft function.
  }
  \label{fig:3WLNA}
  \end{minipage}
  \hfill
  \begin{minipage}[t]{0.48\textwidth}
  \begin{center}
    \includegraphics[width=0.625\textwidth]{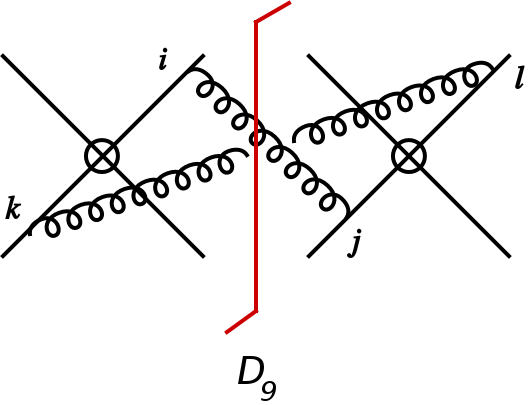}
  \end{center}
  \caption{Four-Wilson-line, double cut diagram required in the calculation of
  the NNLO soft function.
  }
  \label{fig:4WL2cut}
  \end{minipage}
\end{figure}

The abelian graphs depicted in Fig.~\ref{fig:3WL2cut} constitute the only
non-vanishing three-Wilson-line contribution in the group of double-cut diagrams.
This is because the non-abelian, three-Wilson-line graphs cancel when summed
over colour structures, which can be understood by analyzing the graph depicted
in Fig.~\ref{fig:3WLNA}. 
Since it is a double-cut diagram, the corresponding integral is a real-valued
function. And it is multiplied by the colour factor $i f^{abc}\cT^a_i T^b_j
T^c_k$. The complete soft function receives also a contribution from a diagram
which is a complex conjugate of Fig.~\ref{fig:3WLNA} and the complex
conjugation only affects the colour factor, turning it into $-i f^{abc}\cT^a_i
T^b_j T^c_k$.
Hence, the diagram of Fig.~\ref{fig:3WLNA} and its complex
conjugate cancel, and the non-abelian graphs with three distinct Wilson lines
connected by the gluons do not contribute to the soft function.

We emphasize that this happens only because the integrals are real
functions, as they originate from double-cut diagrams. This property will not be
true for single-cut, non-abelian diagrams, and we will see that the
corresponding contributions do not vanish when summed over all diagrams.

Finally, the double-cut part of the soft functions receives contributions from a
subset of diagrams shown in Fig.~\ref{fig:4WL2cut}, where the gluons connect
four distinct Wilson lines. The corresponding expressions take forms of
convolutions of the NLO integrals, similarly to the cases of $D_4$ and $D_8$
from Fig.~\ref{fig:2WL2cut} and \ref{fig:3WL2cut}.  
The four-Wilson-line diagrams come with the colour operator
$\cT_i^a \cT_j^a \cT_k^b \cT_l^b$. 

The complete expression for the NNLO soft function derived using the
diagrammatic approach described above can be obtained alternatively by taking
the soft limit of the relevant matrix elements in QCD. 

In the case of the $gg$ final state, in the limit $k, l \to 0$, the squared
matrix element factorizes as~\cite{Czakon:2014oma}
\begin{align}
\cm^{*(0)}_{g,g,a_1,\dots}(k,l,p_1,\dots)\,
\cm^{(0)}_{g,g,a_1,\dots}(k,l,p_1,\dots)
\hspace{-120pt}
&
\nonumber \\[0.5em]
\simeq \
&
\frac{1}{2}\sum_{ijkl}{\cal S}_{ij}(k){\cal S}_{kl}(l) \,
\la\cm^{(0)}_{a_1,\dots}(p_1,\dots)|
\left\{\cT_i\cdot\cT_j, \cT_k\cdot\cT_l\right\} 
|\cm^{(0)}_{a_1,\dots}(p_1,\dots)\ra
\nonumber \\
&
-C_A\sum_{ij}{\cal S}_{ij}(k,l) \,
\la\cm^{(0)}_{a_1,\dots}(p_1,\dots)|\cT_i\cdot\cT_j
|\cm^{(0)}_{a_1,\dots}(p_1,\dots)\ra\,,
\label{eq:2cutmaster}
\end{align}
where ${\cal S}_{ij}(k)$ is the eikonal integrand
\begin{equation}
  {\cal S}_{ij}(k) = \frac{p_i \cdot p_j}{(p_i \cdot k)\, (p_j\cdot k)}\,,
  \label{eq:eikfun}
\end{equation}
which we had used to construct the NLO soft function integrals in
Eq.~(\ref{eq:soft-func-nlo-mom-space-ddim}), and $p_i$ are the $d$-momenta of
the external partons. 
We note that the functions ${\cal S}_{ij}(k)$ and ${\cal S}_{ij}(k,l)$ are
invariant with respect to rescalings of the momenta of external particles of the
Born process.  Therefore, they can be expressed in terms of velocities
\begin{equation}
  n_i = 
  \left\{
  \begin{array}{l}
     n     \quad \text{for} \quad    i = 1\, \\
     \nbar \quad \text{for} \quad    i = 2\, \\
     \tilde v_i   \quad \text{for} \quad \! i = 3, 4\,
  \end{array}
  \right..
\end{equation}
The function ${\cal S}_{ij}(k,l)$ can be split into two parts
\begin{equation}
  \label{eq:soft}
  {\cal S}_{ij}(k,l) = {\cal S}^{m=0}_{ij}(k,l) + \left( m_i^2
  \; {\cal S}^{m \neq 0}_{ij}(k,l) + m_j^2 \; {\cal S}^{m \neq
    0}_{ji}(k,l) \right) \; ,
\end{equation}
where $m_i$ and $m_j$ are the masses of the external particles.
The first term in Eq.~(\ref{eq:soft}) has been given in
Ref.~\cite{Catani:1999ss} and reads
\begin{eqnarray}
{\cal S}^{m=0}_{ij}(k,l) &=& \frac{(1-\epsilon)}{(k \cdot l)^2}
\frac{p_i \cdot k \; p_j \cdot l + p_i \cdot l \; p_j \cdot
  k}{p_i \cdot (k+l) \; p_j \cdot (k+l)} \nonumber
\\ \nonumber \\
&& - \frac{(p_i \cdot p_j)^2}{2 \; p_i \cdot k \; p_j \cdot l \; p_i
  \cdot l \; p_j \cdot k} \left[ 2 - \frac{p_i \cdot k \; p_j
    \cdot l + p_i \cdot l \; p_j \cdot k}{p_i \cdot (k+l) \;
    p_j \cdot (k+l)} \right] \nonumber \\  \nonumber \\
&& + \frac{p_i \cdot p_j}{2 \; k \cdot l} \left[ \frac{2}{p_i
    \cdot k \; p_j \cdot l} + \frac{2}{p_j \cdot k \; p_i \cdot
    l} - \frac{1}{p_i \cdot (k+l) \; p_j \cdot (k+l)}
  \right. \nonumber \\ \nonumber \\
&& \times \left. \left( 4 + \frac{(p_i \cdot k \; p_j \cdot l +
    p_i \cdot l \; p_j \cdot k)^2}{\; p_i \cdot k \; p_j \cdot
    l \; p_i \cdot l \; p_j \cdot k} \right) \right] \; .
    \label{eq:Smassless}
\end{eqnarray}
In the above equation, the first line comes solely from the gluon and ghost
bubble diagrams of Fig.~\ref{fig:bubble-graphs}. The second line originates from
the $C_A$ part of diagrams which do not involve the triple-gluon vertex, that
is $D_4$ and $D_5$ in Fig.~\ref{fig:2WL2cut}. Finally, the last two lines
receive contributions from the non-abelian diagrams $D_6$, $D_7$ and the gauge
bubble.

The second contribution in Eq.~(\ref{eq:soft}) was derived in
Ref.~\cite{Czakon:2011ve} and represents additional terms generated by
non-vanishing masses. The relevant function is
\begin{eqnarray}
  {\cal S}^{m\neq0}_{ij}(k,l) &=
  &  - \frac{1}{4  \; k \cdot l \; p_i \cdot
  k \; p_i \cdot l} + \frac{p_i \cdot p_j \; p_j
  \cdot (k+l)}{2 \; p_i \cdot k \; p_j \cdot l \; p_i \cdot l \;
    p_j \cdot k \; p_i \cdot (k+l)} \nonumber \\ \nonumber \\
  && - \frac{1}{2 \; k \cdot l \; p_i \cdot (k+l) \; p_j \cdot
    (k+l) } \left( \frac{(p_j \cdot k)^2}{p_i \cdot k \; p_j
    \cdot l} + \frac{(p_j \cdot l)^2}{p_i \cdot l \; p_j \cdot
    k} \right) \; . \nonumber \\
    \label{eq:Smassive}
\end{eqnarray}
Here, the first and the third term come from the $C_A$ part of diagrams
which do not involve the triple-gluon vertex, whereas the second term arises due to
non-abelian contributions, including the gauge bubble.

In the case of the final-state $q{\bar q}$-pair, in the limit $k, l \to 0$, the
matrix element factorizes as~\cite{Czakon:2014oma}
\begin{align}
  \label{qqsoftfac}
  \cm^{*(0)}_{q,{\bar q},a_1,\dots}(k,l,p_1,\dots)\,
  \cm^{(0)}_{q,{\bar q},a_1,\dots}(k,l,p_1,\dots) 
  &  
  \nonumber \\[0.5em]
  &
  \hspace{-100pt}
  \simeq
  T_F \sum_{ij} {\cal I}_{ij}(k,l) \,
  \la \cm^{(0)}_{a_1,\dots}(p_1,\dots)|\cT_i \cdot
  \cT_j|\cm^{(0)}_{a_1,\dots}(p_1,\dots)\ra\; ,
\end{align}
where the function $ {\cal I}_{ij}(k,l)$ has the form
\begin{equation}
  \label{eq:Iij1}
  {\cal I}_{ij}(k,l) = \frac{(p_i \cdot k)\, (p_j \cdot l)
  + (p_j \cdot k)\, (p_i \cdot l) - (p_i \cdot p_j) 
  \,(k \cdot l)}{(k \cdot l)^2 
  \,[p_i\cdot (k+l)]\, [p_j \cdot (k+l)]} \,.
\end{equation}

The above expressions can be used directly to calculate our soft function of
interest by applying the following formula
\begin{align}
  \bfS_{i\bar i}^{(2), f\bar f} (q_T) & = 
   (4\pi)^4
  \left(\frac{\mu^2 e^{\gamma_E}}{4\pi}\right)^{2\epsilon}\!\!\!\!\nu^{2\alpha}
  \frac{(2\pi)^{\frac{d}{2}-1}}{ S_{d-3} \,q_T^{d-3}}
  \int 
  \frac{d^d\, \delta_+(k^2)}{(2\pi)^{d-1} (n\cdot k)^\alpha}\, \frac{d^d l\,
  \delta_+(l^2)}{(2\pi)^{d-1}(n\cdot l)^\alpha}\; 
    \delta(q_T- | k_\perp+l_\perp |)\,
  \nonumber \\
  &
  \hspace{40pt}
  \times
  \langle c_I^{i\bar i} | 
  \cm^{*\, (0)}_{f,\bar f,a_1,\dots}(k,l,p_1,\dots)
  \cm^{(0)}_{f,\bar f,a_1,\dots}(k,l,p_1,\dots)
  |c_J^{i\bar i} \rangle\,,
  \label{eq:S2cutfromM22cut}
\end{align}
where the case with $f=q$ corresponds to the $q\bar q$ final state while
the case with $f=g$ corresponds to the $gg$ final state.

We have checked through direct calculation that the expression for the NNLO soft
function  derived using definitions given in
Eqs.~(\ref{eq:Wdef})-(\ref{eq:Osggdef}), \ie expanding the Wilson lines and
truncating the series at $\order{\as^2}$, matches exactly the 
formula (\ref{eq:S2cutfromM22cut}).

To complete the definition of Eq.~(\ref{eq:S2cutfromM22cut}), we need to specify
the colour matrices $\bm{w}_{ij}^{(1)}$ and $\bm{w}_{ijkl}^{(2S)}$, following
Eq.~(\ref{eq:w1w2def}).  
The matrices $\bm{w}_{ij}^{(1)}$, built from products of two colour operators,
are identical with those found in the calculation of the NLO soft function.
The matrices $\bm{w}_{ijkl}^{(2S)}$, constructed from the anticommutator,
involve products of four colour operators, and appear for the first time at
NNLO.
The complete set of the $\bm{w}_{ij}^{(1)}$ and $\bm{w}_{ijkl}^{(2S)}$ matrices,
in both the $q\bar q$ and the $gg$ channel, is given in
Appendix~\ref{app:colmat}.

\subsubsection*{Single-cut diagrams}

\begin{figure}[t]
  \begin{center}
    \includegraphics[width=0.30\textwidth]{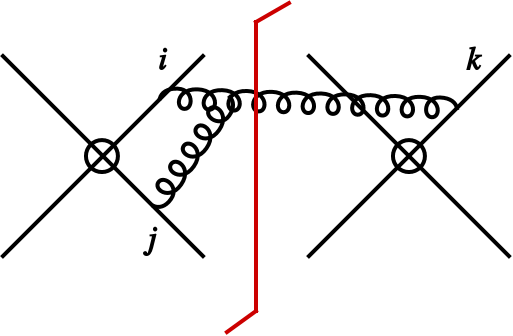}
    \hfill
    \includegraphics[width=0.30\textwidth]{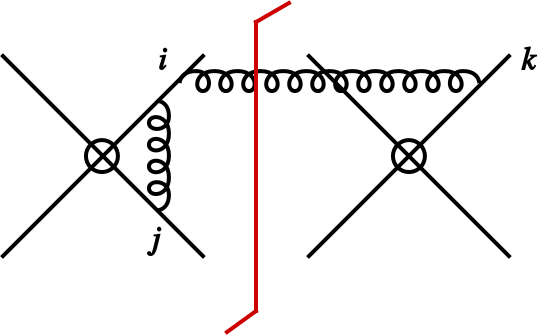}
    \hfill
    \includegraphics[width=0.30\textwidth]{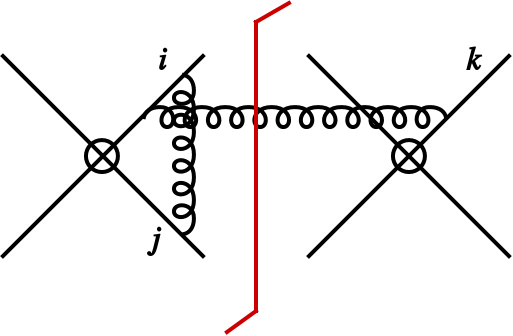}
  \end{center}
  \caption{
  Single-cut diagrams contributing to the NNLO soft function.
  }
  \label{fig:1cutdiagrams}
\end{figure}

The single-cut diagrams required for the calculation of the NNLO soft function
for top pair production are shown in Fig.~\ref{fig:1cutdiagrams}. In this
class of graphs, one gluon is real and the other runs in a loop. Because of the
latter, the integrals from the single-cut graphs produce also an imaginary
contribution.

As in the case of the double-cuts , also the single-cut term derived
using the diagrammatic approach described above can be obtained alternatively by
taking the soft limit of the relevant matrix elements in QCD. This results
in the formula~\cite{Bierenbaum:2011gg}
\begin{align}
  &
  \hspace{-50pt}
  \cm^{*\, (1)}_{g ,a_1,\dots}(k,p_1,\dots)
  \cm^{(0)}_{g ,a_1,\dots}(k,p_1,\dots) + \text{c.c.}
  \nonumber \\[0.5em]
  & = - 
  \Bigg[ 
    2C_A\sum_{i\neq j = 1}^n
    \left(e_{ij}-e_{ii}\right) R_{ij}\langle M^{(0)}(n) 
    \vert \cT_i\cdot \cT_j \vert
    M^{(0)}(n)  \rangle  
    \nonumber \\
    & \hspace{70pt}
    -  4\pi\!\!\sum_{i\neq j\neq k = 1}^n e_{ik}I_{ij} \langle
    M^{(0)}(n)\vert  f^{abc} \cT_i^a \cT_j^b \cT_k^c \vert M^{(0)}(n)  \rangle
  \Bigg]\,,
  \label{eq:1cutmaster}
\end{align}
where
\begin{equation}
  \label{eq:eikprop}
  e_i^\mu = \frac{p_i^\mu}{p_i \cdot k}\,,
\end{equation}
while $ R_{ij}$ and $I_{ij}$ correspond to the real and imaginary parts
from integration over the loop momentum and they were obtained in
Refs.~\cite{Bierenbaum:2011gg, Czakon:2018iev}.  We notice that a new,
antisymmetric colour structure appears in the imaginary part:
$if^{abc} \cT_i^a \cT_j^b \cT_k^c  = [\cT_i \cdot \cT_k, \cT_j \cdot \cT_k]$.
The corresponding colour matrix $\bm{w}_{ijk}^{(2A)}$ is defined in
Eq.~(\ref{eq:w3def}) and its explicit expressions are given in
Appendix~\ref{app:colmat}.

We have checked through direct calculation, based on definitions given in
Eqs.~(\ref{eq:Wdef})-(\ref{eq:Osggdef}), that we reproduce the single-cut
expression of Eq.~(\ref{eq:1cutmaster}). With the latter, we can then
determine the real-virtual part of our soft function
\begin{align}
  \bfS_{i\bar i}^{(2), g} (q_T) & = 
   (4\pi)^4
  \left(\frac{\mu^2 e^{\gamma_E}}{4\pi}\right)^{2\epsilon}
  \!\!\!\!\nu^{\alpha}
  \frac{(2\pi)^{\frac{d}{2}-1}}{ S_{d-3}\,q_T^{d-3}}
  \int \! \, \frac{d^d k \,\delta_+(k^2) }{(2\pi)^{d-1} (n\cdot k)^\alpha} 
   \; \delta(q_T-k_T)\,
  \nonumber \\
  &
  \hspace{40pt}
  \times
  \langle c_I^{i\bar i} | 
  \cm^{*\, (1)}_{g ,a_1,\dots}(k,p_1,\dots)
  \cm^{(0)}_{g ,a_1,\dots}(k,p_1,\dots) + \text{c.c.}\,
  |c_J^{i\bar i} \rangle\,.
  \label{eq:S2cutfromM21cut}
\end{align}
%

\subsubsection*{Zero-cut diagrams}

Purely virtual, two-loop diagrams, do not involve a cut gluon. Therefore, the
measurement function $\delta^{(d-2)}(f(k,l) - q_\perp)$ does not appear in the
corresponding integrals. As a consequence, these integrals are scaleless and
vanish in dimensional regularization. Hence, the NNLO soft function for top
pair production does not receive contributions from two-loop diagrams.

\subsection{Symmetries between integrals}
\label{sec:intsym}

Our double-cut, soft function integrals have the general structure
\begin{align}
  \tilde I(\{p_i, p_j\}) &= \int d^dk\, d^d l\, \delta^{(+)}(k^2)
  \delta^{(+)}(l^2) h(p_i\cdot p_j, p_i\cdot k, p_i \cdot l)\,
  \delta^{(d-2)}(k_\perp+l_\perp -q_\perp)\,,
  \label{eq:Isymmetry}
\end{align}
where $p_i$ are the momenta of external particles.

The above integral can depend only on the scalar productions $p_i \cdot p_j$,
which are invariant with respect to Lorentz transformations 
$p^\mu \to \Lambda^\mu{}_\nu\, p^\nu$. 
To balance the transformation of the
external momenta in the scalar products $p_i \cdot k$ and $p_i \cdot l$ one
needs to transform the gluon momenta $k$ and $l$ with
$\left(\Lambda^{-1}\right)^\mu{}_\nu$. 
This, however, does not leave the integral unchanged for a general Lorentz
transformation because of the transverse delta function appearing in
Eq.~(\ref{eq:Isymmetry}).
Therefore, the integral $\tilde I(\{p_i, p_j\})$ is invariant
only under a subgroup of the Lorentz group which involves:
\begin{itemize}
  \item
  rescaling of the light-cone components of the momenta $k$ and $l$ compensated
  by inverse rescaling of the light-cone components of the external momenta,
  \item
  rotations of the above momenta in the transverse plane such that 
  $|k_\perp + l_\perp| = q_T$.
\end{itemize}
Let us denote
\begin{align}
  p\cdot q & \qquad 
  \text{usual, $d$-dimensional scalar product}, \\
  p * q & = p_0 q_0 - p_3 q_3 = \frac12\left(p_+ q_- + p_- q_+\right)\,.
\end{align}

Given the above reduced Lorentz symmetry, and the fact that the result can
only depend on external momenta, we conclude, that our integral must be a
function of $p_i * p_j$ and $p_{i,\perp} \cdot p_{j,\perp}$ or, equivalently,
$p_i * p_j$ and $p_i \cdot p_j$.

As is clear from Eqs.~(\ref{eq:eikfun})-(\ref{eq:Iij1}), the integrand is
invariant under the rescaling of $p_3 = \lambda_3 p_3$ and $p_4 = \lambda_4
p_4$, for arbitrary $\lambda$, which means that the result can only be a
function of ratios of the above scalar products.

Integrals which do not exhibit $\alpha$ poles are, in addition, invariant under
rescaling of $p_1$. But, here, one cannot form a variable of the type $p_1
\cdot p_3/\sqrt{p_1^2 p_3^2}$ because $p_1$ is massless. As a consequence, the
only possibility is $p_3*p_3/p_3^2 = p_4*p_4/p_4^2 $. 
This is why some abelian integrals are equal. 
In the cases of the integrals with $\alpha$ poles, rescaling of $p_1$ is broken
by the analytic regulator: $I(\lambda p_1) =  \lambda^{-\alpha} I(p_1)$.

\subsection{Differential equations approach and bubble diagrams}
\label{sec:debubble}

\begin{figure}[t]
  \begin{center}
    \includegraphics[width=0.62\textwidth]{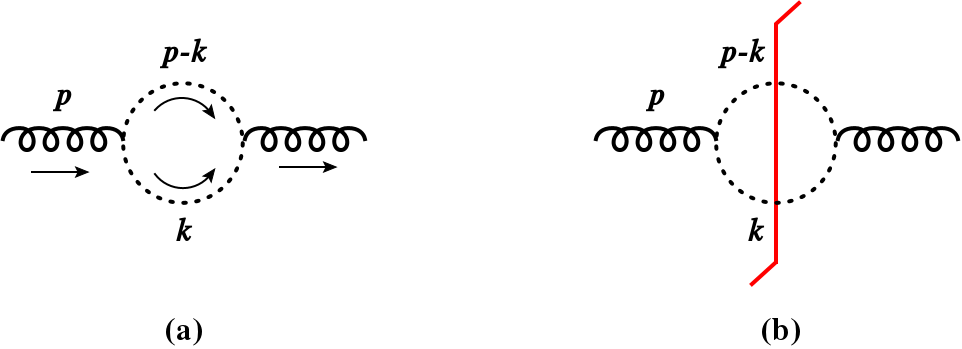}
  \end{center}
  \caption{
  Gluon vacuum polarization diagram (a) and its cut counterpart (b).
  Dashed lines correspond to either a quark, a gluon or a ghost.
  }
  \label{fig:vp-diagram}
\end{figure}

With an exception of the propagators introduced to regularize rapidity
divergences, the quark, gluon and ghost bubble, as well as part of the triple
gluon vertex diagrams depicted in Fig.~\ref{fig:2WL2cut}, which we shall call
``tadpole'', depend only on the
momenta $k$ and $k+l$.
This feature allows one to first integrate over $k+l$ and then solve the
integral over $k$ with help of the differential equation 
approach~\cite{Kotikov:1990kg, Remiddi:1997ny}. The integration
can be performed mostly analytically, with an exception of a few one-dimensional
integrals which we integrate by numerical methods.

\subsubsection{Vacuum polarization tensor}

We consider the process depicted in Fig.~\ref{fig:vp-diagram}~(a), where the
particle running in the loop can be a quark, a gluon or a ghost. The
corresponding vacuum polarization integrals read 
\begin{equation}
  \im \Pi^{\mu\nu}(\alpha) = 
  C_f \frac{g^2}{16}\, e^{\epsilon \gamma_E} \mu^{2\epsilon} \nu^{2\alpha}
  \pi^{-4+\epsilon}
  \int d^d k \frac{N^{\mu\nu}(k,p)}{
  \left(k\cdot n\right)^\alpha\, 
  \left(p\cdot n - k\cdot n\right)^\alpha\, k^2\, (p-k)^2}\,,
  \label{eq:ImPiint}
\end{equation}
where $C_f = C_A$ for the gluon bubble and the tadpole, while $C_f=-T_F$ for
the quark bubble.  The numerator $N^{\mu\nu}(k,p)$ depends on the particle in
the loop.
For the quark loop we have
%
%
\begin{eqnarray}
  N^{\mu\nu} = 
    8\, k^\mu k^\nu
    -4k^\mu p^\nu 
    -4p^\mu k^\nu 
    -4g^{\mu\nu} k^2 
    +4g^{\mu\nu} k\cdot p\,,
\end{eqnarray}
for the gluon+ghost loop
\begin{eqnarray}
  N^{\mu\nu} & =  &
    (2d-4)\, k^\mu k^\nu +
    \frac{2d-3}{2}\, k^\mu p^\nu +
    \frac{2d-5}{2}\, p^\mu k^\nu +
    \frac{d-6}{2}\, p^\mu p^\nu 
  \nonumber \\
  & & 
   + g^{\mu\nu}\, k^2 +
    g^{\mu\nu}\, k\cdot p +
    \frac{5}{2}\, g^{\mu\nu} p^2\,,
\end{eqnarray}
and for the tadpole
\begin{eqnarray}
  N^{\mu\nu} = - 4 g^{\mu\nu} p^2\,.
\end{eqnarray}
The directions of momenta are indicated with arrows in
Fig.~\ref{fig:vp-diagram}~(a).

All propagators are assumed to be defined with the $+i\epsilon$ prescription.
The most generic self-energy tensor that can be formed out the $d$-vectors $p$
and $n$, and the metric tensor $g^{\mu\nu}$ is
\begin{equation}
  \im \Pi^{\mu\nu}(\alpha) = 
  T_{00}\, g^{\mu\nu} +
  T_{pp}\, p^{\mu}p^{\nu} +
  T_{nn}\, n^{\mu}n^{\nu} +
  T_{pn}\, \left(n^{\mu} p^{\nu}+p^{\mu} n^{\nu}\right)\,.
  \label{eq:ImPi}
\end{equation}
The coefficients $T_{ij}$ can be expressed in terms of two scalar integrals,
$A_0$ and $B_0$, through the procedure of Passarino-Veltman
reduction~\cite{Passarino:1978jh}
\begin{align}
  A_0^{(a)}(\alpha_1,\alpha_2) &= 
  \int\frac{d^d k}{(n\cdot k)^{\alpha_1}\, (n\cdot (p+k))^{\alpha_2}\, k^2}\,,
  \\[0.5em]
  B_0^{(a)}(\alpha_1,\alpha_2) &= 
  \int\frac{d^d k}{(-n\cdot k)^{\alpha_1}\, (n\cdot (p+k))^{\alpha_2}\, k^2\,
  (p+k)^2}\,.
\end{align}
The integral $A_0$ can be shown to be scaleless and, therefore, it
vanishes. Hence, it turns out that the only two-point integral that we need to
determine $\im \Pi^{\mu\nu}$ is $B_0$, and this integral can be
calculated exactly by means of Schwinger parametrization. The result takes the
form
%
%
\begin{equation}
  B_0^{(a)}(\alpha_1,\alpha_2) = 
  i \pi ^{\frac{d}{2}}\,
  \frac{\Gamma (\epsilon ) \Gamma (1-\alpha_1 -\epsilon )  
        \Gamma (1-\alpha_2 -\epsilon )}
  {\Gamma (2-\alpha_1 -\alpha_2 -2\epsilon)}\,
  \left(-p^2\right)^{-\epsilon }\, (n \cdot p)^{-\alpha_1 - \alpha_2 }\,.
  \label{eq:B0virt}
\end{equation}
Unitarity allows us to obtain also a version of this function which
corresponds to the cut diagram of Fig.~\ref{fig:vp-diagram}~(b)
%
%
\begin{equation}
  B_0^{(b)}(\alpha_1,\alpha_2) = 
  -2 \pi ^{\frac{d}{2}}\,
  \frac{\Gamma (\epsilon ) \Gamma (1-\alpha_1 -\epsilon )  
        \Gamma (1-\alpha_2 -\epsilon )}
  {\Gamma (2-\alpha_1 -\alpha_2 -2\epsilon)}\,
  \sin (\pi  \epsilon )\,
  p^{-2 \epsilon }\, (n \cdot p)^{-\alpha_1 - \alpha_2 }\,.
  \label{eq:B0real}
\end{equation}
Finally, the coefficients in Eq.~(\ref{eq:ImPi}) take the following forms. For
the quark bubble
\begin{align}
  \label{eq:Tijquark}
  T_{00} &= \frac{2 \left(2 (1-\epsilon)^2-\alpha  (1-2 \epsilon)\right)}
  {(1-\epsilon) (3-2 \alpha -2 \epsilon)}
  \, \tilde B_0(\alpha)\, p^2\,,
  &\quad
  %
  T_{pp} &= -\frac{4 (1-\alpha -\epsilon)}{3-2 \alpha -2 \epsilon} 
  \, \tilde B_0(\alpha)\,, \\
  T_{nn} &= 
  \frac{2 \alpha }{(1-\epsilon) (3-2 \alpha -2 \epsilon)}
  \, \tilde B_0(\alpha)\,
  \frac{p^4}{\left(n\cdot p\right)^2}\,,
  &\quad
  %
  T_{pn} &= 
  -\frac{2 \alpha}{(1-\epsilon) (3-2 \alpha -2 \epsilon) }
  \, \tilde B_0(\alpha)\,
  \frac{p^2}{n\cdot p}
  \,,
  \nonumber
\end{align}
with
\begin{equation}
  \tilde B_0(\alpha) = -\frac{g^2}{16} T_F 
  \frac{e^{\epsilon\gamma_E}\mu^{2\epsilon}\nu^{2\alpha}}{\pi^{4-\epsilon}}
  B_0(\alpha,\alpha)\,.
  \label{eq:tB0quark}
\end{equation}
For the gauge (\ie gluon + ghost) bubble
\begin{align}
  T_{00} &= 
  \frac{5 - 3 \alpha - 3 \epsilon}{3 - 2 \alpha - 2 \epsilon}
  \, \tilde B_0(\alpha)\, p^2\,,
  &\quad
  T_{pp} &= \frac{-5 + 4 \alpha + 3 \epsilon}{3 - 2 \alpha - 2 \epsilon}  
  \, \tilde B_0(\alpha)\,, \\
  T_{nn} &= 
  \frac{\alpha }{3 - 2 \alpha - 2 \epsilon}
  \, \tilde B_0(\alpha)\,
  \frac{p^4}{\left(n\cdot p\right)^2}\,,
  &\quad
  T_{pn} &= 
  -\frac{\alpha}{3 - 2 \alpha - 2 \epsilon }
  \, \tilde B_0(\alpha)\,
  \frac{p^2}{n\cdot p}
  \,,
  \nonumber
\end{align}
with
\begin{equation}
  \tilde B_0(\alpha) = 
  \frac{g^2}{16} C_A 
  \frac{e^{\epsilon\gamma_E}\mu^{2\epsilon}\nu^{2\alpha}}{\pi^{4-\epsilon}}
  B_0(\alpha,\alpha)\,.
  \label{eq:tB0gauge}
\end{equation}
And for the tadpole
\begin{align}
  T_{00} &= 
  -\frac{g^2}{4} C_A 
  \frac{e^{\epsilon\gamma_E}\mu^{2\epsilon}\nu^{2\alpha}}{\pi^{4-\epsilon}}\,
  B_0(\alpha,\alpha)\, p^2\,,
  &\quad
  T_{pp} &= 0 \,, 
  &\quad
  T_{nn} &= 0 \,,
  &\quad
  T_{pn} &= 0 \,.
  \label{eq:Tijtadpole}
\end{align}
The function $B_0(\alpha,\alpha)$ in Eqs.~(\ref{eq:tB0quark}),
(\ref{eq:tB0gauge}) and (\ref{eq:Tijtadpole}) can be given either by
Eq.~(\ref{eq:B0virt}) or by Eq.~(\ref{eq:B0real}). In our soft function
calculation, we will use the latter, cut version of this two-point integral.

One can check that the results given in this section recover standard
expressions for the gluon vacuum polarization tensor in the limit $\alpha \to
0$, where, in particular, the coefficients $T_{nn}$ and $T_{pn}$ vanish.

\subsubsection{Soft function integrals}

The above results for the quark and gauge bubble can now be embedded in the
two-Wilson-line soft function graphs, as depicted in
Fig.~\ref{fig:bubble-graphs}. 
The tadpole integrals correspond to the diagram $D_6$ of
Fig.~\ref{fig:2WL2cut} with the eikonal propagator connecting the two
gluons replaced by a pinch.
The integrals that appear have the following structure 
\begin{equation}
  I = \int\frac{d^d k\, \delta(k_T^2-1)\, \theta(k^2)\, \theta(k_0)}
      {(n\cdot k)^{a_1+2\alpha} (\nbar\cdot k)^{a_2} (v_3 \cdot k)^{a_3} (v_4
      \cdot k)^{a_4} (k^2)^{a_0+\epsilon}}\,.
  \label{eq:sfnf-general-structure}
\end{equation}

We note that, as in the case of the NLO soft function discussed in
Section~\ref{sec:NLOsf}, the integration measure can be written as $d^d k = d
k_+ d k_- d^{d-2} k_\perp$ and the $k_+$ and $k_-$ components are integrated
from minus to plus infinity.
However, the phase space of the integration over the light-cone momenta is
restricted, this time by $\theta(k^2)\, \theta(k^0) = \theta(k_+ k_- -
k_T^2)\, \theta(k_+ + k_-)$.  Because of the first theta function, the
integration is fixed not to a line, as in the case of the NLO soft function
depicted in Fig.~\ref{fig:light-cone-phase-space}~(left), but to the region
defined by the condition $k_+ k_- > k_T^2$, which means that $k_+$ and $k_-$
have to be both positive or negative.  The second theta function chooses
$k_+$ and $k_-$ to be both positive. 
Hence, the $d$-momentum of the off-shell gluon which appears in the bubble and
the tadpole diagrams is integrated over  the region depicted in
Fig.~\ref{fig:light-cone-phase-space}~(right).

We would like to solve the class of integrals from
Eq.~(\ref{eq:sfnf-general-structure}) by means of the method of differential
equations~\cite{Kotikov:1990kg, Remiddi:1997ny}, with help of
reverse unitarity~\cite{Anastasiou:2002yz}, which allows one to turn delta
functions into propagators.
What prevents us from direct use of the latter is the $\theta(k^2)$ function.
However, we can trade this theta function for the Dirac delta function at the
cost of introducing an extra integration over a spurious mass, $m^2$. Namely, we
multiply Eq.~(\ref{eq:sfnf-general-structure}) by 
\begin{equation}
  1 = \int_0^\infty dm^2 \delta(k^2-m^2)\,,
\end{equation}
which leads to
\begin{align}
  I & = \int_0^\infty dm^2 \delta(k^2-m^2)
      \int\frac{d^d k\, \delta(k_T^2-1)\, \theta(k^2)\, \theta(k_0)}
      {(n\cdot k)^{a_1+2\alpha} (\nbar\cdot k)^{a_2} (v_3 \cdot k)^{a_3} (v_4
      \cdot k)^{a_4} (k^2)^{a_0+\epsilon}}
    \nonumber \\
    & = \int_0^\infty dm^2 \delta(k^2-m^2)
      \int\frac{d^d k\, \delta(k_T^2-1)\, \theta(m^2)\, \theta(k_0)}
      {(n\cdot k)^{a_1+2\alpha} (\nbar\cdot k)^{a_2} (v_3 \cdot k)^{a_3} (v_4
      \cdot k)^{a_4} (m^2)^{a_0+\epsilon}}
    \nonumber \\
    & = \int_0^\infty \frac{dm^2}{(m^2)^{a_0+\epsilon}}\
      \int\frac{d^d k\, \delta(k_T^2-1)\,\delta(k^2-m^2)  \theta(k_0)}
      {(n\cdot k)^{a_1+2\alpha} (\nbar\cdot k)^{a_2} (v_3 \cdot k)^{a_3} (v_4
      \cdot k)^{a_4}}\,.
  \label{eq:IandIbardef}
\end{align}
Hence, we obtain
\begin{equation}
  I =\int_0^\infty \frac{dm^2}{(m^2)^{a_0+\epsilon}} \bar I (m^2)\,.
  \label{eq:Im2Ibar}
\end{equation}
Now we can use reverse unitarity to turn delta functions in $\bar I(m^2)$
into propagators, which leads to the following topology
\begin{align}
  \label{eq:topology-bubble}
  \bar I(a_1, a_2, a_3, a_4, a_5, a_6)  
  & =   \\[0.5em] 
  &  \hspace{-70pt}
  \int \frac{d^d k}
  {(n\cdot k)^{a_1+2\alpha} (\nbar\cdot k)^{a_2} (v_3 \cdot k)^{a_3} (v_4
  \cdot k)^{a_4} (k^2-m^2)^{a_5} ((n\cdot k)(\nbar \cdot k)-m^2-1)^{a_6}}\,.
  \nonumber
\end{align}

We observe that all $\epsilon$ poles, hence those of the soft origin,
are generated by performing the integral over~$m^2$ in Eq.~(\ref{eq:Im2Ibar})
while the function $\bar I(m^2)$ is finite in the limit $\epsilon \to 0$.

A set of identities can be derived for the class of integrals defined in
Eq.~(\ref{eq:topology-bubble}). First of all, $\bar I(a_1,a_2,\ldots, a_{6})$
obeys the standard integration by parts~(IBP) identities
\begin{eqnarray}
   \int d^dk\, \frac{\partial}{\partial k^\mu} q^\mu 
   \bar I(a_1,a_2,\ldots, a_{6}) & = & 0\,,
\end{eqnarray}
with $q^\mu = n^\mu, \nbar^\mu, v_3^\mu, v_4^\mu, k^\mu$. The set of IBPs
consists of five relations. In addition, the topology $\bar I$ exhibits certain
redundancies which result in additional identities. One of them comes from the
$q_T$-delta propagator
\begin{equation}
  \bar I\, \frac{(n\cdot k) (\nbar\cdot k) - m^2-1}
         {(n\cdot k) (\nbar\cdot k) - m^2-1}  = \bar I\,,
\end{equation}
which gives
\begin{equation}
  \bar I(a_1-1, a_2-1, a_3, a_4, a_5, a_6+1)  
  - (m^2+1) \bar I(a_1, a_2, a_3, a_4, a_5, a_6+1)   =
  \bar I(a_1, a_2, a_3, a_4, a_5, a_6)\,.
\end{equation}
The last identity arises from  momentum conservation following the
discussion of Section~\ref{sec:kinematics}
\begin{equation}
  n + \nbar = \tilde v_3 + \tilde v_4\,.
\end{equation}
Multiplying the above by $k$ leads to 
%
%
\begin{align}
  \bar I(a_1, a_2, a_3-1, a_4, a_5, a_6) +
  \bar I(a_1, a_2, a_3, a_4-1, a_5, a_6) -
  \nonumber \\
  \bar I(a_1-1, a_2, a_3, a_4, a_5, a_6) -
  \bar I(a_1, a_2-1, a_3, a_4, a_5, a_6) & = 0\,.
\end{align}
Hence, altogether  we have  seven identities which we use to reduce all
the integrals appearing in the problem to a set of master integrals.

While solving the bubble graphs with the above method, it is important to note
that the $d$-vector $k^\mu$ in Eq.~(\ref{eq:topology-bubble}) is now massive.
Hence, the parametrization from Eq.~(\ref{eq:kparam}) does not hold
and it must be replaced with 
\begin{equation}
 k  = 
 (k_0, \ldots,
 |\vec k| \sin\theta_1\sin\theta_2, ,|\vec k| \sin\theta_1\cos\theta_2,
 |\vec k| \cos\theta_1)\,,
 \label{eq:k4v-par-massive}
\end{equation}
where
\begin{equation}
 |\vec k| = \sqrt{k_0^2-m^2}\,.
\end{equation}
Therefore, the relevant inner products now take the form
\begin{align}
  n \cdot k & =  k_0 - |\vec k| \cos\theta_1\,, \\
  \nbar \cdot k & =  k_0 + |\vec k| \cos\theta_1\,, \\
  \tilde v_3 \cdot k & =  k_0- 
  \beta_t\, |\vec k| \sin\theta_1\cos\theta_2\sin\theta -
  \beta_t\, |\vec k| \cos\theta_1\cos\theta\,, \\
  \tilde v_4 \cdot k & =  k_0+
  \beta_t\, |\vec k|\sin\theta_1\cos\theta_2\sin\theta +
  \beta_t\, |\vec k|\cos\theta_1\cos\theta\,.
\end{align}
We also get that
\begin{align}
 k_0 & = \frac12\left(n\cdot k + \nbar \cdot k\right) 
 = \frac12\left(\tilde v_3\cdot k + \tilde v_4 \cdot k\right)\,.
\end{align}

The complete bubble+tadpole-part of the NNLO soft function requires calculation
of the integrals which correspond to the diagrams of
Fig.~\ref{fig:bubble-graphs} with $ij = 13,\ 23,\ 33$ and $34$, and similar
combinations of pinched diagrams $D_6$ of Fig.~\ref{fig:2WL2cut}. 
The remaining integrals can be obtained through the relations:
\begin{equation}
  \bar I_{14}(\beta_t) = \bar I_{13}(-\beta_t), \qquad 
  \bar I_{24}(\beta_t) = \bar I_{23}(-\beta_t), \qquad
  \bar I_{44}(\beta_t) = \bar I_{33}(\beta_t)\,. 
\end{equation}

Since the values of the powers $a_i$ which
appear in the definitions of the integrals are governed by the powers in
the denominator of Eq.~(\ref{eq:ImPiint}) and the powers of the $p^2$, and
$n\cdot p$ propagators in the expressions for the coefficients $T_{ij}$,
Eqs.~(\ref{eq:Tijquark})-(\ref{eq:Tijtadpole}), they are the same the for the
quark and gauge bubble, as well as for the tadpole integrals.
Therefore, obtaining a solution for one type of the bubble, allows us to use it
for the other type.

In our calculation, we also used a certain property of the topology
(\ref{eq:topology-bubble}) that allowed us to express the integrals involved in
the Passarino-Veltman reduction of $\bar I_{23}$ with the integrals coming from
tensor reduction of $\bar I_{13}$.  
Specifically, first of all, we note that the propagator $k^2-m^2$ in
Eq.~(\ref{eq:topology-bubble}) in reality represents the delta function from
Eq.~(\ref{eq:IandIbardef}). Therefore
\begin{equation}
 m^2 = k^2 = (n\cdot k) (\nbar \cdot k) - k_T^2 = 
 (n\cdot k) (\nbar \cdot k) - 1\,,
\end{equation}
where the last equality comes from the transverse delta function in
(\ref{eq:IandIbardef}). The above allows us to write
\begin{equation}
  (n \cdot k) = (m^2 +1) (\nbar\cdot k)^{-1}\,.
  \label{eq:trickrelation}
\end{equation}
Let us now apply this to the class of graphs of the $\bar I_{23}$ type, which
contain linear propagators $\nbar\cdot k$ and $v_3 \cdot k$
\begin{align}
  \bar I(\alpha,\beta_t; 0, a_2, a_3, 0, 1, 1)  
  & =
  \int \frac{d^d k}
  {(n\cdot k)^{2\alpha} (\nbar\cdot k)^{a_2} (v_3 \cdot k)^{a_3} 
  (k^2-m^2) ((n\cdot k)(\nbar \cdot k)-m^2-1)}
  \nonumber \\[0.5em]
  & 
  \hspace{-30pt}
  =
  \frac{1}{(m^2+1)^{2\alpha}}
  \int \frac{d^d k}
  {(\nbar\cdot k)^{-2\alpha+a_2} (v_3 \cdot k)^{a_3} 
  (k^2-m^2) ((n\cdot k)(\nbar \cdot k)-m^2-1)}
  \nonumber \\[0.5em]
  &   
  \hspace{-30pt}
  =
  \frac{1}{(m^2+1)^{2\alpha}}
  \int \frac{d^d k}
  {(n\cdot k)^{-2\alpha+a_2} (v_4 \cdot k)^{a_3} 
  (k^2-m^2) ((n\cdot k)(\nbar \cdot k)-m^2-1)}
  \nonumber \\[0.5em]
  &
  \hspace{-30pt}
  =
  \bar I(-\alpha,-\beta_t; a_2, 0, a_3, 0, 1, 1)\,.
\end{align}
In the second line we simply used the relation (\ref{eq:trickrelation}). In the
third line we changed $\vec k \to - \vec k$, and in the fourth line we used the
property that the propagators $v_3 \cdot k$ and $v_4 \cdot k$  can be turned to
each other by the replacement $\beta_t \to -\beta_t$.

All in all, we see that graphs of the type $\bar I_{23}$ can be expressed as
slightly modified versions of $\bar I_{13}$.
Hence, we effectively need to calculate the following set integrals
\begin{align}
  {\renewcommand{\arraystretch}{1.5}%
    \begin{array}{c@{\qquad \qquad}c}
      \bar I(\alpha, \beta_t; 1, 0, 1, 0, 1, 1)\,,  &
      \bar I(\alpha, \beta_t; 1, 0, 0, 1, 1, 1)\,, 
      \\
      \bar I(\alpha, \beta_t; 0, 0, 2, 0, 1, 1)\,, &
      \bar I(\alpha, \beta_t; 2, 0, 1, 1, 1, 1)\,,
      \\
      \bar I(\alpha, \beta_t; 2, 0, 2, 0, 1, 1)\,, &
      \bar I(\alpha, \beta_t; 0, 0, 1, 1, 1, 1)\,,  
      \\
      \bar I(-\alpha, \beta_t; -1, 0, 0, 1, 1, 1)\,, &
      \bar I(-\alpha, \beta_t; 1, 0, 0, 1, 1, 1)\,.
    \end{array}
  }
  \label{eq:bubble-functions}
\end{align}

We start by reducing them with help of {\tt IdSolver}~\cite{Czakon:idsolver} --
a C++ implementation of the Laporta algorithm~\cite{Laporta:2001dd}, which
depends on { FORM}~\cite{Ruijl:2017dtg} and {Fermat}~\cite{fermat}. 
As a result, we obtain five master integrals, for which we then derive a set of
differential equations with respect to the variable $\beta_t$. 
The structure of
the set is such that the general solutions for the masters can be obtained
iteratively as a series in $\alpha$ and $\epsilon$. 

In fact, we only had to solve two systems of differential equations:
one for three masters from the reduction of $\bar I_{13}$ and the other for two
masters from the reduction of $\bar I_{34}$. The reduction of $\bar I_{33}$ leads to
master integrals identical to those found from reduction of $\bar I_{13}$.
The same masters can be used to determine $\bar I_{23}$ as discussed above.

The expressions for the boundary integrals, corresponding to $\beta_t = 0$, are easily
calculated through direct integration and this allows us to determine the
special solutions.
As a last step, we integrate the expressions for the functions from
Eq.~(\ref{eq:bubble-functions}) over the spurious mass $m^2$,
following~Eq.(\ref{eq:Im2Ibar}).  Except for a few one-dimensional integrals
which appear at order $\epsilon^1$ in momentum space, most of the result is
given in an analytic form.

\subsection{Real-virtual diagrams}
\label{sec:realvirt}

We now turn to the class of single-cut diagrams. As shown in
Fig.~\ref{fig:1cutdiagrams}, these diagrams involve a gluon loop as well as a real
gluon. 
Following Ref.~\cite{Bierenbaum:2011gg}, we introduce the tree-level,
UV-renormalized, one-loop soft currents
\begin{align}
  J_a^{\mu (0)} & = \sum_{i=1}^n T_i^a e_i^\mu\,,
  \\
  J_a^{\mu (1)} & = i f^{abc} \sum_{i\neq j=1}^n T_i^b T_j^c 
                    \left(e_i^\mu-e_j^\mu\right) 
		    g_{ij}^{(1)}(\epsilon, p, p_i, p_j)\,,
\end{align}
where $e_i^\mu$ is the eikonal propagator defined in Eq.~(\ref{eq:eikprop}).

The function $g_{ij}$, which is symmetric under the exchange $i\leftrightarrow
j$, has been obtained in a concise form in Ref.~\cite{Czakon:2018iev}.
The soft current $J_a^{\mu}$ corresponds to the sum of all parts of diagrams
shown in Fig.~\ref{fig:1cutdiagrams} to the left of the cut. It can therefore be
used directly to construct the single-cut contribution to our soft function of
interest by attaching the gluon with momentum $p$ to the Wilson lines in
a Born-level amplitude. As a result, we obtain
\begin{align}
  \bfS^\text{1-cut}_\iibar 
  & = 
  \frac{2\pi}{\alpha_s}
  (4\pi)^4
  \nu^\alpha
  \left(\frac{e^{\gamma_E}}{4\pi}\right)^{3\epsilon}
  \frac{(2\pi)^{\frac{d}{2}-1}}{ S_{d-3}\,q_T^{d-3}}
  \sum_{k=1}^{4} \sum_{i\neq j=1}^4 
  \langle c_I^{i\bar i} | i f^{abc} T_k^a T_i^b T_j^c |c_J^{i\bar i} \rangle\,
  \nonumber \\[0.5em]
  &
  \hspace{20pt}
  \times \int \frac{d^d p}{(2\pi)^{d-1}}\, 
  \frac{\delta^{+}(p^2)\, \delta(p_T-q_T)}{(n\cdot p)^\alpha\; p_i \cdot p}
                    \left(
		    \frac{p_i\cdot p_k}{p_i \cdot p}-
                    \frac{p_j\cdot p_k}{p_j \cdot p}
		    \right)\,
		    g_{ij}^{(1)}(\epsilon,p, p_i, p_j) + \text{c.c.}\,,
  \label{eq:S1cut-main}
\end{align}
where~\cite{Czakon:2018iev}
\begin{equation}
  g^{(1)}_{ij}(\epsilon, p, p_i, p_j) = 
  - \frac{1}{2} \,
  \frac{\alpha_s}{2\pi} \, \left(\frac{4\pi}{e^{\gamma_E}}\right)^\epsilon
  \, \Bigg( \frac{2 \pipj \mu^2}{2
  \pip 2 \pjp} \Bigg)^\epsilon \, \Bigg[ \frac{1}{\epsilon^2} + \sum_{n=-1}^1
  \epsilon^n \Big( R_{ij}^{(n)} + i\pi I_{ij}^{(n)} \Big) \Bigg] \;.
  \label{eq:gij}
\end{equation}
The real and imaginary coefficients, $R_{ij}^{(n)}$ and $ I_{ij}^{(n)}$, were derived
in Refs.~\cite{Bierenbaum:2011gg, Czakon:2018iev}, and they depend solely on the
rescaling-invariant variables
\begin{equation}
  \ai \equiv \frac{m_i^2  \, 2 \pjp}{2 \pipj 2 \pip} \; , 
  \qquad \qquad
  \aj \equiv \frac{m_j^2 \, 2 \pip}{2 \pipj 2 \pjp} \; .
\end{equation}
From Eq.~(\ref{eq:gij}), and taking into account the expansion of the Fourier
Transform coefficient of Eq.~(\ref{eq:FTexpand}), we observe that the
single-cut contributions to the position-space NNLO soft function start at the order
$1/\epsilon^3$ for the real part and $1/\epsilon^2$ for the imaginary part.
As we shall see in Section~\ref{sec:results}, the $1/\epsilon^3$ term cancels
between the single and  the double-cut contributions.

\subsubsection{Diagrammatic configurations}

A range of different configurations of diagrams contribute to
Eq.~(\ref{eq:S1cut-main}).  If the gluons connect three massless Wilson lines,
the corresponding integral is scaleless and vanishes. However, when two
massless and one massive Wilson
lines are connected, the integral does not vanish. 
The soft current for massless Wilson lines takes a very simple form and it has
been calculated in Ref.~\cite{Catani:2000pi}. Connecting this current to a
massive Wilson line leads to an expression which can be integrated analytically.

Another interesting subclass of single-cut diagrams is formed by two-Wilson line
configurations with gluons attached to two massless and one massive leg (\eg
131). These integrals correspond to case~1 of Ref.\cite{Bierenbaum:2011gg} and
they exhibit rapidity singularities. Finally, we also need to include diagrams
with three distinct Wilson lines, which corresponds to case~3 of
Ref.\cite{Bierenbaum:2011gg}.

Altogether, the single-cut part of the soft function receives contributions
from ten independent two-Wilson line and ten independent three-Wilson line
integrals. All the other integrals can be obtained through symmetry relations.
 
In particular, for the two-Wilson line diagrams we observe the symmetry
$D_{i,ji}=D_{i,ij}$, where the comma corresponds to the cut in the diagram.
Swapping $i \leftrightarrow j$ on one side of the cut has no effect as the
colour and the kinematic parts both produce minus signs which balance each
other.  One can also show that $D_{ij,i} = D_{i,{ij}}^*$, which has an important
consequence as it implies that the two-Wilson line, single-cut diagrams are
purely real.

We also find symmetries for the three-Wilson line diagrams. First of all, swapping
particles on one side of the cut has similar effect to the one described above
for the two-Wilson-line case,
namely $D_{k,ji} = D_{k,ij}$, which, again, arises because changes in signs in
the colour and the kinematic part balance each other. 
The most important relation, however, reads $D_{ij,k} = -D_{k,ij}^*$, and it
means that the three-Wilson line, single-cut diagrams are purely imaginary. In
fact, this class of diagrams, constitutes the only source of the imaginary part
of the entire NNLO soft function for top pair production.

\subsubsection{Three-particle diagrams with two massless and one massive Wilson
lines }

\begin{figure}[t]
  \begin{center}
    \includegraphics[width=0.60\textwidth]{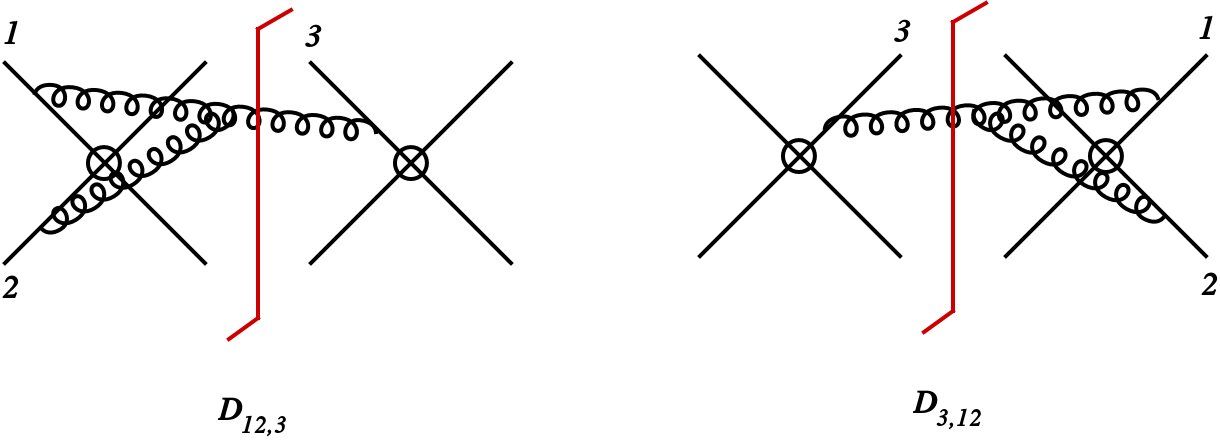}
  \end{center}
  \caption{Example of the three-Wilson-line, single-cut diagrams that yield
  non-vanishing, purely imaginary result.
  }
  \label{fig:diag123}
\end{figure}

Let us consider a special case of three-Wilson-line diagrams which involve two
massless and one massive particle. Two such diagrams are depicted in
Fig.~\ref{fig:diag123}.
Using the notation introduced above, we can write expressions corresponding to
those two diagrams as
\begin{align}
  D_{12,3} & = i f^{abc}\, T_3^a T_1^b T_2^c\, 
             \left(e_1^\mu-e_2^\mu\right) e_3^\mu\, g_{12}^{(1)}\,,
  \\[0.5em]
  D_{3,12} & = -i f^{abc}\, T_3^a T_1^b T_2^c\, 
             \left(e_1^\mu-e_2^\mu\right) e_3^\mu\, g_{12}^{(1)\,*}\,.
\end{align}
Hence, their sum reads
\begin{equation}
  D_{12,3}  + D_{3,12}  = i f^{abc}\, T_3^a T_1^b T_2^c\, 
                          \left(e_1^\mu-e_2^\mu\right) e_3^\mu\,
	                  \left(g_{12}^{(1)}-g_{12}^{(1)\,*}\right)\,.
\end{equation}

We see that the result is purely imaginary and that, contrary to the case of
double-cut diagrams, it does not vanish as, here, the
antisymmetry of the colour factor under the exchange $1 \leftrightarrow 2$ is
compensated by the antisymmetry of the kinematic part.

We note that our case is different from the one of Ref.~\cite{Becher:2009qa},
which considers infrared singularities of QCD amplitudes and where it is
argued that all three-particle structures with two massless and one massive
Wilson lines must vanish. The latter happens because Ref.~\cite{Becher:2009qa}
discusses amplitudes, \ie objects in which all Wilson lines originate from the
same vertex and all soft gluons are virtual. On the contrary, in our real-virtual
diagrams, the Wilson lines meet at two vertices and one gluon is cut.

To understand better why the contribution from the diagrams of
Fig.~\ref{fig:diag123} does not vanish, it is useful to perform an analysis
similar to the one of Ref.~\cite{Aybat:2006wq}. For that purpose, we use the
soft current for massless particles with momenta $p_{1,2} = (1,0,0,\pm 1)$,
which reads~\cite{Catani:2000pi}
\begin{align}
  J_{2P}^{a(1)}(p_1,p_2;p, \epsilon) & = 
  -\frac{1}{16\pi^2}  i f_{abc} T_1^b T_2^c\, \epsilon_\mu(p)
  \left(\frac{p_1^\mu}{p_1\cdot p}- \frac{p_2^\mu}{p_2\cdot p}\right)
  \nonumber \\[0.5em]
  & \quad \quad \times
  \left(\frac{4\pi p_1\cdot p_2}{2\, p_1 \cdot p\ p_2 \cdot p\, e^{-i\pi}}
  \right)^\epsilon
  \frac{1}{\epsilon^2}
  \frac{\Gamma^3(1-\epsilon)\Gamma^2(1+\epsilon)}{\Gamma(1-2\epsilon)}\,.
\end{align}
Embedding the above in the integral over $p$ gives
\begin{align}
  D_{12,3} & \propto i f_{abc} T_3^a T_1^b T_2^c\,
  \frac{1}{\epsilon^2}\,
  \frac{\Gamma^3(1-\epsilon)\Gamma^2(1+\epsilon)}{\Gamma(1-2\epsilon)}
  \nonumber \\
  & \times
  \int d^d p\,
  \frac{\delta^+(p^2)\delta^{(d-2)}(p_\perp-1)}
       {p_+ p_{3-}+p_- p_{3+}-p_\perp \cdot p_{3\perp}}\,
  \frac{1}{p_+^\alpha}
  \left(\frac{p_{3+}}{p_+}- \frac{p_{3-}}{p_-}\right)
  \left(\frac{4\pi}{p_+\, p_-\, e^{-i\pi}}
  \right)^\epsilon.
\end{align}
Let us now apply the change of variables, motivated by Ref.~\cite{Aybat:2006wq}
\begin{equation}
  (p_+,p_-,p_\perp) \to (\xi p_-,\xi^{-1} p_+,p_\perp)
  \qquad
  \text{with} 
  \quad
  \xi = p_{3+}/p_{3-}\,.
  \quad
  \label{eq:xiscaling}
\end{equation}
Our integral becomes
\begin{align}
  D_{12,3} (\alpha) & \propto i f_{abc}  T_3^a T_1^b T_2^c \,
  \frac{1}{\epsilon^2}\,
  \frac{\Gamma^3(1-\epsilon)\Gamma^2(1+\epsilon)}{\Gamma(1-2\epsilon)}
  \nonumber \\
  & \times
  \int d^d p\, 
  \frac{\delta^+(p^2)\delta^{(d-2)}(p_\perp-1)}
       {p_+ p_{3-}+p_- p_{3+}-p_\perp \cdot
  p_{3\perp}}
  \xi^{-\alpha}\frac{1}{p_-^\alpha}
  \left(\frac{p_{3-}}{p_-} - \frac{p_{3+}}{p_+}\right)
  \left(\frac{4\pi}{p_+\, p_-\, e^{-i\pi}}
  \right)^\epsilon
  \nonumber \\
  & = - \xi^{-\alpha} D_{12,3}(-\alpha)\,.
\end{align}
Hence, we see that the above integral would exhibit scaling w.r.t. $\xi$, and
vanish, if only there was no $\alpha$ regulator. However, without the regulator,
the integral is divergent. Hence, we conclude that the contribution from
tree-particle graphs with two massless and one massive Wilson lines does not
vanish.

This result only affects the imaginary part of the soft function. The
tree-particle graphs of the type of 123 do not contribute to the real part 
as there will always be a complex-conjugate diagram with opposite
sign due to colour operator.

The above conclusion does not invalidate the analysis of
Ref.~\cite{Aybat:2006wq}, where it is claimed that the massless-massless-massive
diagrams vanish because of the scaling property~(\ref{eq:xiscaling}). The key
difference between our case and the one discussed there is that, in
Ref.~\cite{Aybat:2006wq}, purely virtual diagrams are considered. In those
diagrams, rapidity divergences are regulated by dimensional
regularization~\cite{Becher:2011dz}.  Hence, no $\alpha$ regulator is required
and the $\xi$ scaling of Eq.~(\ref{eq:xiscaling}) holds.

The key element of our calculation, which prevents the diagram 123 from
vanishing, is the transverse delta function. As explained in
Ref.~\cite{Becher:2011dz} without this function, integration over the
transverse momentum provides a factor $k_-^{-\epsilon}$ which regularizes
rapidity divergences, hence there is no need for the regulator $\alpha$. 
However, when the transverse delta is present, it fixes $q_\perp$ to some
external value and the integration over $q_\perp$ does not provide a regulator
of light-cone singularities. Hence, we need to introduce the analytic
regulator~$\alpha$.

\subsubsection{Method of integration}

To evaluate the integrals in Eq.~(\ref{eq:S1cut-main}) we proceed through the
following set of steps. We start from integration over the $p_T$ and $p_-$
components, which is straightforward as, in the process, we use the two delta
functions $\delta(p^2)$ and $\delta(p_T-1)$. Then, we are left with a nontrivial
integration over $p_+$ and  $\cos \phi$ , where $\phi$ is the azimuthal angle
between $v_{3\perp}$ and $p_\perp$. The remaining angular variables do not
appear in the integrand and they only produce a surface term.

To deal with the nontrivial 2-dimensional integrals, we remap the variables
$p_+$ and $\phi$ to a unit hypercube
\begin{equation}
  p_+ = \frac{x}{1-x},
  \qquad \qquad
  \cos \phi = 1- 2 \cos^2\frac{\pi y}{2}\,.
\end{equation}
The second transformation is introduced for efficiency reasons as it eliminates
integrable singularity in the azimuthal integration. The first transformation
compactifies the plus component of the gluon momentum. This is useful as our
integrals are in general divergent when integrated over $p_+$. The
divergence comes from rapidity singularities and that is why we have introduced
the analytic regulator in Eq.~(\ref{eq:S1cut-main}).
To perform the integral over $x$, we use the Laurent expansion
\begin{equation}
  \frac{1}{x^{1+a\alpha}} = -\frac{1}{a\alpha} \delta(x) + \left[
    \frac{1}{x^{1+a\alpha}} \right]_+ \; ,
  \quad
  \text{where}
  \quad
  \int_0^1\!\! dx \, \left[ \frac{1}{x^{1+a\alpha}} \right]_+\!\! f(x) =
  \int_0^1\!\! dx \, \frac{f(x)-f(0)}{x^{1+a\alpha}} \; ,
\end{equation}
which makes the divergences explicit by turning them in $\alpha$ poles. All
the coefficients of $\alpha$ expansion are finite and take forms of one and
two-dimensional integrals which we perform numerically with help of the {\tt
Cuba} package~\cite{Hahn:2004fe, Hahn:2014fua}. 
The integration is fast, hence, we are able to achieve arbitrary accuracy.

\subsection{Sector decomposition approach and real-real diagrams}
\label{sec:secdec}

The double-cut integrals take the following form
\begin{equation}
  \tilde I = 
  \int \frac{d^d k}{(2\pi)^{d-1}}\, \frac{d^d l}{(2\pi)^{d-1}}\, 
  \frac{\delta^+(k^2) \delta^+(l^2)\,
  \delta(|k_\perp+l_\perp|-1)}
  {(n\cdot k)^{\alpha} (n\cdot l)^\alpha}
  \times \gdp\,.
  \label{eq:int2cutgen}
\end{equation}
Two issues arise when one attempts to evaluate them. First of all, 
they are divergent when integrated over a subset of variables, and the pattern
of divergencies is complex, with many overlapping singularities. Secondly, the
remaining part of the integration, where divergencies do not appear, consists of
complicated azimuthal integrals and care is needed to perform them efficiently.

Because of these two separate challenges, it is convenient to factorize the
graph-dependent part as
\begin{align}
  \gdp & = \underbrace{\left(\gdp\vert_{\beta_t=0}\right)}_\text{boundary part}
  \underbrace{\left(\frac{\gdp}{\gdp\vert_{\beta_t=0}}\right)}_\text{weight}
  \equiv \calI(n_i, k, l)\, \calW(n_i, k, l)\,.
\end{align}
For any graph, the boundary part is not only independent of $\beta_t$ but also
of all the angles except the angle between transverse components of gluons
$d$-momenta $\theta_1 = \sphericalangle(k_\perp, l_\perp)$. In other words, all
divergences are present already in the boundary integrals and the remaining
integration over the  angular variables which appear in the weight is finite.

Our strategy of evaluating the double-cut integrals will therefore consist of
two elements: (i) use of sector decomposition to disentangle overlapping
singularities and cast them into a set of $\alpha$ and $\epsilon$ poles, (ii)
supplementing the sector-decomposed integrals with carefully parameterized
weights and integrating them numerically with {\tt Cuba}.

\subsubsection{Integration of the on-shell and transverse delta functions}

The momenta of the heavy quarks can be written using the following
parametrization
\begin{align}
  \tilde v_3 & = (1, \beta_t \sin \theta\, \hat v_{3\perp}, \beta_t \cos \theta)\,, \\
  \tilde v_4 & = (1, -\beta_t \sin \theta\, \hat v_{3\perp}, -\beta_t \cos \theta) 
               = (1,-\vec v_3)\,,
\end{align}
where $\hat v_{3\perp}$ is a unit vector in $d-2$-dimensional space, \cf
Eq.~(\ref{eq:kparam}).

As discussed in Section~\ref{sec:intsym}, the integrand in
Eq.~(\ref{eq:int2cutgen}) can only depend on the scalar products between
$k_\perp,l_\perp$ and $v_{3\perp}$. Due to rotational invariance of these 
scalar products, one can always change the frame in the transverse space such
that
\begin{subequations}
  \begin{align}
    k_\perp & = |k_\perp|\, (1,0, 0, \vec{0}_{d-5}),\\
    l_\perp & = |l_\perp|\, (\cos\theta_1,\sin\theta_1 , 0, \vec{0}_{d-5}),
    \label{eq:theta1} \\
    v_{3\perp} & = |v_{3\perp}|\, (\cos\phi_1,\sin\phi_1\cos\phi_2, 
                         \sin\phi_1\sin\phi_2,\vec{0}_{d-5}).
  \end{align}
  \label{eq:klv3param}
\end{subequations}

\begin{figure}[t]
  \begin{center}
    \includegraphics[width=0.99\textwidth]{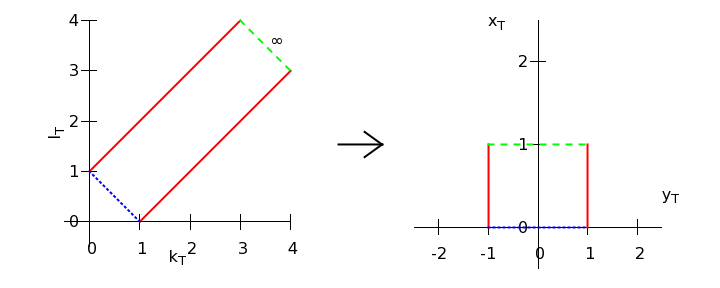}
  \end{center}
  \caption{
  The integration region in the transverse coordinates before (left) and
  after (right) the transformation of variables from Step 1 discussed in
  Section~\ref{sec:mapping}.
  }
  \label{fig:trans-regions}
\end{figure}

Since the momenta of the incoming particles are light-like and $v_{4\perp} =
-v_{3\perp}$, the soft function integrals can only depend on $v_{3\perp}^2$.
Therefore, nothing is sensitive to the position of the versor $\hat v_{3\perp}$.
Hence, the integral~(\ref{eq:int2cutgen}) can be written as
\begin{equation}
  \tilde I = 
  \frac{1}{S_{d-3}}
  \int d\Omega_{d-3} (\hat v_{3\perp})
  \int \frac{d^d k}{(2\pi)^{d-1}}\, \frac{d^d l}{(2\pi)^{d-1}}\, 
  \frac{\delta^+(k^2) \delta^+(l^2)\,
  \delta(|k_\perp+l_\perp|-1)}
  {(n\cdot k)^{\alpha} (n\cdot l)^\alpha}
  \times \gdp\,,
\end{equation}
where the surface of unit-sphere, $S_{d-3}$, is given by Eq.~(\ref{eq:Sd}),
and the differential measures  read
\begin{subequations}
  \label{eq:klv3-measures}
  \begin{align}
   d^{d}k & =  dk_+ dk_- d k_\perp  k_\perp^{d-3}\,
   d\Omega(\chi_1,\chi_2, \dots, \chi_{d-3}),
   \\[0.5em]
   d^{d} l & =  dl_+ dl_- d l_\perp  l_\perp^{d-3}\,
   d\Omega(\theta_1,\theta_2, \dots, \theta_{d-3}),
   \\[0.5em]
   d\Omega_{d-3} (\hat v_{3\perp}) & =  
   d\Omega(\phi_1,\phi_2, \dots, \phi_{d-3})\,.
  \end{align}
\end{subequations}
Because we use the parametrization~(\ref{eq:klv3param}),
all the angles except $\theta_1$, $\phi_1$  and $\phi_2$ can be
integrated trivially. 
Let us now replace the angle between the transverse components of the two
gluons' momenta, $\theta_1$, by
\begin{equation}
  \eta = \frac{1-\cos\theta_1}{2}\,.
  \label{eq:eta-def}
\end{equation}
Then, the whole measure takes the form
%
%
\begin{align}
  d\Omega_{d-3} (\hat v_{3\perp})\,
  d^d k\, d^d l & =
  \\
  &
  \hspace{-50pt}
  4^{-\epsilon}
  S_{1-2\epsilon}\,
  S_{-2\epsilon}\,
  k_T^{1-2\epsilon} \,
  l_T^{1-2\epsilon} 
  \big((1-\eta)\eta \big)^{-\frac12-\epsilon}
  dk_+ dk_- dl_+ dl_-
  d k_T\, 
  d l_T\,
  d\eta\,
  d\Omega(\phi_1, \phi_2, \ldots, \phi_{d-3})\,,
  \nonumber
\end{align}
where we used Eqs.~(\ref{eq:rot-inv-measure}), (\ref{eq:klv3-measures}) and
(\ref{eq:eta-def}).
After performing the graph-independent integrations over $k_-$, $l_-$ and
$\eta$, where the last quantity gets fixed to
\begin{equation}
  \eta = \frac{k_T^2 + l_T^2+ 2 k_T l_T -1}{4 k_T l_T}\,,
\end{equation}
the integral~(\ref{eq:int2cutgen}) turns into
\begin{align}
  \tilde I & =
  \int dk_+\, dl_+\, dk_T\, dl_T\, 
  d\Omega(\phi_1, \phi_2, \ldots, \phi_{d-3})\,
  k_+^{-\alpha}\, l_+^{-\alpha} k_T^{1-2\epsilon} l_T^{1-2\epsilon} \,
  \nonumber \\ &
  \times\,
  \Big\{\big[1-(k_T-l_T)^2\big]\big[1-(k_T+l_T)^2\big]
  \Big\}^{-\frac12-\epsilon}
  {\cal I}(n_i, k_+, l_+, k_T, l_T)\,
  {\cal W}(n_i, k_+, l_+, k_T, l_T, \phi_1, \phi_2, \beta_t)\,,
  \label{eq:S2cutgen-deltaint}
\end{align}
and the delta functions generate the following relations
\begin{equation}
  k_- = \frac{k_T^2}{k_+}\,, \qquad \qquad
  l_- = \frac{l_T^2}{l_+},
  \label{eq:pmrelations}
\end{equation}
and
\begin{equation}
  | k_T - l_T | \leq 1
  \qquad \land \qquad
   k_T + l_T  \geq 1\,.
  \label{eq:kTlT-region}
\end{equation}
The last pair of inequalities defines the integration region in the $(k_T, l_T)$
plane,  which is depicted in Fig.~\ref{fig:trans-regions}~(left).  The
first inequality corresponds to the two (red) solid lines, while the second
inequality corresponds to the (blue) dotted line.

\subsubsection{Divergencies}

The non-integrable divergences of the integral~(\ref{eq:S2cutgen-deltaint}) come
from the graph-dependent part. Since the weight is finite, the divergences
are fully determined by the boundary integrand ${\cal I}(n_i, k_+, l_+, k_T,
l_T)$.
Hence, each integral can in principle diverge due to integration over the four
independent variables:  $k_+, l_+, k_T$ and $l_T$, and some of the limits are
coupled due to the constraint of Eq.~(\ref{eq:kTlT-region}).
The singularities of the integral $\tilde I$ correspond to vanishing of the
propagators in $\cal I$, and this can happen in the following situations.

\subsubsection*{Propagators of the incoming particles}

The incoming-particle propagators, $n\cdot k$, $\nbar\cdot k$, $n\cdot l$ and
$\nbar\cdot l$, produce divergencies in the limits in which the plus or minus
components of the momenta of gluons tend to zero or infinity. This occurs
in three regions of gluons' momenta
\begin{itemize}
  \item
  Rapidity region, where the momenta scale as 
  $(\lambda^{\pm 1},\lambda^{\mp 1},1)$. 
  In terms of the independent variables, the rapidity divergencies occur when
  \begin{equation}
    k_+ \to 0      \qquad \text{or} \qquad
    k_+ \to \infty \qquad \text{or} \qquad
    l_+ \to 0      \qquad \text{or} \qquad
    l_+ \to \infty\,.
  \end{equation}
  \item
  Soft region, with the momenta scaling like
  $(\lambda, \lambda, \lambda)$. This corresponds to either
  \begin{equation}
    k_+ \to 0,\ k_T \to 0 \qquad \text{or} \qquad
    l_+ \to 0,\ l_T \to 0\,.
    \label{eq:softlimit-var}
  \end{equation}
  We note that, when one gluon becomes
  soft, transverse momentum of the other gluon tends to one due to the
  constraint~(\ref{eq:kTlT-region}).
  \item
  Collinear region, characterized by the scaling
  $(1,\lambda^2,\lambda)$. In terms of the independent variables, the collinear
  singularity occurs when
  \begin{equation}
    k_T \to 0 \qquad \text{or} \qquad
    l_T \to 0\,.
  \end{equation}
  Similarly to the case of the soft limit, because of the
  constraint~(\ref{eq:kTlT-region}), vanishing of the transverse momentum of one
  gluon requires that the transverse momentum of the other gluon tends to
  one, \ie the second gluon cannot be soft or collinear to the incoming parton.
  The reason why the collinear limit leads to vanishing of one of the
  incoming-particle propagators is because of the
  relation~(\ref{eq:pmrelations}).
\end{itemize}

\subsubsection*{Propagators of the outgoing particles}

The propagators involving top or anti-top quark, $v_3\cdot k$, $v_4\cdot k$,
$v_3 \cdot l$ and $v_4\cdot l$, vanish only in the soft region.  In terms of the
independent variables, the latter corresponds to the limits given in
Eq.~(\ref{eq:softlimit-var}) above.  

To see that the massive-particle propagators can produce divergencies only in
the soft region, let us use the parametrization of Eqs.~(\ref{eq:klv3param})
and the relation (\ref{eq:pmrelations}). Together they give
\begin{align}
  v_{3} \cdot k & =
  \frac{v_{3+} k_-}{2} + \frac{v_{3-} k_+}{2} - v_{3\perp} \cdot k_\perp
  \nonumber \\[0.2em]
  & =
  \frac{1+\beta_t \cos\theta}{2}\, \frac{k_T^2}{k+} + 
  \frac{1-\beta_t \cos\theta}{2}\, k_+ - 
  \beta_t \sin\theta\, k_T\,.
  \label{eq:v3knonvanish}
\end{align}
Since $\beta_t \leq \sqrt{1-4 m_t^2/\shat} < 1$ and $-1 \leq \cos\theta \leq 1$, the
coefficients in front of the components of gluon's $d$-momentum in the first two
terms in Eq.~(\ref{eq:v3knonvanish}) can never vanish. The collinear region
corresponds to $k_T \to 0$ and finite $k_+$, and we see that the propagator
stays finite in this limit. Similarly, the propagator does not vanish when $k_+
\to 0$ or $k_+ \to \infty$, while $k_T \neq 0$, which is the region of
small/large rapidities of the gluon.  Hence, the only way to make the above
propagator vanish is to send $k_+$ and $k_T$ to zero simultaneously, 
and this corresponds to the soft limit.
The above proof can be repeated for the other propagators of the outgoing
particles.

\subsubsection*{Exact gluon propagator}

Double-cut diagrams with the triple gluon vertex contain the propagator $(k + l)^2$
in which the momenta of the gluons are commensurate.  After the integration over
$k_-$, $l_-$ and $\eta$ is performed, this exact propagator
reads
%
%
\begin{equation}
  (k+l)^2  = 2\,k\cdot l  =   
  -1 + 
  \frac{k^+ + l^+}{k^+}\, k_\perp^2  + 
  \frac{k^+ + l^+}{l^+}\, l_\perp^2\,.
\end{equation}
Within the domain of integration defined in
Fig.~\ref{fig:trans-regions} (left), the above propagator vanishes when
\begin{equation}
  k_T \to \frac{k^+}{k^+ + l^+} 
  \qquad \text{and} \qquad
  l_T \to \frac{l^+}{k^+ + l^+}\,.
  \label{eq:colllimit}
\end{equation}
This limit corresponds to the two gluons with momenta $k$ and $l$ becoming
collinear. 
Note that, due to condition~(\ref{eq:kTlT-region}), only one of the two gluons
can have vanishing transverse momentum.
One notices that the limit (\ref{eq:colllimit}) is different with respect to the
cases discusses earlier. While the divergencies of the eikonal propagators
happen only at the edges of the integration domain, \ie $0$ or $\infty$
(endpoint singularities), the exact propagator vanishes inside the integration
region, on the manifold defined by Eq.~(\ref{eq:colllimit}).

\subsubsection{Mapping procedure}
\label{sec:mapping}

We would like to transform the integration region in the $(\kp, \lp, k_T, l_T)$
space into a unit hyper-cube in the space of variables $\{x_i\}$, and, if
necessary, split it such that each of the resulting integrals $\int \prod_i d
x_i f(\{x_i\})$ has singularities only when one or more variables $x_i \to 0$.
We achieve the above in the following four steps:

\subsubsection*{Step 1}

\begin{figure}[t]
  \begin{center}
    \includegraphics[width=0.49\textwidth]{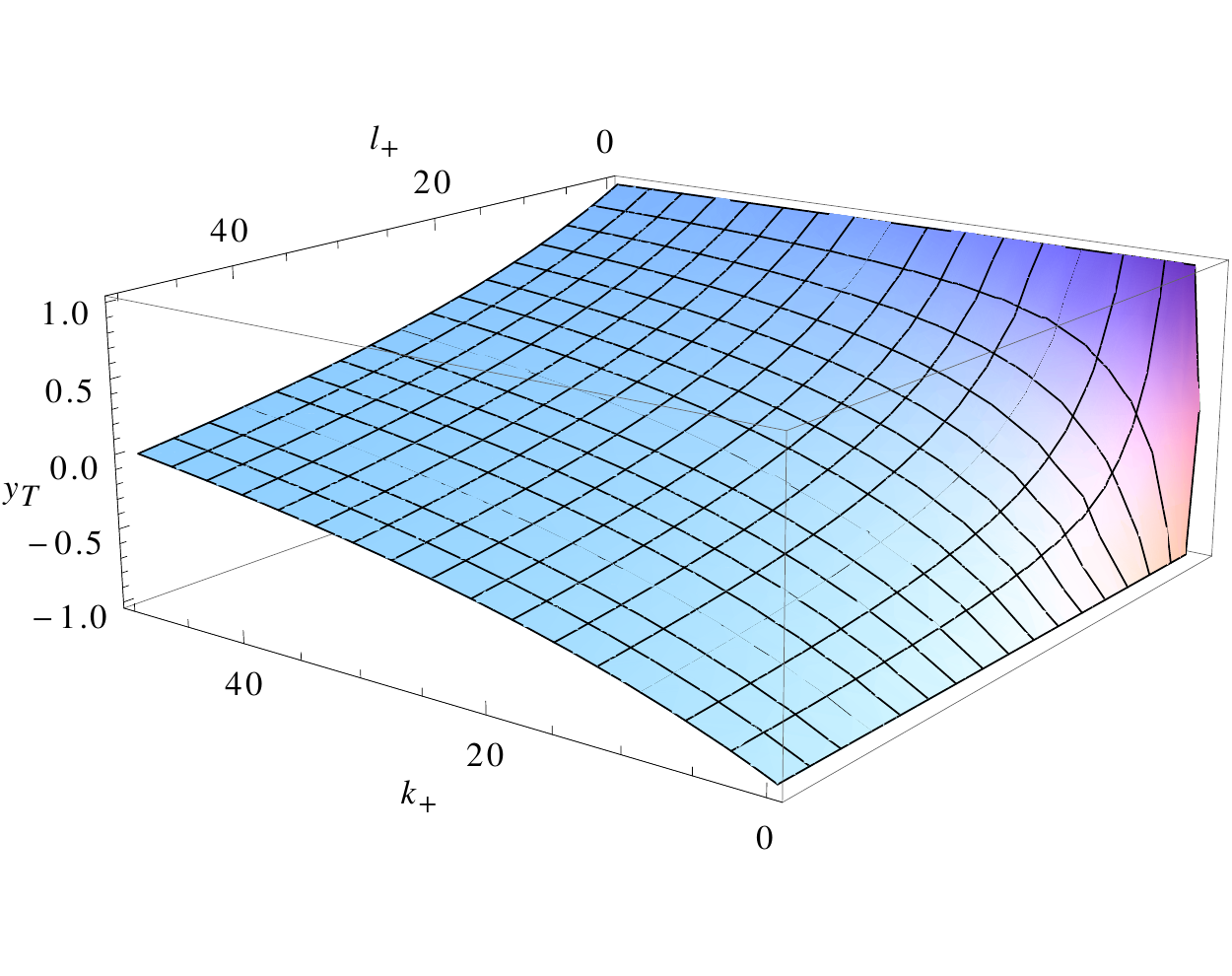}
  \end{center}
  \caption{
  Manifold where the double-cut integrals with triple-gluon vertex become
  divergent. It corresponds to the limit of the two gluons $k$ and $l$ becoming
  collinear to each other.
  }
  \label{fig:yTmanifold}
\end{figure}

The transverse variables $(k_T, l_T)$ are transformed to new variables 
$(x_T, y_T)$ with the replacements
\begin{equation}
  k_T = \frac{1+y_T-x_T y_T}{2(1-x_T)}\,,
  \qquad \qquad
  l_T = \frac{1-y_T+x_T y_T}{2(1-x_T)}\,,
\end{equation}
and the inverse transformation reads
\begin{equation}
  x_T = 1-\frac{1}{k_T+l_T}\,,
  \qquad \qquad
  y_T = k_T-l_T\,.
\end{equation}
This results in change of the integration region from the one shown in
Fig.~\ref{fig:trans-regions}~(left) to that of
Fig.~\ref{fig:trans-regions}~(right). The latter corresponds to
\begin{equation}
  0 \leq x_T \leq 1
  \qquad \land \qquad
  -1 \leq y_T \leq 1\,.
\end{equation}
The collinear divergences now occur in the limits
\begin{eqnarray}
  x_T \to 0,\, y_T \to -1 \qquad \text{or} \qquad
  x_T \to 0,\, y_T \to 1\,, 
\end{eqnarray}
and the soft divergences correspond to
\begin{eqnarray}
  k_+ \to 0,\,  x_T \to 0,\, y_T \to -1 \qquad \text{or} \qquad
  l_+ \to 0,\,  x_T \to 0,\, y_T \to 1\,. 
\end{eqnarray}
The limit of two gluons becoming collinear, Eq.~(\ref{eq:colllimit}), in the new
variables happens when
%
%
\begin{equation}
  x_T \to 0,\ y_T \to \displaystyle \frac{\kp-\lp}{\kp+\lp}\,.
  \label{eq:manifoldsing1}
\end{equation}
The last divergence occurs on a manifold inside the integration region as
depicted in Fig.~\ref{fig:yTmanifold}.

\subsubsection*{Step 2}

In order to transform the manifold singularity (\ref{eq:manifoldsing1}) into
an endpoint singularity, we split the integration region in the variable $y_T$
precisely on the manifold of Fig.~\ref{fig:yTmanifold}
%
%
\begin{eqnarray}
  I & =  & 
  \int_0^\infty\!\!\! d\kp
  \int_0^\infty\!\!\! d\lp
  \int_0^1\!\!\! d x_T
  \int_{-1}^1\!\!\! d y_T\
  \bar I(\kp, \lp, x_T, y_T)
  \nonumber \\
  & = &
  \int  d\kp d\lp d x_T
  \int_{-1}^{c(\kp, \lp)}\!\!\! d y_T\
  \bar I(\kp, \lp, x_T, y_T)
  +
  \int  d\kp d\lp d x_T
  \int_{c(\kp, \lp)}^{1}\!\!\! d y_T\
  \bar I(\kp, \lp, x_T, y_T)
  \nonumber \\
  & \equiv &
  I_d + I_u\,,
\end{eqnarray}
where
\begin{equation}
  c(\kp, \lp) = \frac{\kp-\lp}{\kp+\lp}\,.
\end{equation}
Now, we use two different parametrizations
\begin{equation}
  y_T = \frac{\kp-\lp + 2 \lp \bar y_T}{\kp+\lp}\,,
  \qquad 
  \bar y_T = \frac{y_T-c(\kp,\lp)}{1-c(\kp,\lp)}
           = \frac{k_+ (-1 + y_T) + l_+ (1+ y_T)}{2 l_+}
  \qquad 
  \text{for}\quad I_u\,,
\end{equation}
and
\begin{equation}
  y_T = \frac{\kp-\lp - 2 \kp \bar y_T}{\kp+\lp}\,,
  \qquad
  \bar y_T = \frac{y_T-c(\kp,\lp)}{-1-c(\kp,\lp)}
           = \frac{k_+ (1 - y_T) - l_+ (1+ y_T)}{2 k_+}
  \qquad
  \text{for}\quad I_d\,.
\end{equation}
Hence, we obtain
\begin{subequations}
  \begin{align}
    I_d  & =  
     \int_0^\infty\!\!\! d\kp
     \int_0^\infty\!\!\! d\lp
     \int_0^1\!\!\! d x_T
     \int_{0}^1\!\!\! d \bar y_T\
     \bar I_u(\kp, \lp, x_T, \bar y_T)\,,
     \\[0.5em]
    I_u  & =  
     \int_0^\infty\!\!\! d\kp
     \int_0^\infty\!\!\! d\lp
     \int_0^1\!\!\! d x_T
     \int_{0}^1\!\!\! d \bar y_T\
     \bar I_d(\kp, \lp, x_T, \bar y_T)\,.
  \end{align}
  \label{eq:IuIdstep2}
\end{subequations}
With the above changes, the soft singularities happen at
\begin{equation}
  k_+ \to 0,\ x_T \to 0
  \qquad \text{or} \qquad
  l_+ \to 0,\ x_T \to 0\,,
\end{equation}
in both integrals, $I_u$ and $I_d$, and for any value of $\bar y_T$. 
The case in which a gluon is collinear to the incoming parton corresponds to
\begin{equation}
  \bar y_T \to 1\,,
  \label{eq:yTbarlimit}
\end{equation}
for both integrals. We note that in the above limit, for $I_u$, it is the gluon
with momentum $l$ that becomes collinear, whereas, for $I_d$, the limit
(\ref{eq:yTbarlimit})  corresponds to the gluon with momentum $k$ being
collinear to the beam.

Finally, the limit of two gluons becoming collinear to each other,
Eq.~(\ref{eq:colllimit}), in the new variables corresponds to 
\begin{equation}
  x_T \to 0,\ \bar y_T \to 0\,,
\end{equation}
for both $I_u$ and $I_d$.

\subsubsection*{Step 3}

We see that  the integrals (\ref{eq:IuIdstep2}) can be divergent at both ends in
the variable $\bar y_T$.
In order to move all singularities to the limit $x_i \to 0$, we split $I_u$ and
$I_d$ at $\bar y_T = \frac12$
\begin{equation}
 I_{\{u,d\}} = \int_0^1 d \bar y_T\, \tilde I_{\{u,d\}} =
 \int_0^\frac12 d \bar y_T\, \tilde I_{\{u,d\}} +
 \int_\frac12^1 d \bar y_T\, \tilde I_{\{u,d\}} \equiv 
 I_{\{u,d\}0} + I_{\{u,d\}1}\,.
\end{equation}
Then, we apply the following transformations
\begin{equation}
  \tilde y_T = 2(1-\bar y_T)\,,
  \qquad \qquad
  \bar y_T = 1-\frac{\tilde y_T}{2}\,,
  \qquad 
  \text{for}\quad I_{\{u,d\}1}\,,
\end{equation}
and
\begin{equation}
  \tilde y_T = 2 \bar y_T\,,
  \qquad \qquad
  \bar y_T = \frac{\tilde y_T}{2}\,,
  \qquad 
  \text{for}\quad I_{\{u, d\}0}\,.
\end{equation}
At this point, we have the following singularities:
\begin{equation}
  {\renewcommand{\arraystretch}{1.5}%
  \begin{array}{c@{\qquad \quad}c@{\qquad \quad}l}
   \text{soft}     &  
   k_+ \to 0,\  x_T \to 0 \quad \text{or} \quad l_+ \to 0,\  x_T \to 0 &
   I_{u0}, I_{u1}, I_{d0}, I_{d1}\,,              \\
   \text{collinear} & 
   x_T \to 0,\ \tilde y_T \to 0 & 
   I_{u1}, I_{d1}\,, \\
   k \parallel l & 
   x_T \to 0,\ \tilde y_T \to 0 & 
   I_{u0}, I_{d0}\,. 
  \end{array}
  }
\end{equation}

\subsubsection*{Step 4}

In the last step, we compress the ranges of the $\kp$ and $\lp$ integrals by the
transformation
\begin{equation}
  \kp = \frac{x}{1-x}\,, 
  \qquad \qquad
  \lp = \frac{y}{1-y}\,, 
\end{equation}
whose inverse reads
\begin{equation}
  x = \frac{\kp}{1+\kp}\,, 
  \qquad \qquad
  y = \frac{\lp}{1+\lp}\,.
\end{equation}
Hence, finally, our integral is a sum of four contributions
\begin{equation}
  I = I_{d0} + I_{d1} + I_{u0} + I_{u1}\,.
\end{equation}

\subsubsection{Integrating the weight}

As a final step, we integrate the weight over the azimuthal angles of  $v_3$:
$\phi_1$ and $\phi_2$. In order to map the angular variables into a unit
hypercube, we first write the integration measure as

\begin{align}
  \int_{S_1^{d-3}}\!\!\! d\Omega(\phi_1,\phi_2,\ldots,\phi_{d-3}) 
  & =
  \int_{S_1^{1-2\epsilon}}\!\!\!
  d\Omega(\phi_1,\phi_2,\ldots,\phi_{1-2\epsilon}) 
  \nonumber \\
  & =
  \int_{-1}^1 d\cos\phi_1 \sin^{-1-2\epsilon}\phi_1
  \int_{S_1^{-2\epsilon}}\!\!\! d\Omega(\phi_2,\ldots,\phi_{1-2\epsilon})\,.
\end{align}
Then, we represent the remaining measure as~\cite{Czakon:2014oma}
%
%
\begin{align}
  \int_{S_1^{-2\epsilon}} {d}{\Omega}(\phi_2, \ldots, \phi_{1-2\epsilon}) & =
  \frac{(4\pi)^{-\epsilon}\Gamma(1-\epsilon)}{\Gamma(1-2\epsilon)} 
  \\
  &
  \hspace{-70pt}
  \times \int_{-1}^{+1} {d}\cos\phi_2 \left( \delta(1-\cos\phi_2)
  + \delta(1 + \cos\phi_2) - 2 \epsilon \, \frac{4^\epsilon
    \Gamma(1-2\epsilon)}{\Gamma^2(1-\epsilon)} \left[
    \frac{1}{(1-\cos^2\phi_2)^{1+\epsilon}} \right]_+\right)\,,
  \nonumber
\end{align}
where we used the fact that our integrand is independent of the angles $\phi_3,
\ldots, \phi_{1-2\epsilon}$. The plus distribution is defined as

\begin{align}
  \int_{-1}^{+1} {d}\cos\rho \left[
  \frac{1}{(1-\cos^2\rho)^\alpha} \right]_+ f(\cos\rho) 
  & 
  \\[0.5em]
  & \hspace{-80pt} =
  \int_{-1}^{0} {d}\cos\rho \,
  \frac{f(\cos\rho)-f(-1)}{(1-\cos^2\rho)^\alpha} 
  + \int_{0}^{+1} {d}\cos\rho \,
  \frac{f(\cos\rho)-f(+1)}{(1-\cos^2\rho)^\alpha}
  \; .
  \nonumber
\end{align}
The integral over the weight now reads
\begin{align}
  & \frac{1}{S_{1-2\epsilon}}\int_{S_1^{1-2\epsilon}}
  d\Omega (\phi_1,\phi_2,\ldots,\phi_{1-2\epsilon})   
  \calW(\beta_t,\theta,\phi_1,\phi_2)=
  \frac{(4\pi)^{-\epsilon}\Gamma(1-\epsilon)}{S_{1-2\epsilon}\, \Gamma(1-2\epsilon)}
  \int_{-1}^{1}d\!\cos{\phi_1}\,\sin^{-1-2\epsilon}\phi_1 
  \nonumber \\[0.5em]
  & \times \left[ 
  \int_{0}^{1}d\!\cos{\phi_2} \left( \delta(1-\cos{\phi_2})
  \calW(\beta_t,\theta,\phi_1,0)
  -2\epsilon \frac{4^\epsilon \Gamma(1-2\epsilon)}{\Gamma^2(1-\epsilon)} 
  \frac{\calW(\beta_t,\theta,\phi_1,\phi_2)-
  \calW(\beta_t,\theta,\phi_1,0)}{\left( 1-\cos^2\phi_2\right)^{1+\epsilon}}  \right)
  \right.
  \nonumber \\[0.5em]
  &+ \left.
  \int^{0}_{-1}d\!\cos{\phi_2} \left( \delta(1+\cos{\phi_2})
  \calW(\beta_t,\theta,\phi_1,\pi)
  -2\epsilon \frac{4^\epsilon \Gamma(1-2\epsilon)}{\Gamma^2(1-\epsilon)} 
  \frac{\calW(\beta_t,\theta,\phi_1,\phi_2)-\calW(\beta_t,\theta,
  \phi_1,\pi)}{\left( 1-\cos^2\phi_2\right)^{1+\epsilon}}  \right)
  \right]\,.
\end{align}
To map this integral to a unit hypercube, we use the following changes 
of variables
\begin{equation}
  \cos\phi_1= 1-2 \cos^2(\chi \pi /2) , \qquad 
  \cos\phi_2= 1-\eta_2, \qquad
  \sin^2\phi_2=\eta_2(2-\eta_2)\,,
\end{equation}
in the second 
\begin{equation}
  \cos\phi_1= 1-2 \cos^2(\chi \pi /2) , \qquad 
  \cos\phi_2= \eta_2-1, \qquad
   \sin^2\phi_2=\eta_2(2-\eta_2)\,,
\end{equation}
and the third line, respectively.
From these, one gets 
\begin{align}
  & \frac{1}{S_{1-2\epsilon}}\int_{S_1^{1-2\epsilon}}
  d\Omega (\phi_1,\phi_2,\ldots,\phi_{1-2\epsilon})   
  \calW(\beta_t,\theta,\phi_1,\phi_2)=
  \calR
  \int_{0}^{1}  d \chi \, \sin^{-2\epsilon}(\pi \chi)
   \nonumber\\
  & \times \left[ 
  \int_{0}^{1}d \eta_2 
  \left( 4^{-\epsilon}
  \delta(1-\cos{\phi_2}) \calW(\beta_t,\theta,\phi_1,0)
  -\frac{\epsilon}{\calR} 
  \frac{\calW(\beta_t,\theta,\phi_1,\phi_2)-\calW(\beta_t,\theta,
  \phi_1,0)}{\left( 1-\cos^2\phi_2\right)^{1+\epsilon}}  \right)
  \right.+\nonumber \\
  &\left.
  \int_{0}^{1}d \eta_2 \left( 
  4^{-\epsilon} \delta(1+\cos{\phi_2}) \calW(\beta_t,\theta,
  \phi_1,\pi)
  -\frac{\epsilon}{\calR} 
  \frac{\calW(\beta_t,\theta,\phi_1,\phi_2)-\calW(\beta_t,\theta,
  \phi_1,\pi)}{\left( 1-\cos^2\phi_2\right)^{1+\epsilon}}  \right) 
  \right]\,,
  \label{eq:v3average}
\end{align}
where we define the prefactor
\begin{equation}
  \calR = \frac{\Gamma(1-\epsilon)^2}{2\Gamma(1-2\epsilon)}.
\end{equation}

\section{NNLO soft function: results}
\label{sec:results}

\begin{figure}[t]
  \begin{center}
    \includegraphics[width=0.99\textwidth]{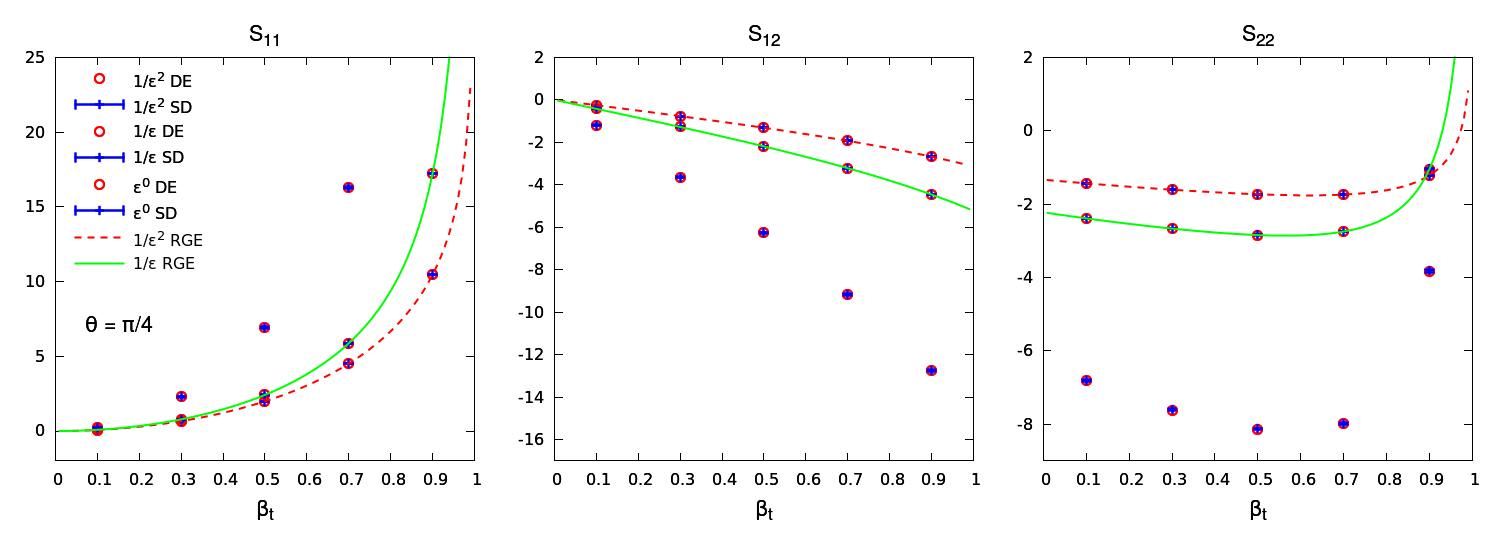}
  \end{center}
  \caption{
  Quark bubble contribution to the NNLO soft function in the \qqbar channel.
  The plots correspond to three independent entries of the $2\times 2$ matrix
  determined for the fixed values of $\theta = \pi/4$ and $\LT = 0$.
  We see perfect agreement between RG prediction and the direct calculation
  obtained with two independent methods: differential equations~(DE) and sector
  decomposition~(SD).
  }
  \label{fig:quarkbubbleqq}
\end{figure}

\begin{figure}[!h]
  \begin{center}
    \includegraphics[width=0.99\textwidth]{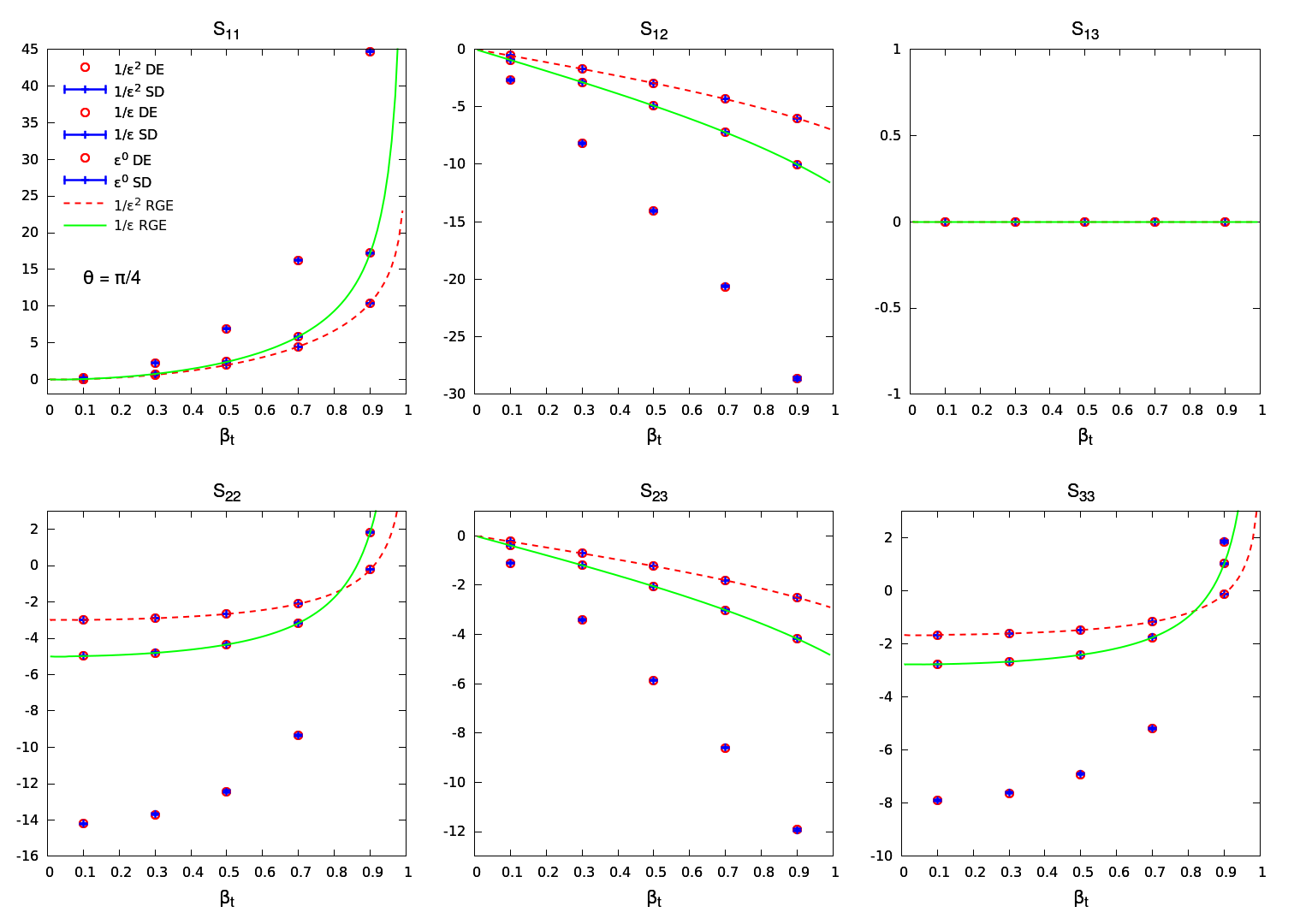}
  \end{center}
  \caption{
  Quark bubble contribution to the NNLO soft function in the $gg$ channel.
  The plots correspond to six independent entries of the $3\times 3$ matrix
  determined for the fixed values of $\theta = \pi/4$ and $\LT = 0$.
  We see perfect agreement between RG prediction and the direct calculation
  obtained with two independent methods: differential equations~(DE) and sector
  decomposition~(SD).
  }
  \label{fig:quarkbubblegg}
\end{figure}

The bare NNLO soft function for the top pair production has the following
structure
\begin{equation}
  \bfS^{(2)}_\text{bare}(\LT, \beta_t,\theta) =
  \frac{1}{\epsilon^2} \bfS^{(2,-2)}(\LT, \beta_t,\theta) +
  \frac{1}{\epsilon} \bfS^{(2,-1)}(\LT, \beta_t,\theta)
  + \bfS^{(2,0)}(\LT, \beta_t,\theta)\,.
  \label{eq:bareSFNNLOpoles}
\end{equation}
Because it encodes single and double-soft limit, it exhibits at most
$1/\epsilon^2$ singularity. However, higher order $\epsilon$ poles, as well as
$\alpha$ poles, appear at intermediate stages of the calculation, but they
cancel in the final combination.

We have checked that, in our calculation, all $\alpha$ poles, including
$\epsilon/\alpha$, as well as $1/\epsilon^4$ vanish within each colour structure
defined in Eq.~(\ref{eq:w1w2def}). As for the $1/\epsilon^3$ pole, it comes out
with a non-zero coefficient in the single-cut and in the double-cut
contributions. The value of the $1/\epsilon^3$ coefficient is however identical,
in those two pieces up to a sign. Hence, in the final combination, this pole
does not appear in our result.
We have demonstrated that, when all contributions to the bare NNLO soft
function are included, following Eq.~(\ref{eq:bareSFNNLOdef}) the soft function
indeed shows at most $1/\epsilon^2$ singularity.

As discussed in Section~\ref{sec:renormalization}, the pole part of the soft
function, \ie the functions $S^{(2,-2)}(\LT, \beta_t,\theta)$ and
$S^{(2,-1)}(\LT, \beta_t,\theta)$ defined in Eq.~(\ref{eq:bareSFNNLOpoles}), as
well as the $\LT$-dependent part of the finite contribution, $S^{(0,0)}(\LT,
\beta_t,\theta)$, can be completely determined from the renormalization group. 
The only term that has to be obtained through direct calculation is the
$\LT$-independent part of $S^{(2,0)}(\LT, \beta_t,\theta)$. 
In spite of the above, it is worth calculating all the terms appearing in
Eq.~(\ref{eq:bareSFNNLOpoles}) and use the
redundant ones for cross checks of our framework against RG prediction.

As much as the comparison of the results from direct calculations to the
prediction of the renormalization group is valuable, it is limited by the fact
that RG misses the constant part of $S^{(2,0)}(\LT, \beta_t,\theta)$. In our
calculation, we are however able to cross-check this missing component as well
since part of our soft function, namely the one corresponding to the bubble and
tadpole diagrams, can be determined with two completely different methods:
differential equations and sector decomposition.

Figs.~\ref{fig:quarkbubbleqq} and \ref{fig:quarkbubblegg} show the results for
the quark bubble part of the soft function, respectively in the \qqbar and the
$gg$ channel, obtained from the renormalization group and from the direct
calculation with differential equations and with sector decomposition. 
The quark bubble contribution can be singled out from the RG prediction as it is
proportional to $n_f$.
The plots correspond to three independent entries of the $2\times 2$ matrix and
six independent entries of the $3\times 3$ matrix, determined for the fixed
values of $\theta = \pi/4$ and $\LT = 0$, and shown as a function of $\beta_t$.
We see perfect agreement between the two sets of points as well as between the
points and the RG prediction.  The error bars in the result from sector
decomposition come from the uncertainties of numerical integration. 
We have performed similar comparison of the results from differential equations
and sector decomposition for the gauge bubble and for the tadpole, in both
channels, each time finding an excellent agreement.

This constitutes a very strong validation of our computational framework. It
tests all the elements of the sector decomposition method and tools, discussed
in Section~\ref{sec:secdec}, up to the finite order in $\epsilon$. Even though
the bubble graphs are easier to solve analytically than the rest of the
double-cut graphs, they pose a serious challenge to the sector decomposition
approach due to non-trivial numerators. Therefore, the comparison shown in
Figs.~\ref{fig:quarkbubbleqq} and \ref{fig:quarkbubblegg} makes up a
highly-nontrivial test of the entire framework used in the calculation of the
NNLO soft function.

\begin{figure}[t]
  \begin{center}
    \includegraphics[width=0.99\textwidth]{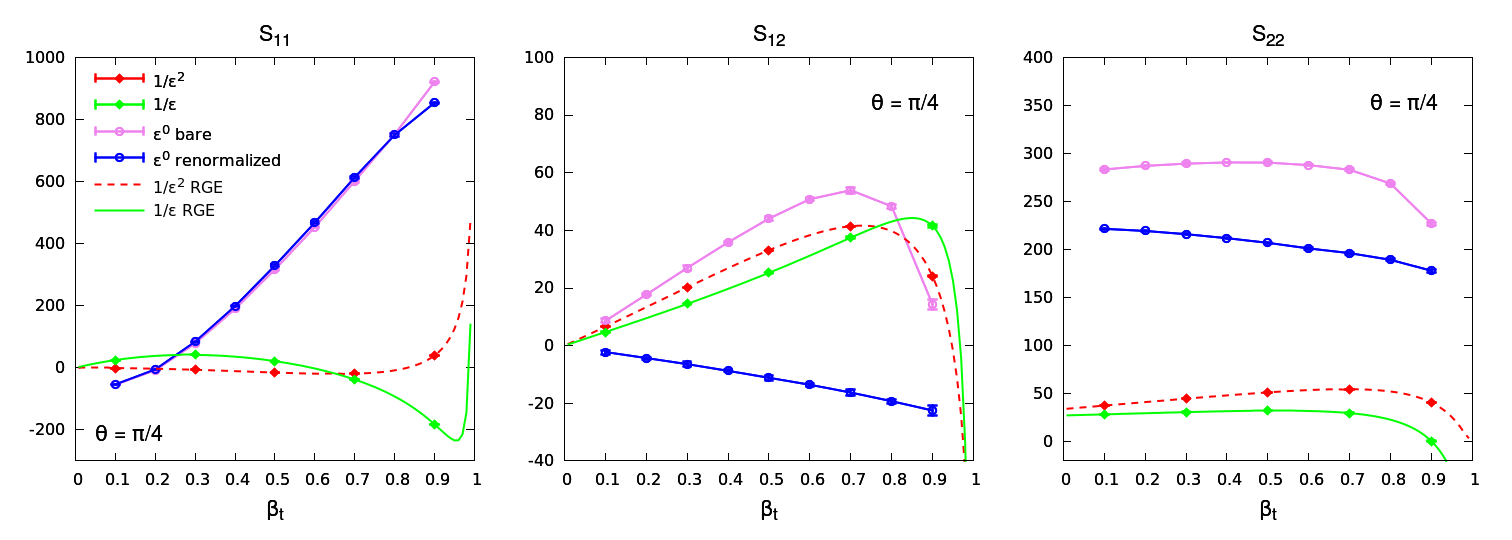}
  \end{center}
  \caption{
  Real part of the NNLO soft function in the \qqbar channel.
  The figure shows three independent entries of the $2\times 2$ matrix
  determined for the fixed values of $\theta = \pi/4$ and $\LT = 0$.
  All the points come from our direct calculation at different orders in
  $\epsilon$ expansion. The two results at order $\epsilon^0$ correspond to the
  finite part of the NNLO soft function before and after renormalization.
  The RG predictions for the pole part are shown as
  red~(dashed) and green~(solid) lines.
  The points for order $\epsilon^0$ are connected by straight lines for better
  visibility.
  }
  \label{fig:SFresqqreal}
\end{figure}

\begin{figure}[!t]
  \begin{center}
    \includegraphics[width=0.99\textwidth]{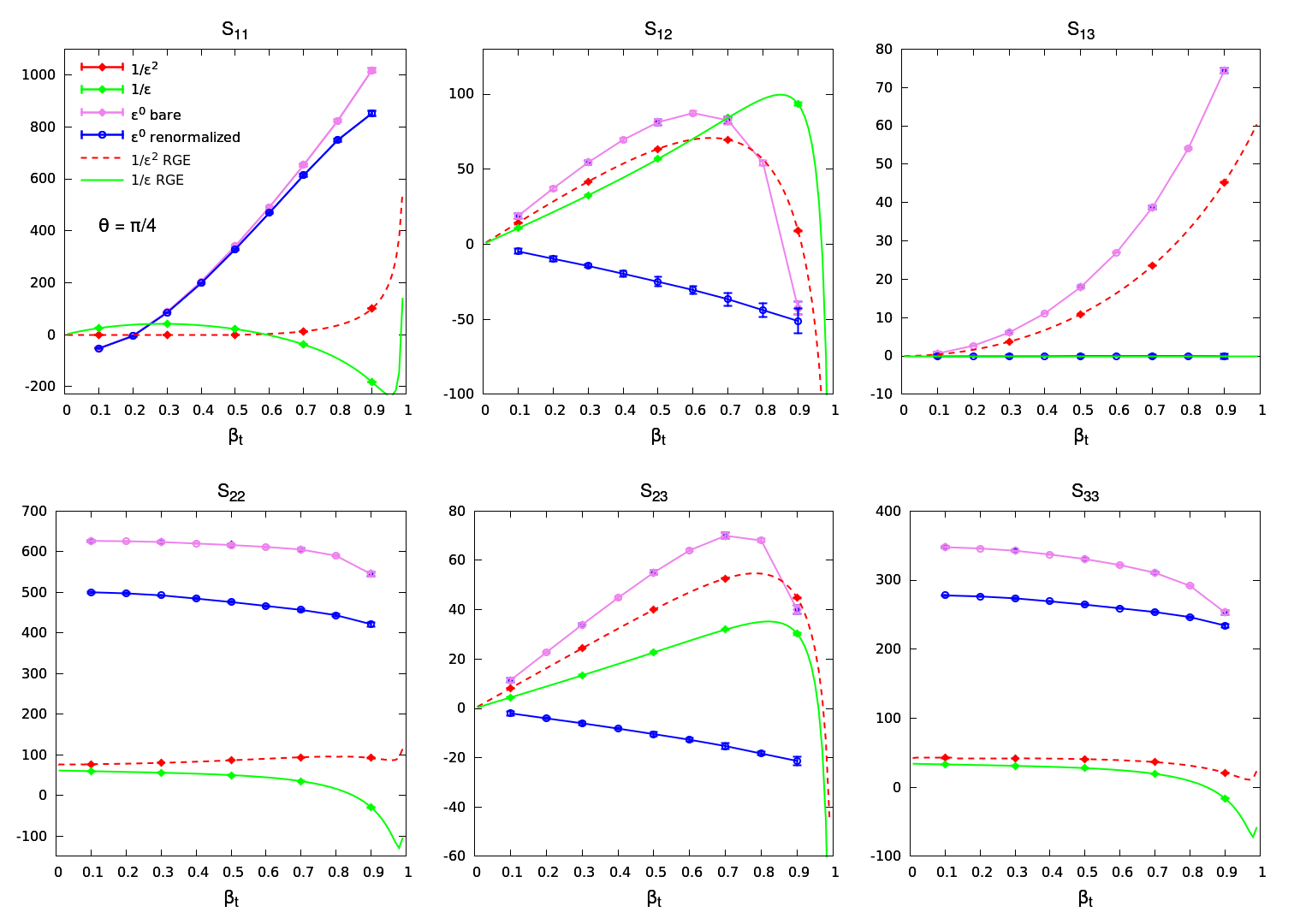}
  \end{center}
  \caption{
  Real part of the NNLO soft function in the $gg$ channel.  The figure shows six
  independent entries of the $3\times 3$ matrix. All details as in
  Fig.~\ref{fig:SFresqqreal}.
  }
  \label{fig:SFresggreal}
\end{figure}

We are now in a position to present the complete NNLO soft function for top
quark pair production at small $q_T$.  The results for the real part of the
independent entries of the matrices in the \qqbar and $gg$ channel, at fixed
values of $\theta = \pi/4$ and $\LT = 0$, are given in
Figs.~\ref{fig:SFresqqreal} and \ref{fig:SFresggreal}, respectively.
The points correspond to our direct calculation at orders $1/\epsilon^2$,
$1/\epsilon$ and $\epsilon^0$. 

For the poles, we also show, as lines,
predictions from the renormalization group. We see that the two sets of
predictions agree perfectly.

For the finite term, we show two sets of points: the one
before and the one after renormalization. They differ by a finite function
following the formula~(\ref{eq:renS2}). We notice that the difference is often
substantial.

\begin{figure}[!t]
  \begin{center}
    \includegraphics[width=0.33\textwidth]{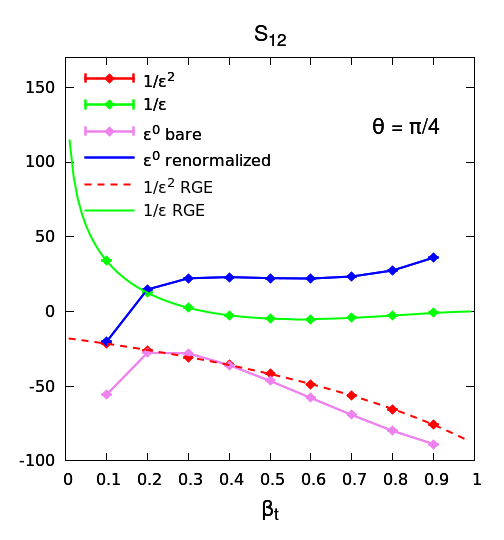}
    \hspace{35pt}
    \includegraphics[width=0.33\textwidth]{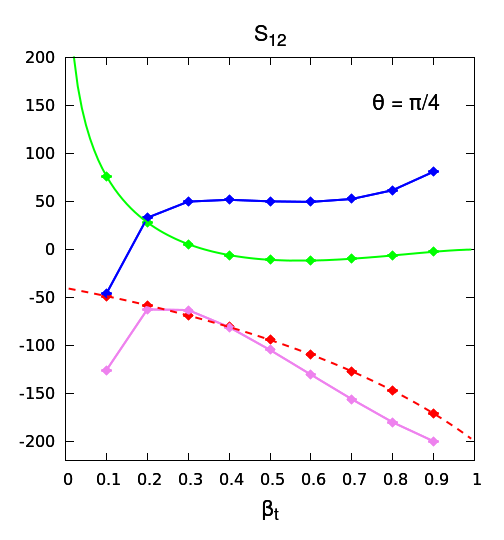}
  \end{center}
  \caption{
  Imaginary part of the NNLO soft function in the \qqbar channel (left) and in
  the $gg$ channel (right).
  The figures show the only one independent structure in each matrix. All
  details as in Fig.~\ref{fig:SFresqqreal}.
  }
  \label{fig:SFresim}
\end{figure}

As discussed in Section~\ref{sec:realvirt}, due to the real-virtual
contributions, the NNLO soft function contains also an imaginary part.  As
follows from Eqs.~(\ref{eq:w2Aqq}) and (\ref{eq:w2Agg}), discussed in
Appendix~\ref{app:colmat}, the imaginary part of the soft function matrix has
only one independent element.  The corresponding results in the \qqbar and $gg$
channel are given in Fig.~\ref{fig:SFresim}. As in the previous cases, the
predictions were obtained for the values of $\theta = \pi/4$ and $\LT = 0$, as
functions of $\beta_t$. 
Like in the case of the real part,  we see excellent agreement
between the renormalization group predictions and the direct calculation for the
pole part.

\section{Summary}

A second, independent calculation of the NNLO correction for top quark pair
production in proton-proton collisions would be of great value.
One of the methods that could be used to obtain it, is the $q_T$ slicing
approach.
In this article, we presented 
the result for the complete NNLO, small-$q_T$ soft function for \ttbar
production, which  forms a significant step on the way towards
this goal.
Now, all the ingredients needed to construct the NNLO cross section for top pair
production valid at small-$q_T$ are available.

To obtain our results, we have constructed a framework based on sector
decomposition and differential equations. The framework has been extensively
validated. In particular, we have checked that all the $\alpha$
poles, including $\epsilon/\alpha$, and all the $\epsilon$ poles beyond
$1/\epsilon^2$, vanish in the complete result for the NNLO soft function.
We also checked that we recover all pieces predicted by the renormalization
group, both the pole terms and the $\LT$-dependent part of the finite term.
The strongest validation of our framework and tools comes from 
finding a perfect agreement between numerical results from
the sector decomposition-based framework and analytic results obtained with the
method of differential equations, for the graphs
involving gauge, ghost or quark bubble.

The NNLO, small-$q_T$ soft function  can now be used to obtain full $t\bar t$
cross section at NNLO by means of the $q_T$-slicing method, as well as for
small-$q_T$ resummation at NNLL'.

\label{sec:summary}

\section*{Acknowledgements}

This work has been supported by the National Science Centre, Poland grant
POLONEZ  
%
No. 2015/19/P/ST2/03007. 
The project has received funding from the
European Union's Horizon 2020 research  and  innovation  programme  under  the
Marie Sk\l{}odowska-Curie grant agreement No. 665778.
The work of M.C. was supported in part by a grant of the BMBF.
R.A.M and S.S acknowledge hospitality of the
Institute for Theoretical Particle Physics and Cosmology, RWTH Aachen
University, where substantial part of this work has been done.
We are also grateful to Andreas van Hameren for illuminating discussions
throughout the course of this work.
We thank Thomas Hahn for assistance with the Cuba library.
This research was supported in part by PLGrid infrastructure.
We are grateful for computing resources of ACC Cyfronet AGH where our numerical
calculations were performed on the Prometheus cluster.
S.S. would like to acknowledge the Mainz Institute for Theoretical Physics
(MITP) for hospitality and support during the completion of this work.

\appendix


%
%
\section{Colour matrices}
\label{app:colmat}

In this appendix, we collect all colour matrices that appear in the formulae
for the soft function at NLO and NNLO, in the basis defined in 
Eqs.~(\ref{eq:colour-basis-qq}) and~(\ref{eq:colour-basis-gg}).

Let us start by determining the set of independent dot products of the operators
$\cT_i$. For the top pair production in the $\iibar$ channel, we have
\begin{eqnarray}
  \begin{array}{ll}
    \cT_i^2 =  C_i \unitop\,,  &
    \qquad \text{for the incoming partons } \iibar\,,
    \\
    \cT_i^2 =  C_F \unitop\,, &
    \qquad \text{for the outgoing } \ttbar \text{ pair}\,.
    \label{eq:Ti2}
  \end{array}
\end{eqnarray}
The colour conservation 
\begin{equation}
  \sum_{i=1}^4 \cT_i |\calM\rangle = 0\,,
  \label{eq:colconsdef}
\end{equation}
allows us to write the following set of equations
\begin{equation}
 \{\cT_i \cdot \cT_j  = 
 \sum_{\substack{k \neq i \\ m \neq j}} \cT_k \cdot \cT_m \}\,,
  \label{eq:colconservTT}
\end{equation}
for $i,j \in \{1,\ldots, 4\}$.
By combining~(\ref{eq:Ti2}) and (\ref{eq:colconservTT}), we arrive at the
relations
\begin{subequations}
  \label{eq:TiTjrelations}
  \begin{align}
    \cT_1 \cdot \cT_4  & = \cT_2 \cdot \cT_3\,,\\ 
    \cT_2 \cdot \cT_4  & = \cT_1 \cdot \cT_3\,,\\ 
    \cT_1 \cdot \cT_1  & = \cT_2 \cdot \cT_2  =
    -\cT_1 \cdot \cT_3  - \cT_2 \cdot \cT_3  - \cT_1 \cdot \cT_2  \,,\\ 
    \cT_3 \cdot \cT_3  & = \cT_4 \cdot \cT_4  =
    -\cT_1 \cdot \cT_3  - \cT_2 \cdot \cT_3  - \cT_3 \cdot \cT_4  \,.
  \end{align}
\end{subequations}
Hence, in general, we have four independent structures: 
$\cT_1 \cdot \cT_3$, $\cT_2 \cdot \cT_3$, $\cT_1 \cdot \cT_2$, $\cT_3 \cdot
\cT_4$. For the \qqbar channel, this number is further reduced to
three, as all $\cT_i\cdot \cT_i$ are equal, which leads to the relation
$\cT_1 \cdot \cT_2 = \cT_3 \cdot \cT_4$.

In the case of the \qqbar channel, the three independent matrices
$\bfw_{ij}^{(1)}$, defined in Eq.~(\ref{eq:w1def}), read
\begin{subequations}
  \allowdisplaybreaks
  \begin{align}
  \bfw_{13}^{(1)} & = 
  - \frac{C_F}{2}
  \left( \begin{array}{cc}
    0 & 1 \\
    1 & 2 C_F -\frac{N}{2}
  \end{array} \right)\,,
  \\[0.5em]
  \bfw_{23}^{(1)} & = 
  - \frac{C_F}{2N}
  \left( \begin{array}{cc}
    0 & -N \\
    -N & 1 
  \end{array} \right)\,,
  \\[0.5em]
  \bfw_{34}^{(1)} & = 
  - \frac{C_F}{4N}
  \left( \begin{array}{cc}
    4N^2 & 0 \\
    0 & -1 
  \end{array} \right)\,,
  \end{align}
\end{subequations}
while, for the $gg$ channel, we have the following four independent
$\bfw_{ij}^{(1)}$ matrices
\begin{subequations}
  \allowdisplaybreaks
  \begin{align}
   \bfw_{13}^{(1)} & = 
   -\frac{1}{8}
   \left( \begin{array}{ccc}
     0  & 4N     & 0 \\
     4N & N^2    & N^2-4 \\
     0  & N^2-4  & N^2-4
   \end{array} \right)\,,
   \\[0.5em]
   \bfw_{23}^{(1)} & = 
   -\frac{1}{8}
   \left( \begin{array}{ccc}
     0   & -4N     & 0 \\
     -4N & N^2     & -N^2+4 \\
     0   & -N^2+4  & N^2-4
   \end{array} \right)\,,
   \\[0.5em]
   \bfw_{12}^{(1)} & = 
   - \frac{1}{4}
   \left( \begin{array}{ccc}
     4N^2 & 0   & 0 \\
     0    & N^2 & 0 \\
     0    & 0   & N^2-4
   \end{array} \right)\,,
   \\[0.5em]
   \bfw_{34}^{(1)} & = 
   - 
   \left( \begin{array}{ccc}
     C_F N & 0            & 0 \\
     0     & -\frac{1}{4} & 0 \\
     0     & 0            & -\frac{N^2-4}{4N^2}
   \end{array} \right)\,.
  \end{align}
\end{subequations}

The $\bfw_{ij}^{(2S)}$ matrices, which enter the abelian part of the double-cut
contributions to the soft function (\ref{eq:2cutmaster}), were defined in
Eq.~(\ref{eq:w2def}). In the case of the $\qqbar$ channel, they take the
following explicit forms
\begin{subequations}
\allowdisplaybreaks
\begin{align}
    \bfw_{1333}^{(2S)}  & = 
    \left(
    \begin{array}{cc}
     0 & -\frac{\left(N^2-1\right)^2}{4 N^2} \\
     -\frac{\left(N^2-1\right)^2}{4 N^2} & -\frac{\left(N^2-2\right)
    \left(N^2-1\right)^2}{8 N^3} \\
    \end{array}
    \right)
    \,,\\[0.5em]
    \bfw_{1433}^{(2S)}  & = 
\left(
\begin{array}{cc}
 0 & \frac{\left(N^2-1\right)^2}{4 N^2} \\
 \frac{\left(N^2-1\right)^2}{4 N^2} & -\frac{\left(N^2-1\right)^2}{4 N^3} \\
\end{array}
\right)
    \,,\\[0.5em]
    \bfw_{1314}^{(2S)}  & = 
\left(
\begin{array}{cc}
 -\frac{N^2-1}{2 N} & \frac{1}{8} \left(-N^2+5-\frac{4}{N^2}\right) \\
 \frac{1}{8} \left(-N^2+5-\frac{4}{N^2}\right) & \frac{N^4-4 N^2+3}{8 N^3} \\
\end{array}
\right)
    \,,\\[0.5em]
    \bfw_{3334}^{(2S)}  & = 
\left(
\begin{array}{cc}
 -\frac{\left(N^2-1\right)^2}{2 N} & 0 \\
 0 & \frac{\left(N^2-1\right)^2}{8 N^3} \\
\end{array}
\right)
    \,,\\[0.5em]
    \bfw_{1334}^{(2S)}  & = 
\left(
\begin{array}{cc}
 0 & \frac{N^4-3 N^2+2}{8 N^2} \\
 \frac{N^4-3 N^2+2}{8 N^2} & -\frac{N^4-3 N^2+2}{8 N^3} \\
\end{array}
\right)
    \,,\\[0.5em]
    \bfw_{1434}^{(2S)}  & = 
\left(
\begin{array}{cc}
 0 & -\frac{N^4-3 N^2+2}{8 N^2} \\
 -\frac{N^4-3 N^2+2}{8 N^2} & -\frac{N^2-1}{4 N^3} \\
\end{array}
\right)
    \,,\\[0.5em]
    \bfw_{3434}^{(2S)}  & = 
\left(
\begin{array}{cc}
 \frac{\left(N^2-1\right)^2}{2 N} & 0 \\
 0 & \frac{N^2-1}{8 N^3} \\
\end{array}
\right)
    \,,\\[0.5em]
    \bfw_{1414}^{(2S)}  & = 
\left(
\begin{array}{cc}
 \frac{N^2-1}{2 N} & \frac{1}{2} \left(\frac{1}{N^2}-1\right) \\
 \frac{1}{2} \left(\frac{1}{N^2}-1\right) & \frac{N^4+2 N^2-3}{8 N^3} \\
\end{array}
\right)
    \,,\\[0.5em]
    \bfw_{1313}^{(2S)}  & = 
\left(
\begin{array}{cc}
 \frac{N^2-1}{2 N} & \frac{N^4-3 N^2+2}{4 N^2} \\
 \frac{N^4-3 N^2+2}{4 N^2} & \frac{N^6-4 N^4+6 N^2-3}{8 N^3} \\
\end{array}
\right)
    \,,\\[0.5em]
    \bfw_{3333}^{(2S)}  & = 
\left(
\begin{array}{cc}
 \frac{\left(N^2-1\right)^2}{2 N} & 0 \\
 0 & \frac{\left(N^2-1\right)^3}{8 N^3} \\
\end{array}
\right)
\,.
\end{align}
\end{subequations}
In the $gg$ channel, these colour matrices read
\begin{subequations}
\allowdisplaybreaks
\begin{align}
    \bfw_{1333}^{(2S)}  & = 
\left(
\begin{array}{ccc}
 0 & \frac{1}{2} \left(1-N^2\right) & 0 \\
 \frac{1}{2} \left(1-N^2\right) & -\frac{1}{8} N \left(N^2-1\right) &
-\frac{N^4-5 N^2+4}{8 N} \\
 0 & -\frac{N^4-5 N^2+4}{8 N} & -\frac{N^4-5 N^2+4}{8 N} \\
\end{array}
\right)
    \,,\\[0.5em]
    \bfw_{1433}^{(2S)}  & = 
\left(
\begin{array}{ccc}
 0 & \frac{1}{2} \left(N^2-1\right) & 0 \\
 \frac{1}{2} \left(N^2-1\right) & -\frac{1}{8} N \left(N^2-1\right) &
\frac{N^4-5 N^2+4}{8 N} \\
 0 & \frac{N^4-5 N^2+4}{8 N} & -\frac{N^4-5 N^2+4}{8 N} \\
\end{array}
\right)
    \,,\\[0.5em]
    \bfw_{1314}^{(2S)}  & = 
\left(
\begin{array}{ccc}
 -N & 0 & 1-\frac{N^2}{4} \\
 0 & -\frac{N}{4} & 0 \\
 1-\frac{N^2}{4} & 0 & \frac{N}{4}-\frac{1}{N} \\
\end{array}
\right)
    \,,\\[0.5em]
    \bfw_{3334}^{(2S)}  & = 
\left(
\begin{array}{ccc}
 -\frac{\left(N^2-1\right)^2}{2 N} & 0 & 0 \\
 0 & \frac{N^2-1}{4 N} & 0 \\
 0 & 0 & \frac{N^4-5 N^2+4}{4 N^3} \\
\end{array}
\right)
    \,,\\[0.5em]
    \bfw_{1334}^{(2S)}  & = 
\left(
\begin{array}{ccc}
 0 & \frac{1}{4} \left(N^2-2\right) & 0 \\
 \frac{1}{4} \left(N^2-2\right) & -\frac{N}{8} & -\frac{N^2-4}{8 N} \\
 0 & -\frac{N^2-4}{8 N} & -\frac{N^2-4}{8 N} \\
\end{array}
\right)
    \,,\\[0.5em]
    \bfw_{1434}^{(2S)}  & = 
\left(
\begin{array}{ccc}
 0 & \frac{1}{4} \left(2-N^2\right) & 0 \\
 \frac{1}{4} \left(2-N^2\right) & -\frac{N}{8} & \frac{N^2-4}{8 N} \\
 0 & \frac{N^2-4}{8 N} & -\frac{N^2-4}{8 N} \\
\end{array}
\right)
    \,,\\[0.5em]
    \bfw_{3434}^{(2S)}  & = 
\left(
\begin{array}{ccc}
 \frac{\left(N^2-1\right)^2}{2 N} & 0 & 0 \\
 0 & \frac{1}{4 N} & 0 \\
 0 & 0 & \frac{N^2-4}{4 N^3} \\
\end{array}
\right)
    \,,\\[0.5em]
    \bfw_{1414}^{(2S)}  & = 
\left(
\begin{array}{ccc}
 N & -\frac{N^2}{4} & \frac{1}{4} \left(N^2-4\right) \\
 -\frac{N^2}{4} & \frac{1}{8} N \left(N^2+2\right) & -\frac{1}{8} N
\left(N^2-4\right) \\
 \frac{1}{4} \left(N^2-4\right) & -\frac{1}{8} N \left(N^2-4\right) &
\frac{N^3}{8}-\frac{3 N}{4}+\frac{1}{N} \\
\end{array}
\right)
    \,,\\[0.5em]
    \bfw_{1313}^{(2S)}  & = 
\left(
\begin{array}{ccc}
 N & \frac{N^2}{4} & \frac{1}{4} \left(N^2-4\right) \\
 \frac{N^2}{4} & \frac{1}{8} N \left(N^2+2\right) & \frac{1}{8} N
\left(N^2-4\right) \\
 \frac{1}{4} \left(N^2-4\right) & \frac{1}{8} N \left(N^2-4\right) &
\frac{N^3}{8}-\frac{3 N}{4}+\frac{1}{N} \\
\end{array}
\right)
    \,,\\[0.5em]
    \bfw_{3333}^{(2S)}  & = 
\left(
\begin{array}{ccc}
 \frac{\left(N^2-1\right)^2}{2 N} & 0 & 0 \\
 0 & \frac{\left(N^2-1\right)^2}{4 N} & 0 \\
 0 & 0 & \frac{\left(N^2-4\right) \left(N^2-1\right)^2}{4 N^3} \\
\end{array}
\right)
    \,.
\end{align}
\end{subequations}
All the other $\bfw_{ijkl}^{(2S)}$ matrices can be derived from the above set
by using the relations~(\ref{eq:TiTjrelations}).

As for the antisymmetric configurations, $\bm{w}_{ijk}^{(2A)}$, there is only
one independent matrix per channel, which we choose to be $\bm{w}_{123}^{(2A)}$. In the $\qqbar$ channel it reads
\begin{eqnarray}
  \bm{w}_{123}^{(2A)} & =
  \frac{1}{8} \left(
  \begin{array}{cc}
  0     & -(N^2-1) \\
  N^2-1 & 0 \\
  \end{array}
  \right)\,,
  \label{eq:w2Aqq}
\end{eqnarray}
and in the $gg$ channel
\begin{eqnarray}
  \bm{w}_{123}^{(2A)} & =
  \frac{1}{4} \left(
  \begin{array}{ccc}
  0    & -N^2 & 0 \\
  N^2 & 0   & 0 \\
  0    & 0   & 0
  \end{array}
  \right)\,.
  \label{eq:w2Agg}
\end{eqnarray}
All the other antisymmetric matrices can be obtained  as combinations of the
above and $\bfw_{ij}^{(1)}$, using colour conservation (\ref{eq:colconsdef}),
together with the relation
\begin{equation}
  f^{abc} T^b_i T^c_i = \frac{i}{2} C_A T^a_i\,.
\end{equation}

\section{Cusp angles}

The amplitudes are functions of Lorentz invariants~\cite{Becher:2009kw}
\begin{equation}
  s_{ij} \equiv 2 \sigma_{ij} p_i \cdot p_j + \iep\,,
  \label{eq:sij-def}
\end{equation}
and
\begin{equation}
  p_i^2 = m_i^2\,,
\end{equation}
where
\begin{eqnarray}
  \sigma_{ij} = \left\{
    \begin{array}{cl} 
      + 1 &  \text{if } p_i, p_j \text{ are both incoming or outgoing} \\
      - 1 &  \text{otherwise}
    \end{array}
    \right.\,.
\end{eqnarray}
For massive particles we define velocities
\begin{equation}
  v_i \equiv  \frac{p_i}{m_i}\,,
  \qquad \qquad v_i^2 = 1\,.
\end{equation}
To this end,
we label massless particles with lowercase $i, j, \ldots$ and massive
ones with capital indices $I, J, \ldots$.

The soft anomalous dimension of Eq.~(\ref{eq:softAD}) is a function of
\emph{cusp angles}, $\beta_{ij}$, $\beta_{Ij}$ and $\beta_{IJ}$, formed by
massless and/or massive Wilson lines. They read~\cite{Becher:2009kw}
\begin{subequations}
  \begin{align}
    \beta_{ij} & = 
    \ln \frac{-2 \sigma_{ij}\, p_i \cdot p_j\, \mu^2}{(-p_i^2)(-p_j^2)} =
    L_i + L_j - \ln\frac{\mu^2}{-s_{ij}}\,,
    \\
    \beta_{Ij} & = 
    \ln \frac{-2 \sigma_{Ij}\, v_I \cdot p_j\, \mu}{(-p_j^2)} =
    L_j - \ln\frac{m_I \mu}{-s_{Ij}}\,,
    \\
   \label{eq:betaIJ-def}
    \beta_{IJ} & =  {\rm arccosh} \left(w_{IJ}\right) = 
                    {\rm arccosh} \left(\frac{-s_{IJ}}{2 m_I m_J}\right)\,,
  \end{align}
\end{subequations}
where we have introduced
\begin{equation}
  L_i = \ln \frac{\mu^2}{(-p_i^2)}\,,
\end{equation}
which comes from regularization of the IR divergences in the effective theory by
taking massless partons slightly off-shell $(-p_i^2) > 0$~\cite{Becher:2009qa}.
The logarithms $L_i$ need to cancel in the final result for the anomalous
dimension matrix.

\subsection*{$\mathbold{\betaIJ}$ in space-like kinematics}

For space-like kinematics of heavy particles (\eg a top quark incoming and a top
quark outgoing), the cusp angle $\beta_{IJ}$ is real. It can still, however, be
chosen to be positive or negative, as the function $\text{arccosh}$ has two
branches, even for real argument. 

The choice of the positive $\betaIJ$ has at least two advantages: (i)
functions that appear in the $\order{\as^2}$ contribution in
Eq.~(\ref{eq:vel-dep-ad}), which we shall discuss in Appendix~\ref{app:AD}, are real (in the space-like case) and do not require
analytic continuation, as $0 < x < 1$ for $\Li_{2,3}(x)$ and for $\ln(x)$, (ii),
$\coth(\betaIJ)$ has a physical interpretation of an inverse of a relative
velocity between particles $I$ and $J$, $\vIJ$, which reads
\begin{equation}
  \coth(\betaIJ) = \frac{1}{\vIJ}\,,
  \label{eq:cothbetaIJ}
\end{equation}
where
\begin{equation}
  v_{IJ} = \sqrt{1- \frac{4 m_t^4}{s_{IJ}^2}}\,.
\end{equation}

The choice $\betaIJ>0$ implies $\vIJ > 0$ and this leads to the following
relation between the latter and the $\beta_t$ function introduced in
Eq.~(\ref{eq:beta-def})
%
\begin{equation}
  \vIJ = \frac{2 \beta_t}{1+\beta^2_t}\,.
  \label{eq:vIJ-beta-rel}
\end{equation}
We notice that
%
\begin{equation}
  \beta_t = \sqrt{\frac{s-1}{s+1}}\,, \qquad \text{where} \qquad 
  s = \frac{\sIJ}{2 m_t^2}\,,
\end{equation}
which means that
\begin{equation}
  \begin{array}{lll}
    \text{space-like:}\quad & s < -1 \quad & \beta_t \in [1,\infty)\,, \\[0.5em]
    \text{time-like:}\quad & s > 1 \quad & \beta_t \in [0,1]\,.
  \end{array}
\end{equation}
We also note that $\vIJ \in [0,1]$, both in the space-like and in the time-like
case.

Eq.~(\ref{eq:cothbetaIJ}) has a unique solution for the space-like case with
$\betaIJ >0$ which reads
%
\begin{equation}
  \betaIJ =  -\frac12 \ln \frac{1-\vIJ}{1+\vIJ}\,,
  \label{eq:betaIJ-vIJ}
\end{equation}
and, with help of Eq.~(\ref{eq:vIJ-beta-rel}) it can be expressed as
\begin{equation}
 \betaIJ = -\frac12 \ln\left[\left(\frac{\beta_t-1}{\beta_t+1}\right)^2\right]
         = \ln\left(\frac{\beta_t+1}{\beta_t-1}\right)\,,
  \qquad \beta_t \in [1,\infty).
  \label{eq:betaIJ-beta}
\end{equation}

\subsection*{$\mathbold{\betaIJ}$ in time-like kinematics}
\label{sec:beta34TL}

We would like to analytically continue $\betaIJ$ to the time-like region which
corresponds to $\sIJ > 0$. The function in Eq.~(\ref{eq:betaIJ-beta}) has a cut
at imaginary axis and, depending whether we cross it in the upper or the
lower half-plane, we get a different result.

The convention is already established in Eq.~(\ref{eq:sij-def}).
It implies that the invariant $\sIJ$  has a small
positive part which corresponds to analytic continuation over the upper
half-plane. And this means that $\betaIJ$ acquires an imaginary part $- i \pi$
in the time-like region.
The real part is identical to the one of the space-like kinematics and, at the
end, we get
\begin{equation}
  \betaIJ = \ln\left(\frac{1+\beta_t}{1-\beta_t}\right) - i \pi\,,
  \qquad \beta_t \in [0,1]\,.
  \label{eq:beta34TL}
\end{equation}

\section{Anomalous dimensions and IR renormalization constant}
\label{app:AD}

The cusp anomalous dimension is  given by the following perturbative series
\begin{align}
  \label{eq:gammaexp}
  \Gamma_{\text{cusp}}^i(\alpha_s) &= 
  \Gamma_0^i \, \frac{\alpha_s}{4\pi} + 
  \Gamma_1^i \left( \frac{\alpha_s}{4\pi} \right)^2 + 
  \Gamma_2^i \left( \frac{\alpha_s}{4\pi} \right)^3 +
  \ldots \,, 
\end{align}
where
\begin{equation}
  \Gamma_{n}^i  =  C_i\, \gamma_{n}^\text{cusp}
  \qquad
  \text{ with }
  \qquad
  C_i =  \left\{
  \begin{array}{cc}
    C_F \quad \text{for the $q\bar q$ channel}\,, \\
    C_A \quad \text{for the $gg$ channel}\,,
  \end{array}
  \right.
\end{equation}
and the coefficients to three loops read
\cite{Moch:2004pa}
\begin{align}
  \gamma^{\text{cusp}}_0 &= 4 \, ,
  \nonumber
  \\
  \gamma^{\text{cusp}}_1 &= \left( \frac{268}{9} - \frac{4\pi^2}{3} \right) C_A -
  \frac{80}{9} T_F n_f \, ,
  \nonumber
  \\
  \gamma^{\text{cusp}}_2 &= C_A^2 \left( \frac{490}{3} - \frac{536\pi^2}{27} +
    \frac{44\pi^4}{45} + \frac{88}{3} \zeta_3 \right) + C_A T_F n_f \left( -
    \frac{1672}{27} + \frac{160\pi^2}{27} - \frac{224}{3} \zeta_3 \right) \nonumber
  \\
  &\quad + C_F T_F n_f \left( - \frac{220}{3} + 64 \zeta_3 \right) - \frac{64}{27} T_F^2
  n_f^2 \, .
\end{align}
The QCD $\beta$ function is given by
\begin{align}
  \label{eq:betaexp}
  \beta(\alpha_s) &= -2\alpha_s \left[ \beta_0 \, \frac{\alpha_s}{4\pi} 
  + \ldots \right] ,
\end{align}
where the leading coefficient for $n_f$ flavours of the active, massless quarks,
is
\begin{equation}
  \beta_0 = \frac{11}{3} C_A - \frac{4}{3} T_F n_f \,.
  \label{eq:beta0def}
\end{equation}

The soft renormalization factor $\bfZ_s$ can be obtained from the hard
renormalization factor $\bfZ$, determined in Ref.~\cite{Becher:2009cu}, by
discarding all terms proportional to derivatives of the anomalous dimension and
by making the replacement $\bfGamma_n \to \bfgamma^s_n$. 
Up to the order $\alpha_s^2$, it reads 

\begin{align}
  \bfZ_s & = 1 + \frac{\as}{4\pi} 
  \frac{\bm{\bfgamma^s}_0}{2\epsilon}
  + \left( \frac{\as}{4\pi} \right)^2 \left[
  \frac{\bm{\bfgamma^s}_0}{8\epsilon^2} 
  \left( \bm{\bfgamma^s}_0 -2\beta_0 \right) 
  + \frac{\bm{\bfgamma^s}_1}{4\epsilon} \right] + \mathcal{O}(\alpha_s^3) \,.
\end{align}
The soft anomalous dimension
\begin{equation}
  \bfgamma^s  = \sum_{n=0}^\infty 
  \left(\frac{\as}{4\pi}\right)^{n+1}\bfgamma^s_n\,,
\end{equation}
has been introduced in
Section~\ref{sec:renormalization}, and its explicit result for processes
involving massive particles reads~\cite{Ahrens:2010zv}
\begin{align}
  \bfgamma^s_{q\bar{q}} &= 
  \left[ 
  C_F \,
  \gamma_{\text{cusp}}(\beta_{34},\alpha_s) + 
  2\gamma^Q(\alpha_s) \right] \unitop \nonumber
  \\
  &\quad\mbox{} + \frac{N}{2} \left[ \gamma_{\text{cusp}}(\alpha_s) \, \left(
      \ln\frac{t_1^2}{M^2m_t^2} + i\pi \right) - \gamma_{\text{cusp}}(\beta_{34},\alpha_s)
  \right]
  \begin{pmatrix}
    0 & 0
    \\
    0 & 1
  \end{pmatrix}
  \nonumber
  \\
  &\quad\mbox{} + \gamma_{\text{cusp}}(\alpha_s) \, \ln\frac{t_1^2}{u_1^2} \left[
    \begin{pmatrix}
      0 & \frac{C_F}{2N}
      \\
      1 & -\frac{1}{N}
    \end{pmatrix}
    + \frac{\alpha_s}{4\pi} \, g(\beta_{34})
    \begin{pmatrix}
      0 & \frac{C_F}{2}
      \\
      -N & 0
    \end{pmatrix}
  \right] , 
\end{align}
and 
\begin{align}
  \bfgamma^s_{gg} &= \left[
  C_F \,
  \gamma_{\text{cusp}}(\beta_{34},\alpha_s) +
  2\gamma^Q(\alpha_s)
  \right] \unitop \nonumber
  \\
  &\quad\mbox{} + \frac{N}{2} \left[ \gamma_{\text{cusp}}(\alpha_s) \, \left(
  \ln\frac{t_1^2}{M^2m_t^2} + i\pi\right) -
  \gamma_{\text{cusp}}(\beta_{34},\alpha_s) \right]
  \begin{pmatrix}
    0 & 0 & 0
    \\
    0 & 1 & 0
    \\
    0 & 0 & 1
  \end{pmatrix}
  \nonumber
  \\
  &\quad\mbox{} + \gamma_{\text{cusp}}(\alpha_s) \, \ln\frac{t_1^2}{u_1^2}
  \left[ \begin{pmatrix}
  0 & \frac{1}{2} & 0
  \\
  1 & -\frac{N}{4} & \frac{N^2-4}{4N}
  \\
  0 & \frac{N}{4} & -\frac{N}{4}
  \end{pmatrix} 
  + \frac{\alpha_s}{4\pi}\,g(\beta_{34})
  \begin{pmatrix}
  0 & \frac{N}{2} & 0
  \\
  -N & 0 & 0 \\
  0 & 0 & 0
    \end{pmatrix}
  \right]\,.
\end{align}

In order to fully determine the above objects, we need the single-particle
massive anomalous dimensions for heavy quarks~\cite{Becher:2009qa,
Becher:2009kw}
\begin{align}
  \gamma^Q_0 &= - 2 C_F \, ,
  \nonumber
  \\
  \gamma^Q_1 &= C_F C_A \left( -\frac{98}{9} + \frac{2\pi^2}{3} - 4 \zeta_3 \right) +
  \frac{40}{9} C_F T_F n_f \, ,
\end{align}
as well as the velocity-dependent anomalous dimensions~\cite{Becher:2009kw,
Ferroglia:2009ep, Ferroglia:2009ii, Korchemsky:1987wg, Korchemsky:1991zp,
Kidonakis:2009ev}
\begin{align}
  \label{eq:vel-dep-ad}
  \gamma^{\text{cusp}}_0(\beta) &= \gamma^{\text{cusp}}_0 \beta \coth \beta
  \, ,
  \nonumber
  \\
  \gamma^{\text{cusp}}_1(\beta) &= \gamma^{\text{cusp}}_1 \beta \coth \beta
  + 8 C_A \Bigg\{ \frac{\pi^2}{6} + \zeta_3 + \beta^2 \nonumber
  \\
  &\quad + \coth^2\beta \bigg[ \Lithree(e^{-2\beta}) + \beta
  \Litwo(e^{-2\beta}) - \zeta_3 + \frac{\pi^2}{6} \beta + \frac{\beta^3}{3}
  \bigg]
  \nonumber
  \\
  &\quad + \coth\beta \bigg[ \Litwo(e^{-2\beta}) - 2\beta
  \ln(1-e^{-2\beta}) - \frac{\pi^2}{6} (1+\beta) - \beta^2 -
  \frac{\beta^3}{3} \bigg] \Bigg\} \,,
\end{align}
and the function
\begin{align}
  \label{eq:gfunc}
  g(\beta) &= \coth\beta \bigg[ \beta^2 + 2\beta
  \ln(1-e^{-2\beta}) - \Litwo(e^{-2\beta}) + \frac{\pi^2}{6} \bigg] -
  \beta^2 - \frac{\pi^2}{6} \, .
\end{align}

The velocity-dependent cusp anomalous dimensions
$\gamma^{\text{cusp}}(\beta_{34})$ requires careful treatment. It is
unambiguously defined for space-like kinematics, in which case the function
given in Eq.~(\ref{eq:vel-dep-ad}) is real.  In the time-like kinematics,
however, $\gamma^{\text{cusp}}(\beta_{34})$ has to be analytically continued to
the region $\beta \in [0,1]$, as discussed in Appendix~\ref{sec:beta34TL}.

It turns out that the functions $\coth$, $\ln$ and $\Li_{2,3}$, appearing in
Eqs.~(\ref{eq:vel-dep-ad}) and~(\ref{eq:gfunc}), do not develop discontinuities
when going from the space-like to the time-like case. 
%
%
Hence, in those functions, we just substitute $\beta_{34} \to \displaystyle
\ln\left(\frac{1+\beta_t}{1-\beta_t}\right)$. And then, wherever $\beta_{34}$
appears in a coefficient, we use Eq.~(\ref{eq:beta34TL}). At the end, we obtain
%
%
\allowdisplaybreaks
\begin{align}
  \label{eq:vel-dep-ad-explicit}
  \gamma^{\text{cusp}}_0(\beta_t) &= 
  -2\, \frac{1+\beta_t ^2}{\beta_t}
  \left\{\ln \left(\frac{1-\beta_t}{1+\beta_t}\right) + i \pi\right\}\,,
  \\
  \gamma^{\text{cusp}}_1(\beta_t) &= 
  \frac{1}{9 \beta_t ^2}\,
  \Bigg\{30\, C_A\, \pi ^2 \beta_t\,  (1-\beta_t)^2 +
  \ln \left(\frac{1-\beta_t}{1+\beta_t}\right) 
  \bigg[72\, C_A\, \beta_t \left(1+ \beta_t ^2\right) 
  \ln \left(\frac{4 \beta_t }{(1+ \beta_t)^2}\right)
  \nonumber \\
  &
  \hspace{30pt}
  +\left(1+\beta_t ^2\right)
  \Big[C_A \big(3\, \pi ^2 (\beta_t  (5 \beta_t -8)+5)-134\, \beta_t \big)+
  40\,  T_F n_f \beta_t \Big]
  \nonumber \\
  &
  \hspace{30pt}
  -6\, C_A (1-\beta_t)^2 \ln \left(\frac{1-\beta_t}{1+\beta_t}\right) 
  \Big[6\, \beta_t +\left(1+\beta_t ^2\right)
  \ln \left(\frac{1-\beta_t}{1+\beta_t}\right)\Big]\bigg]
  \nonumber \\
  & 
  \hspace{30pt}
  +18 C_A \Bigg(\left(1+\beta_t ^2\right)^2 
  \text{Li}_3\left[\left(\frac{1-\beta_t}{1+\beta_t}\right)^2\right]
  -\left(1-\beta_t ^2\right)^2 \zeta (3)
  \nonumber \\
  &
  \hspace{30pt}
  -\left(1+\beta_t ^2\right) 
  \Big[\left(1+\beta_t ^2\right) \ln
  \left(\frac{1-\beta_t}{1+\beta_t}\right)- 2 \beta_t \Big]
  \text{Li}_2\left[\left(\frac{1-\beta_t}{1+\beta_t}\right)^2\right]
  \Bigg)
  \Bigg\}
  \nonumber \\
  &
  + \frac{i \pi}{9 \beta_t ^2}
  \Bigg\{72\, C_A\, \beta_t  \left(1+\beta_t ^2\right) 
  \ln \left(\frac{4 \beta_t }{(1+\beta_t)^2}\right)
  -18\, C_A\, \left(1+\beta_t ^2\right)^2
  \text{Li}_2\left[\left(\frac{1-\beta_t}{1+\beta_t}\right)^2\right]
  \nonumber \\
  &
  \hspace{30pt}
  -18\, C_A\, (1-\beta_t)^2\ln 
  \left(\frac{1-\beta_t}{1+\beta_t}\right) \left[4 \beta_t
  +\left(1+\beta_t ^2\right) \ln
  \left(\frac{1-\beta_t}{1+\beta_t}\right)\right]
  \nonumber \\
  &
  \hspace{30pt}
  + \left(1+ \beta_t ^2\right) \Big[C_A  \left(3\,\pi ^2 
  \left(1+ \beta_t ^2\right)-134\, \beta_t \right) +40\,  T_F n_f\, \beta_t\Big]
  \Bigg\}
  \,,
\end{align}
and
\begin{align}
  g(\beta_t) &=
  -\frac{1}{\beta_t}
  \Bigg\{\left(1+\beta_t ^2\right) 
  \ln \left(\frac{4 \beta_t }{(1+\beta_t)^2}\right) 
  \ln \left(\frac{1-\beta_t }{1+\beta_t}\right)+
  (1-\beta_t)^2 \left(\frac{5 \pi ^2}{12}- 
  \ln\left(\frac{1-\beta_t }{1+\beta_t}
  \right)\right)
  \nonumber \\
  & +\frac12 \left(1+\beta_t ^2\right)
  \Li_2\left[\left(\frac{1-\beta_t}{1+\beta_t}\right)^2\right]
  \Bigg\}
  + \frac{i\pi}{\beta_t} 
  \left\{(1-\beta_t)^2 \ln \left(\frac{1-\beta_t }{1+\beta_t}\right)
  -\left(1+\beta_t ^2\right) \ln
  \left(\frac{4 \beta_t }{(1+\beta_t)^2}\right)\right\}\,.
\end{align}

The functions given explicitly in this Appendix, together with the result for
the NLO soft function, Eq.~(\ref{eq:NLOSFres}), allow for the complete
determination of all the poles and all $L_\perp$-dependent terms of the NNLO
soft function, following the discussion in Section~\ref{sec:renormalization}.

\bibliographystyle{unsrt}

\end{document}